\documentclass[useAMS,usenatbib]{mn2e}
\usepackage{natbib}
\usepackage{subfigure}
\usepackage{gensymb}
\usepackage{amssymb}
\usepackage{amsmath}
\usepackage{wasysym}
\usepackage{graphicx}
\usepackage{color}
\usepackage{textcomp}
\usepackage{mathrsfs}

\DeclareMathSymbol{\varOmega}{\mathord}{letters}{"0A}
\DeclareMathSymbol{\varSigma}{\mathord}{letters}{"06}

\DeclareMathSymbol{\varPsi}{\mathord}{letters}{"09}

\newcommand{\lsim}{\mathrel{\rlap{\lower4pt\hbox{\hskip1pt$\sim$}}
   \raise1pt\hbox{$<$}}}                
\newcommand{\gsim}{\mathrel{\rlap{\lower4pt\hbox{\hskip1pt$\sim$}}
   \raise1pt\hbox{$>$}}}                

\begin{document}
\title[Resonant capture of dust]{Capture and evolution of dust in planetary mean-motion resonances: a fast, semi-analytic method for generating resonantly trapped disk images}
\author[Shannon et al.]{Andrew Shannon$^{1}$, Alexander J Mustill$^{2,3}$, \& Mark Wyatt$^{1}$ \\
$^{1}$Institute of Astronomy, University of Cambridge, Madingley Road, Cambridge, UK, CB3 0HA \\
$^{2}$Departamento de F\'isica Te\'orica, Universidad Aut\'onoma de Madrid, Cantoblanco, 28049 Madrid, Spain \\
$^{3}$Lund Observatory, Department of Astronomy and Theoretical Physics, Lund University, Box 43, SE-221 00 Lund, Sweden}

\maketitle

\begin{abstract}
Dust grains migrating under Poynting-Robertson drag may be trapped in mean-motion resonances with planets.  Such resonantly trapped grains are observed in the solar system.  In extrasolar systems, the exozodiacal light produced by dust grains is expected to be a major obstacle to future missions attempting to directly image terrestrial planets.  The patterns made by resonantly trapped dust, however, can be used to infer the presence of planets, and the properties of those planets, if the capture and evolution of the grains can be modelled.  This has been done with N-body methods, but such methods are computationally expensive, limiting their usefulness when considering large, slowly evolving grains, and for extrasolar systems with unknown planets and parent bodies, where the possible parameter space for investigation is large.  In this work, we present a semi-analytic method for calculating the capture and evolution of dust grains in resonance, which can be orders of magnitude faster than N-body methods.  We calibrate the model against N-body simulations, finding excellent agreement for Earth to Neptune mass planets, for a variety of grain sizes, initial eccentricities, and initial semimajor axes.  We then apply the model to observations of dust resonantly trapped by the Earth.  We find that resonantly trapped, asteroidally produced grains naturally produce the `trailing blob' structure in the zodiacal cloud, while to match the intensity of the blob, most of the cloud must be composed of cometary grains, which owing to their high eccentricity are not captured, but produce a smooth disk.
\end{abstract}

\section{Introduction}
Small dust grains are released into the Solar system in collisions between asteroids and by the outgassing of comets.  If they are not so small as to be unbound, they will spiral towards the Sun due to Poynting-Robertson (PR) drag.  These dust grains reflect, absorb, and re-radiate the solar light, producing the zodiacal light and zodiacal thermal emission.  Around other stars, similar processes may produce exozodiacal light.  Such exozodiacal light is expected to be a significant barrier to future missions attempting to directly image terrestrial planets if the level of dust is greater than $\sim 10$~times that of the Solar system \citep{1998exdu.work.....B,2006ApJ...652.1674B}, a level that may be present in many, perhaps even most, systems \citep{2013MNRAS.433.2334K}.

Small grains undergoing PR drag can become trapped in mean motion resonances with planets \citep{1975Icar...25..489G,1982Icar...51..633G}.  This produces clumpy circumstellar rings of grains, which are known to exist around the Sun due to trapping by the Earth \citep{1994Natur.369..719D,1995Natur.374..521R, 2010Icar..209..848R}, and Venus \citep{2007A&A...472..335L,2013Sci...342..960J}.  The dust detector aboard New Horizons may soon detect dust in resonance with Neptune \citep{2014AJ....147..154V}.  As similar processes might be expected to occur in extrasolar planetary systems, this raises the interesting possibility that structures in extrasolar debris disks are created by this phenomenon, and can be used to infer the presence of planets and their properties.

In the outer part ($>> 10~\rm{au}$) of a stellar system, debris disks can be spatially resolved by facilities and instruments such as HST \citep{2005Natur.435.1067K}, Subaru \citep{2011ApJ...743L...6T}, SMA \citep{2011ApJ...740...38H}, VLT \citep{2010A&A...524L...1B}, Herschel \citep{2014ApJ...784..148D,2014ApJ...780...97M}, Gemini \citep{2014A&A...567A..34W}, and ALMA \citep{2012ApJ...750L..21B}.  This could allow the inference of resonantly trapped dust from the spatial distribution \citep{2003ApJ...588.1110K,2003ApJ...598.1321W}, although in these bright disks collisions are far more important than PR drag \citep{2005A&A...433.1007W}, which may wash out such patterns \citep{2010AJ....140.1007K}.  Possible detections of clumpy structures which may be associated with planetary resonances have already been reported \citep{1998Natur.392..788H,2005ApJ...619L.187G,2009ApJ...690L..65C}.  In the inner parts ($ << 10~\rm{au}$) of systems, bright dust disks are much rarer \citep{2013A&A...550A..45F}, and transient in nature \citep{2007ApJ...658..569W}.  However, bright outer disks may leak dust inwards due to PR drag \citep{MENINPEN}, where resonant capture may occur.  PR drag can also be relatively more important in dimmer disks.  The Keck Nulling Interferometer \citep{2011ApJ...734...67M,2013ApJ...763..119M} and the Large Binocular Telescope Interferometer \citep{2013AAS...22140306H,2014arXiv1412.0675K} can allow for the detection of these dimmer disks.  These interferometers detect the flux leak from a dust disk through an interference pattern.  With repeated measurements of the same system, rotating structures, such as resonantly trapped dust, may be able to be inferred.

To model resonantly trapped dust, various groups have used N-body simulations of dust grains.  \citet{1994Natur.369..719D} simulated the inspiral of 912 dust grains with diameters of 12 $\mu m$~from the asteroid belt, and showed that the pattern produced by the capture and persistence of those grains in resonance is compatible with that observed by the Infrared Astronomical Satellite.  \citet{2005ApJ...625..398D} integrated 500 dust grains under PR drag for 120 different systems, generating a catalogue of images for different planet mass, planet eccentricity, dust size, dust parent body eccentricity, and dust parent body semimajor axis.  \citet{2008ApJ...686..637S} required a 420 core cluster to perform simulations of 5000 particles spiralling past a planet under PR drag for 120 different parameter combinations.  Similarly, \citet{2014ApJ...780...65R} used a suite of 160 simulations to produce a fitting formula for disk appearances.  In these cases, the long computation times of N-body simulations makes their use to generate potential disk images on a system-by-system case somewhat impractical, hence the aforementioned groups have generated catalogues of disk images and fitting formulae to be matched with disks once discovered.  

These high computation times are set by the need to resolve individual orbital periods of the planet(s) and dust grains.  However, the evolution under PR drag and the evolution in resonance occur on much longer time-scales than the orbital period.  Thus, in this paper we develop a model of resonant capture and evolution, with which dust grain trajectories can be calculated on the PR drag time-scale, rather than the orbital time-scale, resulting in orders of magnitude decreases in the computation time.

In \textsection \ref{sec:singleparticlephysics}, we outline the physics of a single dust grain undergoing PR drag around a star where it may be captured into mean motion resonance with a planet.  In \textsection \ref{sec:nbodyintegrations} we perform a suite of $N$-body simulations, and use them to validate and calibrate our model.
In \textsection \ref{section:makedisk}, we describe how to use the equations of evolution to produce disk images, and compare them to the N-body simulations images.  In \textsection \ref{sec:earthicandisk}, we compare the output of the model to measurements of the structure of Earth's resonantly trapped ring.  In \textsection \ref{sec:conclusions}, we discuss other possible applications of the model.

\section{Single Particle Physics}
\label{sec:singleparticlephysics}

Consider a star of mass $m_*$, orbited by a dust grain, the orbit of which is described by six parameters: semimajor axis $a$, eccentricity $e$, inclination $i$, longitude of pericentre ($\varpi$), ascending node ($\Omega$), and mean anomaly ($M$).  The orbital evolution of the dust grain will be dictated by the parameter $\beta_{\rm{PR}}$, the ratio of the force of radiation pressure to the gravitational force.  If the star is also orbited by a planet of mass $m_p$, and orbital parameters $a_p$, $e_p$, $i_p$, $\varpi_p$, $\Omega_p$, $M_p$, resonant interactions between the planet and dust grain may also be important to its orbital evolution.  In this section we consider these effects analytically.  These expectations will then be compared to simulations in \textsection \ref{section:results}.

\subsection{Poynting-Robertson Drag}
\label{sec:prdrag}

Assuming dust grains to be spherical, they experience a force from the stellar radiation of the form

\begin{equation}
 \vec{F}_{\rm{radiation}} = \beta_{\rm{PR}} \left|\vec{F}_{\rm{gravity}}\right|\left[\left(1-2\frac{\dot{r}}{c}\right)\hat{r} -\left(\frac{r\dot{\theta}}{c}\right)\hat{\theta}\right],
 \label{eq:radiationforce}
\end{equation}
where $\beta_{\rm{PR}}$~is the ratio of the force of radiation pressure to gravity acting on the grain, which will depend on its size, shape and composition.  For example, a spherical blackbody grain of size $s$~and density $\rho$~has
\begin{eqnarray}
\label{eq:stobeta}
\beta_{\rm{PR}} = \frac{3 L_{*}}{8\pi c\rho Gm_* s},
\end{eqnarray}
where $L_*$~is the stellar luminosity.

Averaged over an orbit, this causes dust particles to evolve in semimajor axis and eccentricity as \citep{1950ApJ...111..134W}
\begin{eqnarray} \nonumber
 \left.\frac{da}{dt}\right|_{\rm{PR}} &=& -\frac{Gm_{*}}{a}\frac{\beta_{\rm{PR}}}{c} \frac{\left(2+3e^{2}\right)}{\left(1-e^{2}\right)^{\frac{3}{2}}} \\ 
 \label{eq:dadtpr}
  &=& -0.624 \frac{\rm{au}}{\rm{kyr}} \beta_{\rm{PR}} \frac{m_*}{m_\odot} \frac{1 \rm{au}}{a}\frac{\left(2+3e^{2}\right)}{\left(1-e^{2}\right)^{\frac{3}{2}}} ,
\end{eqnarray}
\begin{eqnarray} \nonumber
 \left.\frac{de}{dt}\right|_{\rm{PR}} &=& -\frac{5}{2}\frac{Gm_{*}}{a^{2}}\frac{\beta_{\rm{PR}}}{c} \frac{e}{\left(1-e^{2}\right)^{\frac{1}{2}}} \\
  &=& -1.56 \rm{kyr}^{-1} \beta_{\rm{PR}} \frac{m_*}{m_\odot} \left(\frac{1 \rm{au}}{a}\right)^2\frac{e}{\left(1-e^{2}\right)^{\frac{1}{2}}} .
 \label{eq:dedtpr}
\end{eqnarray}
 Integrating equation \ref{eq:dadtpr}~from an initial semimajor axis $a_0$~and eccentricity $e_0 = 0$, a dust particle will reach the star in time 
\begin{eqnarray} \nonumber
\tau_{\rm{PR}} &=& \frac{a_0^2c}{4Gm_*\beta_{\rm{PR}}} \\
 &=& 0.40 \left(\frac{a_0}{ \rm{1 au}}\right)^2 \frac{m_{\odot}}{m_*}\frac{1}{\beta_{\rm{PR}}} \rm{kyrs}.
\label{eq:lifetimepredict}
\end{eqnarray}
Particles with higher $e_0$~reach the star more quickly, but the problem does not lend itself to a compact analytic form.

During the orbital evolution, there is a constant of integration which can be obtained from equations \ref{eq:dadtpr} and \ref{eq:dedtpr} \citep{1950ApJ...111..134W}
\begin{equation}
 K = a\left(1-e^2\right)e^{-\frac{4}{5}}.
 \label{eq:prk}
\end{equation}

The value of $K$~is constant over many orbits, but small variations occur during an orbit.  Differentiating equation \ref{eq:prk}~with respect to time, substituting in $\dot{a}$~and $\dot{e}$~from \citet{1979Icar...40....1B}, and taking the expression to first order in $e$, we find
\begin{equation}
 \frac{1}{K}\frac{dK}{df} \approx \frac{8}{5}\frac{\beta_{\rm{PR}}}{e}\frac{v_{\rm{kepler}}}{c}\cos{f},
\end{equation}
and thus a libration in the constant $K$~with a magnitude:
\begin{equation}
 \label{eq:kevolves}
 \frac{\Delta K}{K} \approx \frac{8}{5}\frac{\beta_{\rm{PR}}}{e}\frac{v_{\rm{kepler}}}{c}
\end{equation}
over the course of an orbit.  

\subsection{Mean motion resonance}
\label{sec:res}

A dust grain can be in a $k$-th order mean motion resonance with a planet if their mean motions are close to the ratio $j:j+k$, where j and k are positive integers.  From the mean longitudes $\lambda = M + \varpi$~we construct a resonant angle
\begin{equation}
 \label{eq:varphidefined}
 \varphi = j\lambda_{\rm{p}} - \left(j+k\right)\lambda + k \varpi,
\end{equation}
where $\lambda$~and $\lambda_p$~are the mean longitude of the dust grain and the planet respectively.  For our purposes, $\varpi$~evolves slowly compared to $\lambda$~and $\lambda_{\rm{p}}$, so the $j:j+k$~resonance occurs at
\begin{equation}
 \label{eq:ares}
 a_{j:j+k} \approx \left(1-\beta_{\rm{PR}}\right)^{\frac{1}{3}}\left(\frac{j+k}{j}\right)^{\frac{2}{3}}a_{\rm{p}},
\end{equation}
which differs from the usual expression due to the effects of radiation pressure.  Here the match is only approximate as the $k\varpi$~term allows for a small offset in $a$, and there is a small libration of $a$~about the nominal resonance location corresponding to the libration in $\varphi$.  While the dust grain is near resonance, the perturbations from the planet can be approximated by taking the terms in the disturbing function that are first order in eccentricity, and using Lagrange's equations for the evolution of the orbital elements to give \citep{2000ssd..book.....M}
\begin{equation}
 \left.\frac{da}{dt}\right|_{\rm{j:j+k}} = -2j\mathcal{C}_r ae^k\sin{\varphi},
 \label{eq:dadtres}
\end{equation}
\begin{equation}
 \left.\frac{de}{dt}\right|_{\rm{j:j+k}} = k\mathcal{C}_r e^{k-1}\sin{\varphi},
 \label{eq:dedtres}
\end{equation}
where
\begin{eqnarray}
  \label{eq:crdefined}
  \mathcal{C}_r &=& - \frac{G m_{\rm{p}}}{n a^{2}a_{\rm{p}}}\left(\alpha f_d\left(\alpha\right) + f_i\left(\alpha\right)\right) \\
  & = & \frac{-\sqrt{G}m_p\left(\alpha f_d\left(\alpha\right) + f_i\left(\alpha\right)\right)}{\sqrt{m_*}\left(\frac{j+k}{j}\right)^{\frac{1}{2}}\left(1-\beta_{\rm{PR}}\right)^{\frac{2}{3}}a_p^2},
\end{eqnarray}
where $n = \sqrt{Gm_*\left(1-\beta_{\rm{PR}}\right)a^{-3}}$, $\alpha = a_p/a \approx \left(1-\beta_{\rm{PR}}\right)^{-1/3}\left(\left(j+k\right)/j\right)^{-2/3}$, $f_d\left(\alpha\right)$~is the direct term of the disturbing function, and $f_i\left(\alpha\right)$~is the indirect term of the disturbing function.

Neglecting PR drag, and using equation \ref{eq:ares} for the dust grain's semimajor axis, taking the second time derivative of $\varphi$, neglecting the contribution from $\ddot{\lambda_p}$~and $\ddot{\varpi}$~terms, one finds $\varphi$~evolves as 
\begin{eqnarray}
 \label{eq:ddotvarphi}
 \ddot{\varphi} &=& 3j^2\mathcal{C}_rne^k\sin{\varphi}, \\
 & = & 3\frac{j^3}{j+k}\sqrt{Gm_*\left(1-\beta_{\rm{PR}}\right)}a_p^{-1}\mathcal{C}_re^k\sin{\varphi}.
\end{eqnarray}
As $\mathcal{C}_r < 0$, this is the equation for a pendulum with frequency
\begin{eqnarray}
 \label{eq:pendulum}
 \omega^2 &=& 3j^2\left|\mathcal{C}_r\right|ne^k \\
& = & 3\frac{j^3}{j+k}\sqrt{Gm_*\left(1-\beta_{\rm{PR}}\right)}a_p^{-1}\left|\mathcal{C}_r\right|e^k.
\end{eqnarray}
Thus, $\varphi$~can librate or circulate, with libration in the case that the dust grain is in resonance with the planet.  This is similar to the case without radiation pressure, but $\left|\mathcal{C}_r\right|$~contains dependence on $\beta_{\rm{PR}}$, and $\alpha$, which also depends on $\beta_{\rm{PR}}$.

As noted, for a fixed $a_p$~there is a small range of possible $a$~for the dust grains for which it is still possible to construct a resonant angle that can librate.  Defining $\mathcal{C}_r$~as equation \ref{eq:crdefined}, and otherwise following the derivation from \citet{2000ssd..book.....M}, the maximum libration width of a particle in a $k = 1$~resonance is 
\begin{equation}
 \frac{\Delta a_{\rm{max}}}{a_{j:j+1}} = \pm \left(\frac{16}{3}\frac{\left|\mathcal{C}_r\right|}{n}e\right)^{\frac{1}{2}}\left(1 + \frac{1}{27j^2e^{3}}\frac{\left|\mathcal{C}_r\right|}{n}\right)^{\frac{1}{2}} - \frac{2}{9je}\frac{\left|\mathcal{C}_r\right|}{n}
 \label{eq:deltaa1}
\end{equation}
and for a $k = 2$~resonance the width is 
\begin{equation}
 \frac{\Delta a_{\rm{max}}}{a_{j:j+2}} = \pm \left(\frac{16}{3}\frac{\left|\mathcal{C}_r\right|}{n}e^{2}\right)^{\frac{1}{2}}.
 \label{eq:deltaa2}
\end{equation}
We do not investigate resonances with higher $k$, as our N-body simulations produce very few captures at $k > 1$.  

\subsection{Resonance capture}
\label{sec:rescap}
As particles migrate past a planet's resonances due to PR drag, the question of whether they become caught in resonance is deterministic. However, as it can depend on the relative orbital phases of the planet and dust grain as resonance is approached, it can be treated probabilistically \citep{1982CeMec..27....3H,2006MNRAS.365.1367Q}.  \citet{2011MNRAS.413..554M} use a simple Hamiltonian model which uses the lowest order term of the disturbing function (as we did in \textsection \ref{sec:res}) to calculate the capture rates and initial libration widths for massless particles that encounter first and second order mean motion resonances during migration, as well as the eccentricity kicks during resonant crossings that do not result in capture.  While the model is developed in the context of the capture of planetesimals by a migrating planet, the calculations depend only on the rate of change of the bodies' orbital separation, and so can be applied to the capture of migrating dust particles by a fixed planet.  However, that model assumes $\beta_{\rm{PR}} = 0$, which requires a slight correction for our context.  This is because when $\beta_{\rm{PR}} > 0$, the reduction in effective stellar mass seen by the dust grains due to radiation pressure means that resonances occur closer to the planet than when particles do not feel radiation forces, and therefore the strength of a given resonance is different for particles of different $\beta_{\rm{PR}}$.

\begin{figure}
  \includegraphics[width=0.5\textwidth,trim = 0 0 110 340, clip]{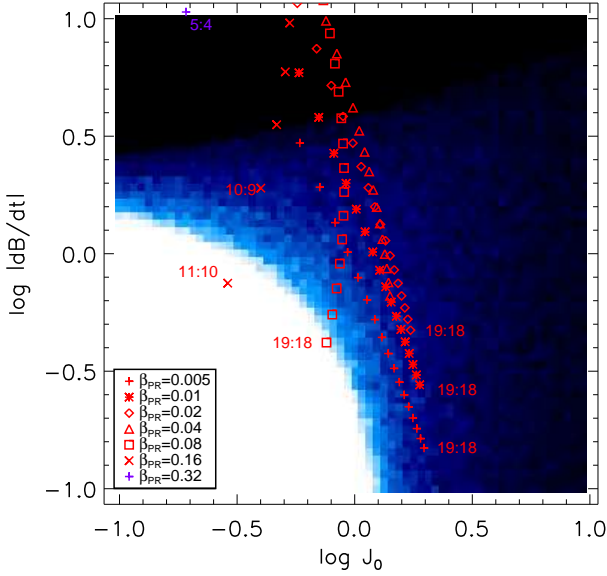}
  \caption{Capture probabilities calculated by integrating the Hamiltonian model (white = 100\%, black = 0\%), including eccentricity damping at $k=5\times10^{-5}$. The x-axis measures a dimensionless ``eccentricity'' $\left( J \propto e^2\right)$, and the y-axis a dimensionless migration rate $\left(dB/dt \propto da/dt\right)$.  Overplotted are the dimensionless momentum and migration rate corresponding to the comparison integrations described in the text.}
  \label{fig:grid}
\end{figure}

We present details of the required changes to the Hamiltonian model, including changes to the resonant strengths and eccentricity damping effects, in the Appendix.  Based on these changes, we now predict capture probabilities, as well as initial libration widths and eccentricity kicks during non-capture resonance crossings, for real scenarios. We adopt a stellar mass of $1M_\odot$, planet mass of $1m_\oplus$, and planet semi-major axis of 1 au. We integrate trajectories of dust particles with $\beta_\mathrm{PR}=0.005,0.01,0.02,0.04,0.08,0.16$ and $0.32$, with initial eccentricities of $0.01$. Encounters with resonances from the 2:1 up to the 19:18 are integrated, unless said resonance lies inside the planet's orbit. 
In figure \ref{fig:grid}, we show the capture probabilities for first-order resonances as a function of dimensionless parameters $J \left(\propto e^2\right)$~and $dB/dt \left(\propto da/dt\right)$.\footnote{We use $B$~for the parameter denoted $\beta$~in MW11, to avoid confusion.}  In this figure we overplot the parameters corresponding to capture in the first-order resonances of a $1 M_\oplus$~planet for dust particles with $\beta_{\rm{PR}} = 0.005$~up to $\beta_{\rm{PR}} = 0.32$.  We can see that many of these resonances lie in the regime of possible but not certain capture, meaning that a population of drifting dust grains will populate a range of resonances.

\subsection{Maximum j of capture}
\label{sec:maxjguess}
\citet{1980AJ.....85.1122W} showed that first order resonances overlap when they are at semimajor axes less than $\sim 1.3\left(m_p/m_*\right)^{2/7}a_p$~greater than that of the planet.  Bodies move chaotically between the overlapping resonances, which makes them unstable.  Consequently, we would expect such resonances to not capture migrating dust grains.  Here we employ a different approach to that of \citet{1980AJ.....85.1122W} to calculate the width of this chaotic zone, but which recovers their result, at the same time accounting for the effects of radiation pressure.

In the inertial frame of a planet with semimajor axis $a_p$, a dust particle with radiation pressure co-efficient $\beta_{\rm{PR}}$~on a circular orbit with semimajor axis $a_p\left(1+\epsilon\right)$~approaches from infinity with velocity
\begin{eqnarray}
 v_{\infty}
  & \approx & \sqrt{\frac{Gm_*}{a_p}}\left(1-\sqrt{1-\beta_{\rm{PR}}}\left(1 - \frac{\epsilon}{2}\right)\right) ,
\end{eqnarray}
assuming $\epsilon << 1$, an assumption we make throughout this derivation.  The particle's path is deflected by an angle $\theta$, given by
\begin{eqnarray}
 \theta & \approx & 2\frac{m_p}{m_*}\frac{1}{\epsilon}\left(1-\sqrt{1-\beta}\left(1-\frac{\epsilon}{2}\right)\right)^{-2}
\end{eqnarray}
assuming the angle is small.  Converting back to the frame of the star, and considering the final radial velocity ($v_r$) and azimuthal velocity ($v_{\theta}$), 
the change in specific energy of the dust grain is
\begin{eqnarray}
 \delta E 
 & \approx & \frac{2}{\epsilon^2}\frac{Gm_*}{a_p} \left(\frac{m_p}{m_*}\right)^2 \left(1-\sqrt{1-\beta}\left(1-\frac{\epsilon}{2}\right)\right)^{-3} .
\end{eqnarray}


For an initial $\epsilon_1$ and a final $\epsilon_2$, the specific orbital energy change is also
\begin{eqnarray}
 \delta E 
 & \approx & \frac{Gm_*\left(1-\beta\right)}{2a_p}\left(\epsilon_1 - \epsilon_2\right) ,
\end{eqnarray}
and equating the two energies yields
\begin{equation}
 \label{eq:deltaep1}
 \epsilon_1 - \epsilon_2 = \frac{4}{1-\beta_{\rm{PR}}} \left(\frac{m_p}{m_*}\right)^2 \frac{1}{\epsilon^2}\left(1-\sqrt{1-\beta_{\rm{PR}}}\left(1-\frac{\epsilon}{2}\right)\right)^{-3} ,
\end{equation}
which assumes $\epsilon_1 - \epsilon_2 << \epsilon$.  

After one conjunction, the planet and dust grain move apart in longitude, meaning that 
the next conjunction will happen at 
\begin{equation}
 \lambda_c =  2\pi\left(1-\sqrt{1-\beta_{\rm{PR}}}\left(1-\frac{3}{2}\epsilon\right)\right)^{-1}.
\end{equation}

If we assume that interactions between the dust grain and the planet will be dephased if the change in the longitude of conjunction $\delta \lambda_c$~is greater than some amount $\Delta$,
this yields a constraint that
\begin{eqnarray} \nonumber
 \label{eq:critjfinal}
 \frac{12\pi}{\sqrt{1-\beta_{\rm{PR}}}} \left(\frac{m_p}{m_*}\right)^2\frac{1}{\epsilon^2}\left(1-\sqrt{1-\beta_{\rm{PR}}}\left(1-\frac{\epsilon}{2}\right)\right)^{-3} \\ 
 \times \left(1-\sqrt{1-\beta_{\rm{PR}}}\left(1-\frac{3}{2}\epsilon\right)\right)^{-2} \geq \Delta ,\\ \nonumber
\end{eqnarray}
which reduces to the scaling of \citet{1980AJ.....85.1122W} in the limit $\beta_{\rm{PR}} \rightarrow 0$~of 

\begin{equation}
 \epsilon \leq \left(\frac{128\pi}{3\Delta}\right)^{\frac{1}{7}}\left(\frac{m_p}{m_*}\right)^\frac{2}{7} .
\end{equation}
A reasonable expectation is that $\Delta = 2\pi$, which results in a leading coefficient of $\approx 1.5$, slightly larger than the value obtained by \citet{1980AJ.....85.1122W}, though numerically, \citet{1989Icar...82..402D} found this coefficient to be $\approx 1.5$, and \citet{2009ApJ...693..734C}~preferred a coefficient of $\approx 2.0$~to fit their simulations of Fomalhaut's disk.  Recent results have also shown some dependence on system age and collision time \citep{2014arXiv1410.7784N,2014arXiv1411.1378M}.  For our own simulations, we will fit this coefficient in \textsection \ref{sec:maxjfit}.

\subsection{Resonant evolution}
\label{sec:resevolve}

In the absence of drag a dust grain resonating with an interior planetary perturber librates around $\varphi = \pi$~(Equation \ref{eq:ddotvarphi}).  With the addition of PR drag, the centre of libration changes.  Momentum and energy loss to PR drag must be offset by a gain of momentum and energy from the resonant interaction.   Equating equations \ref{eq:dadtpr} and \ref{eq:dadtres}, we can estimate that $\varphi$~librates around $\varphi_0$, given by
\begin{eqnarray}
 \label{eq:varphi0}
 \varphi_0 &=& \sin^{-1}{\left(\frac{n_{j:j+k}}{\mathcal{C}_r}\frac{v_{\rm{kepler,j:j+k}}}{c}\frac{\beta_{\rm{PR}}}{2je^{k}}\frac{\left(2+3e^2\right)}{\left(1-e^2\right)^{\frac{3}{2}}}\right)}, \\ \nonumber
 &=& \sin^{-1}\left(\frac{\left(1-\beta_{\rm{PR}}\right)^{\frac{1}{3}}}{2e^k\mathcal{C}_r}\frac{\beta_{\rm{PR}}}{c}\frac{2+3e^2}{\left(1-e^2\right)^{\frac{3}{2}}}\frac{Gm_*}{a_p^2}\frac{j^{\frac{1}{3}}}{\left(j+k\right)^{\frac{4}{3}}}\right) .
\end{eqnarray}
Note that this expression does not hold for the $2:1$ resonance, or other resonances which exhibit asymmetric libration for which higher order terms in the disturbing function are important \citep{1958AJ.....63..443M}.  For those resonances there are two possible centres of libration with differing capture probabilities \citep{2005ApJ...619..623M}.  In this case, a numerically calibrated prescription from \citet{2003ApJ...598.1321W}~may be employed, which we do in \textsection \ref{section:makedisk}.  They found that the centre of libration is given by $\cos{\varphi_0^{2:1}} = 0.39 - 0.061/e$.  Using the PR-drag migration rate of equation \ref{eq:dadtpr}, their results would predict that the relative chance of capture into the lower $2:1$~resonance is $1/2 - 0.85\sqrt{Gm_*\beta_{\rm{PR}}/c}\left(m_pa\right)^{-0.25}\left(2+3e^2\right)^{0.5}\left(1-e^2\right)^{-0.75}$.
  

While the dust particle is in resonance, it undergoes only small variations in $a$.  The eccentricity, however, may evolve substantially.  Taking the value of $\varphi_0$~from equation \ref{eq:varphi0} and substituting it into equation \ref{eq:dedtres}~gives an average $\dot{e}$~from resonant interaction.  Combined with the $\dot{e}$~from PR drag (equation \ref{eq:dedtpr}), the eccentricity evolves according to
\begin{equation}
 \frac{de}{dt} = \frac{1}{2}\frac{Gm_*}{a_{j:j+k}^2}\frac{\beta_{\rm{PR}}}{c}\left(\frac{k\left(2+3e^2\right)}{\left(j+k\right)e\left(1-e^2\right)^{\frac{3}{2}}}-\frac{5e}{\sqrt{1-e^2}}\right) .
\end{equation}
Expanding to second order in $e$~and solving yields \citep{1993Icar..104..244W}
\begin{equation}
 e\left(t\right) = \sqrt{\frac{2k}{5\left(j+k\right)}\left(1-e^{-\frac{t}{\tau_e}}\right)} .
 \label{eq:wandsgrowth}
\end{equation}
Note that the expression in \citet{1993Icar..104..244W} has a typographical error: it is missing the square root.  Here the growth time of the particle's eccentricity is
\begin{equation}
 \tau_e = 0.2 a_{j:j+k}^2c/\left(Gm_*\beta_{\rm{PR}}\right).
\end{equation}

\section{N-body integrations}

\label{sec:nbodyintegrations}

\subsection{Single particle example: Evolution under PR drag}

\label{sec:onesim}

\begin{figure*}
  \centering
  \subfigure{\includegraphics[width=0.48\textwidth,trim = 50 50 0 50, clip]{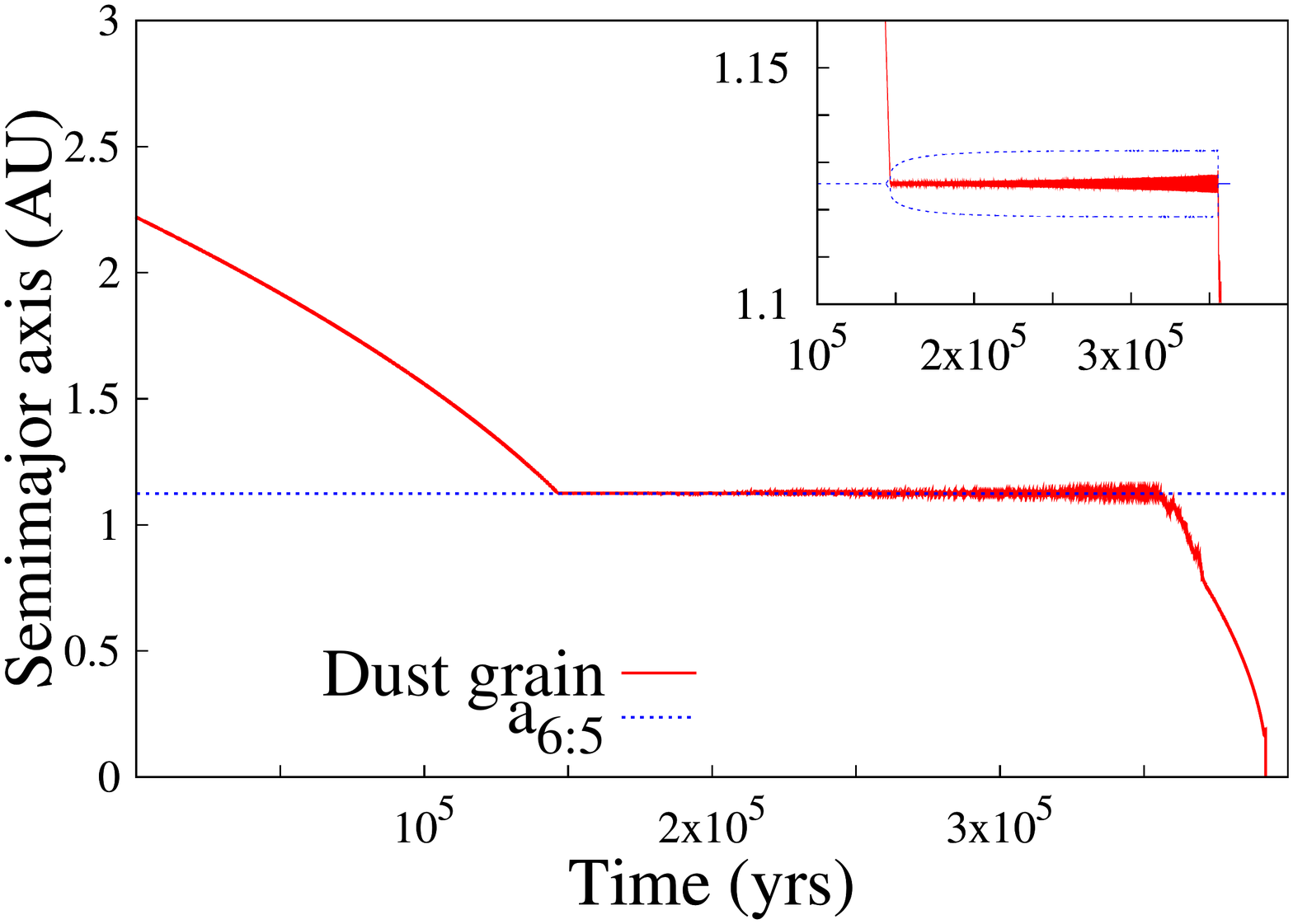}}
  \subfigure{\includegraphics[width=0.48\textwidth,trim = 50 50 0 50, clip]{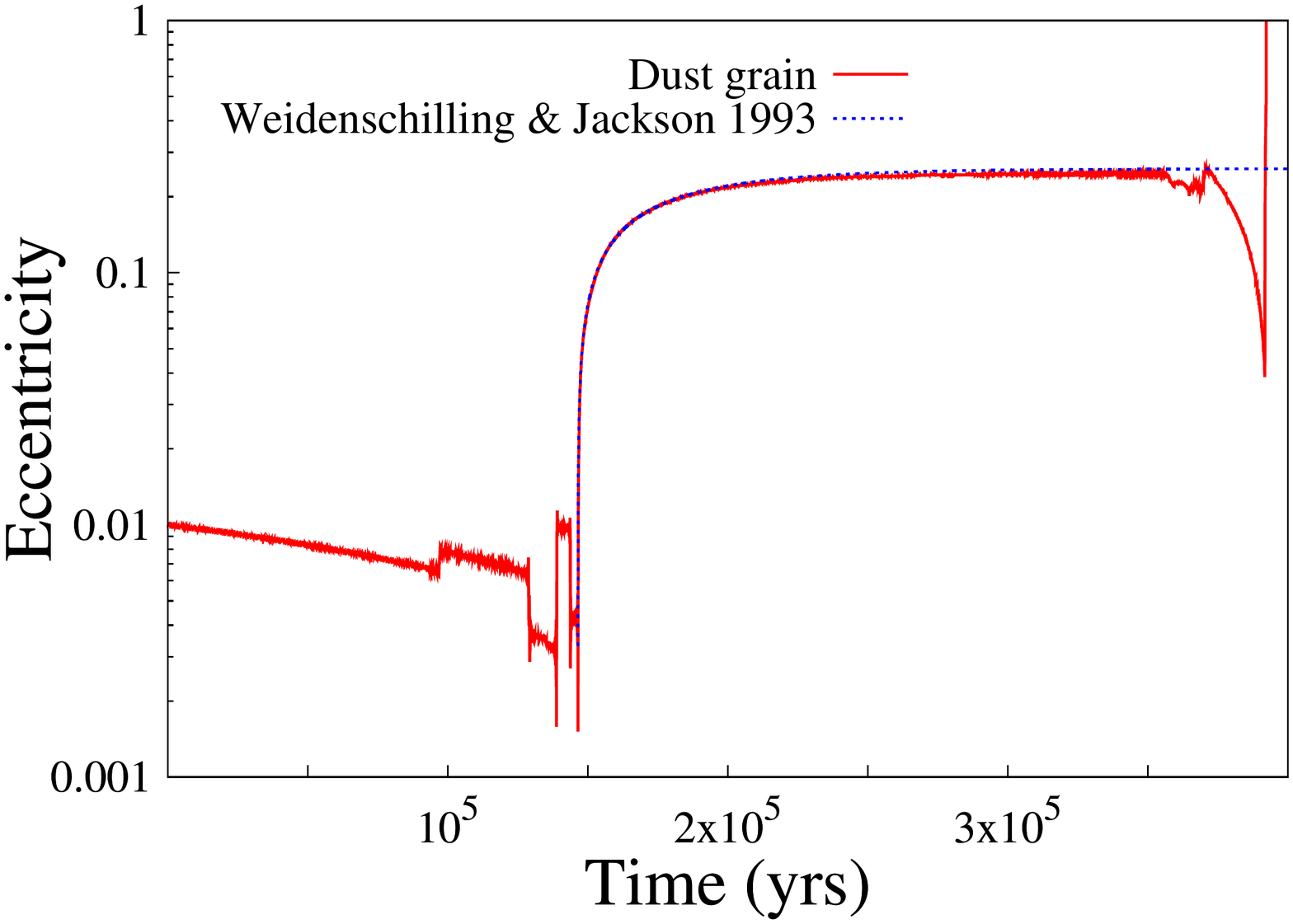}} \\
  \subfigure{\includegraphics[width=0.48\textwidth,trim = 50 50 0 50, clip]{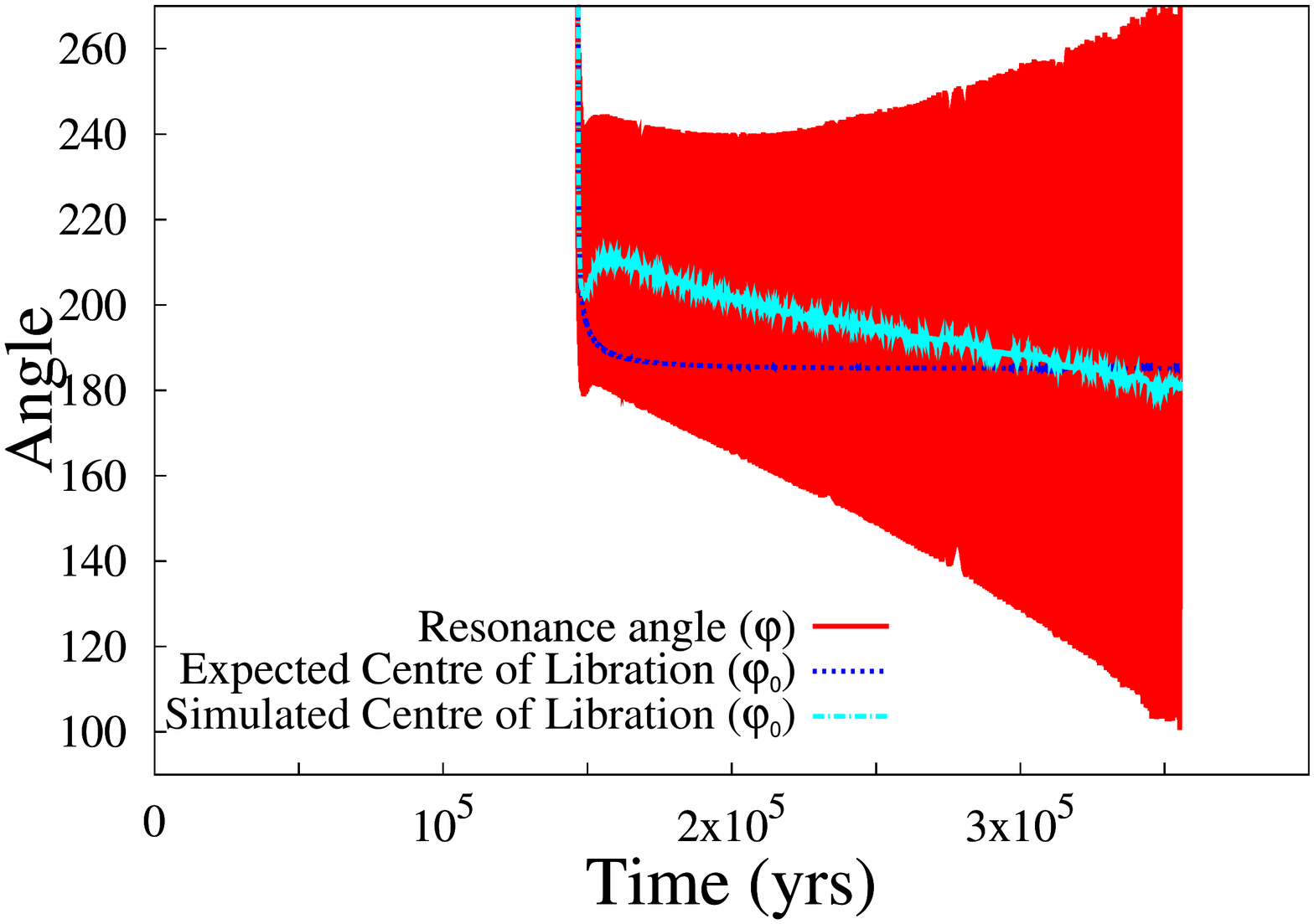}}
  \subfigure{\includegraphics[width=0.48\textwidth,trim = 50 50 0 50, clip]{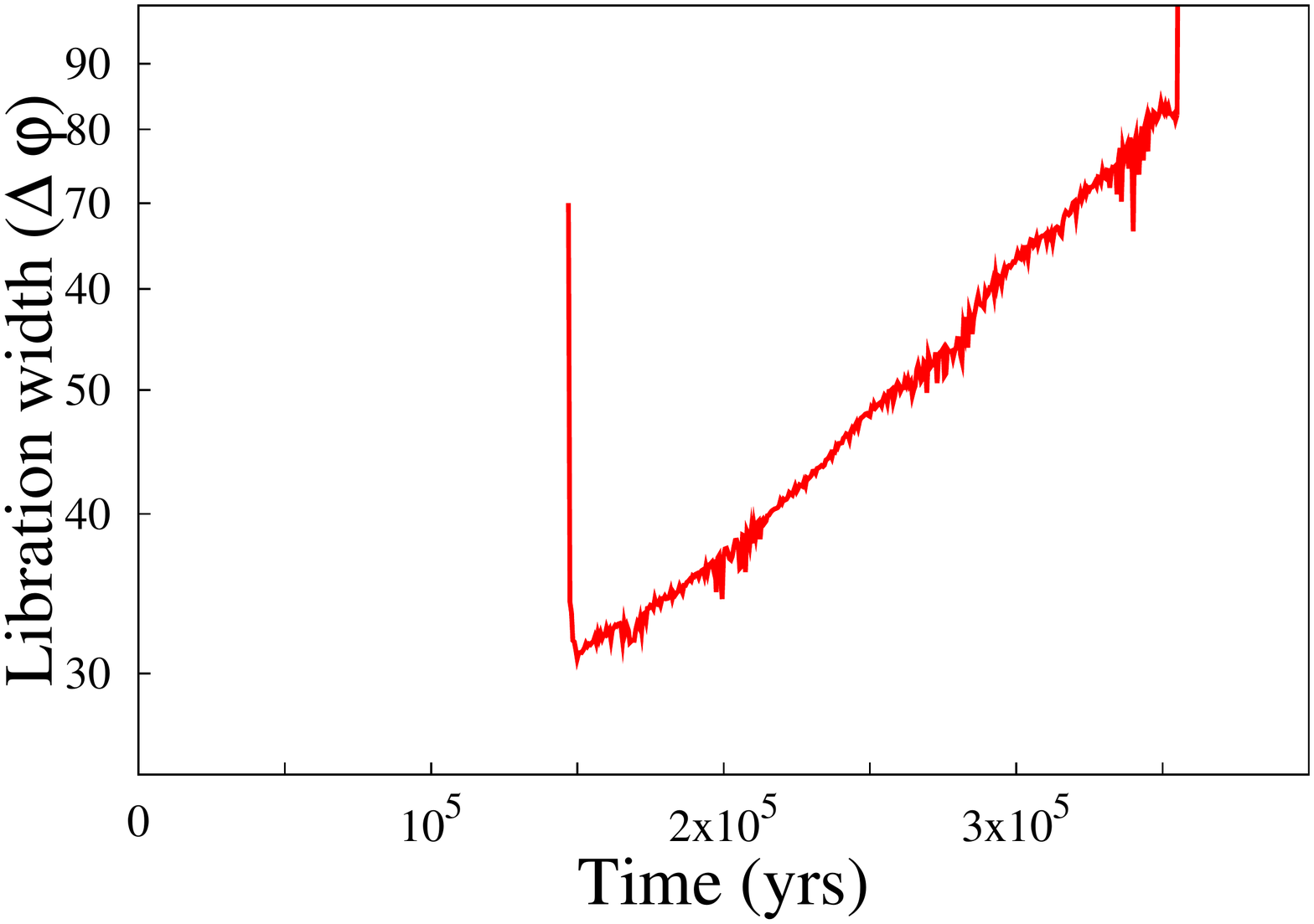}}
  \caption{Evolution of a dust grain in semimajor axis (top left), eccentricity (top right), resonance angle (bottom left), and resonance width (bottom right).  The dust is affected by PR drag and radiation pressure, and perturbed by a planet with $a_p = 1 \rm{au}$~and $m_p = m_{\oplus}$.  At $10^5 \rm{yrs}$, the dust grain crosses the 2:1 resonance and is perturbed to a higher $e$.  Similar jumps are discernible when it crosses the 3:2, 4:3, and 5:4 mean motion resonances.  At $1.5\times 10^5~\rm{yrs}$, the particle is captured in a 6:5 mean motion resonance with the planet.  The eccentricity grows quickly in while in resonance, flattening out at a constant value, closely following equation \ref{eq:wandsgrowth}.  The dust grain escapes the resonance at $3.6 \times 10^5 \rm{yrs}$.  After escape PR drag brings the particle into the star at $3.8 \times 10^{5} \rm{yrs}$.}
\label{fig:singledust}
\end{figure*}

To enhance qualitative understanding, we begin by presenting the evolution of a single dust grain (Figure \ref{fig:singledust}).  We perform this simulation, and all other simulations, using the hybrid integrator of MERCURY suite of N-body integrators \citep{1999MNRAS.304..793C}, which we have modified to include the effect of radiation of dust grains, as given in equation \ref{eq:radiationforce}.  To test that it was implemented properly, we confirmed that our N-body code produces a $K$~that is constant on long timescales with short timescale variations of size $\Delta K$ (equation \ref{eq:kevolves}).

We present here a particle from a simulation with a solar mass star, an earth mass planet on a circular orbit at 1 au, and a dust grain with $\beta_{\rm{PR}} = 0.01$, $e_0 = 0.01$, $a_0 = 2.22 \rm{au}$, $i_0 = 0.0628$.  This particle is from the simulation labelled B in table \ref{tab:sims}, where we describe the total range of system parameters we consider in section \ref{section:allsims}.  The particle evolves initially under only PR drag, decreasing its semimajor axis as per equation \ref{eq:dadtpr}~and its eccentricity as per equation \ref{eq:dedtpr}.  At $\sim 10^5$~years, it encounters the 2:1 mean motion resonance, which does not capture the particle, but provides an eccentricity kick.  Similar jumps are discernible when it crosses the 3:2, 4:3, and 5:4 mean motion resonances.  

At $1.5\times 10^5~\rm{yrs}$, the particle is captured in a 6:5 mean motion resonance with the planet.  For a particle not captured in lower $j$~resonances, the method of \textsection \ref{sec:rescap} calculates a probability of capture into the 6:5 resonance of $\sim 0.3$.  Allowing for capture into lower $j$~resonances, the overall capture probability into the 6:5 mean motion resonance for particles with similar initial conditions is $\sim 0.24$.  Thus, we expect this is a typical outcome.  After capture, the semimajor axis of the dust particle oscillates, but remains within the possible values for a resonant dust grain as outlined in equation \ref{eq:deltaa1} (top left panel of figure \ref{fig:singledust}).  After capture, the evolution of the eccentricity is well described by equation  \ref{eq:wandsgrowth} (top right panel of figure \ref{fig:singledust}).  The expected centre of libration of the resonance angle ($\varphi_0$)~changes quickly at early times, as expected from equation \ref{eq:varphi0}, while the centre of libration of the resonance angle in the simulation catches up on a much slower timescale (bottom left panel of figure \ref{fig:singledust}). The libration width grows exponentially with time (bottom right panel of figure \ref{fig:singledust}).  At $3.6 \times 10^5$~years, the particle escapes from resonance.  Later on we will quantify when and how dust grains escape from resonance.

After escaping from resonance, the particle continues to evolve under PR drag decreasing its semimajor axis as equation \ref{eq:dadtpr}~and its eccentricity as equation \ref{eq:dedtpr}.  At about $3.9 \times 10^5$~years, the particle is removed from the simulation once $a \sim 0.05$, at which point the integration timestep is too large to resolve the orbit.

From this example, we can see that we have some understanding of the capture process.  However, as we have noted, some elements are not well understood (libration width growth, resonance escape), and others (capture probability, resonance widths at capture) would benefit from additional validation.  To this end, we perform a suite of N-body simulations varying system parameters of interest.  

\subsection{Suite of simulations: Setup}
\label{section:allsims}

The central star is characterised by one property, its mass ($m_*$).  In all simulations, $m_* = m_{\odot}$.  

We include one planet of mass $m_{\rm{p}}$, with semimajor axis $a_{\rm{p}}$, with eccentricity $e_p$, inclination $i_p$, longitude of pericentre ($\varpi_p$), ascending node ($\Omega_p$), and mean anomaly ($M_p$).  In all cases, $e_p = 0$, $i_p = 0$.  In all cases $\varpi_p$, $\Omega_p$, and the initial $M_p$~are chosen randomly and uniformly between $0$~and $2\pi$.  Unless otherwise stated $m_p = m_{\oplus}$~and $a_p = 1 \rm{au}$.

The dust grains are assumed to have zero mass, and the orbits are characterised by the same properties as the planet: semimajor axis $a$, with eccentricity $e$, inclination $i$, longitude of pericentre ($\varpi$), ascending node ($\Omega$), and mean anomaly ($M$).  As with the planet, $\varpi$, $\Omega$, and the initial $M$~are chosen randomly and uniformly between $0$~and $2\pi$.  Unless otherwise stated, the dust grains begin with eccentricity\footnote{This is the initial eccentricity of the dust grain, and neglects the consideration that high $\beta_{\rm{PR}}$~are created at higher eccentricities than their parent bodies (\textsection \ref{subsec:collisionalproduction}).  We incorporate that only when considering a full disk model (section \ref{section:makedisk})} of 0.01, inclination\footnote{Comparisons to runs at $i = 0.0314$, and $i = 0.0157$~show no significant differences.} of $2\pi \times 10^{-2}$, and semimajor axis~chosen randomly and uniformly between 2.20 $a_p$~and 2.25 $a_p$.  We choose this initial semimajor axis~as we are interested in first and second order mean motion resonances, the most distant of which is the 3:1, which for a dust particle with $\beta_{\rm{PR}} = 0$~is located at 2.08 $a_p$, and closer for $\beta_{\rm{PR}} > 0$ (equation \ref{eq:ares}).  Thus, $a = 2.20-2.25~a_p$~is effectively an arbitrarily large distance for all choices of $\beta_{\rm{PR}}$.\footnote{Because PR drag decreases $e$~and $a$~for particles with larger initial $a$, this isn't strictly true.  We can use equations \ref{eq:dadtpr} and \ref{eq:dedtpr}~to evolve dust grains at higher $a$~to an $e$~at $2.2 \rm{au}$, so this choice is effective for our purposes.}


We aim to be able to simulate disks over a wide parameter space in $\beta_{\rm{PR}}$, $m_p$, $a_p$, and $e_0$.  A complete survey of the parameter space is computationally prohibitive.  We choose a canonical simulation $\beta_{\rm{PR}} = 0.01$, $m_p = m_{\oplus}$, $a_p = 1 \rm{au}$, $e_0 = 0.01$~(simulation B), and explore these dimensions by holding three of the parameters constant while varying the fourth one\footnote{Non-zero initial $e$~and $i$~are more physical than exactly zero, and avoid possible degeneracies/singularities that might occur when they're precisely zero; in particular, the resonant angle $\varphi$~and its components $\left(\lambda_{\rm{p}}, \lambda, \varpi\right)$~are not well defined for zero eccentricity and inclination}.  We consider $\beta_{\rm{PR}} = 0.005...0.32$, $m_p = 1 m_{\oplus}...256 m_{\oplus}$, $a_p = 1 \rm{au}...16\rm{au}$, $e_0=0.01...0.64$, in all cases spacing trial points by factors of two in the relevant parameter (table \ref{tab:sims}). Thus we perform a total of 25 simulations, each with $10^4$~particles.  Simulations are run with an 8 $(0.01/\beta_{PR})$ $(a/1 \rm{au})^{1.5}$ day timestep, with data outputs every $10^4 (0.01/\beta_{PR})$ $(a/1 \rm{au})^{1.5}$~days.  Particles are removed from the simulation when they are more than 100 au, or less than 0.05 au, from the star.

As all results are for $m_* = M_{\odot}$, any fits to results may have unknown stellar mass dependence.  Because our simulations generate a negligible amount of $k \geq 2$~captures, we only analyse the $j+1:j$~captures in our simulations.

\begin{table}

\begin{tabular*}{\textwidth}{ | l r r r r | }
 \cline{1-5}
 Simulation & $e_0$ & $\beta_{\rm{PR}}$ & $m_p$ & $a_p$ \\
 \cline{1-5}
  A & 0.01 & 0.005 & $m_\oplus$ & $1~\rm{au}$ \\
  B & 0.01 & 0.01 & $m_\oplus$ & $1~\rm{au}$ \\
  C & 0.01 & 0.02 & $m_\oplus$ & $1~\rm{au}$ \\
  D & 0.01 & 0.04 & $m_\oplus$ & $1~\rm{au}$ \\
  E & 0.01 & 0.08 & $m_\oplus$ & $1~\rm{au}$ \\
  F & 0.01 & 0.16 & $m_\oplus$ & $1~\rm{au}$ \\
  G & 0.01 & 0.32 & $m_\oplus$ & $1~\rm{au}$ \\
  H & 0.02 & 0.01 & $m_\oplus$ & $1~\rm{au}$ \\
  I & 0.04 & 0.01 & $m_\oplus$ & $1~\rm{au}$ \\
  J & 0.08 & 0.01 & $m_\oplus$ & $1~\rm{au}$ \\
  K & 0.16 & 0.01 & $m_\oplus$ & $1~\rm{au}$ \\
  L & 0.32 & 0.01 & $m_\oplus$ & $1~\rm{au}$ \\
  M & 0.64 & 0.01 & $m_\oplus$ & $1~\rm{au}$ \\
  N & 0.01 & 0.01 & 2$m_\oplus$ & $1~\rm{au}$ \\
  O & 0.01 & 0.01 & 4$m_\oplus$ & $1~\rm{au}$ \\
  P & 0.01 & 0.01 & 8$m_\oplus$ & $1~\rm{au}$ \\
  Q & 0.01 & 0.01 & 16$m_\oplus$ & $1~\rm{au}$ \\
  R & 0.01 & 0.01 & 32$m_\oplus$ & $1~\rm{au}$ \\
  S & 0.01 & 0.01 & 64$m_\oplus$ & $1~\rm{au}$ \\
  T & 0.01 & 0.01 & 128$m_\oplus$ & $1~\rm{au}$ \\
  U & 0.01 & 0.01 & 256$m_\oplus$ & $1~\rm{au}$ \\
  V & 0.01 & 0.01 & 1$m_\oplus$ & $2~\rm{au}$ \\
  W & 0.01 & 0.01 & 1$m_\oplus$ & $4~\rm{au}$ \\
  X & 0.01 & 0.01 & 1$m_\oplus$ & $8~\rm{au}$ \\
  Y & 0.01 & 0.01 & 1$m_\oplus$ & $16~\rm{au}$ \\
 \cline{1-5}
\end{tabular*}

 \caption{A table of all the simulations we performed. Simulation B forms a centre, with other simulations varying by factors of $2$~along axes of initial eccentricity, particle size $\left(\beta_{\rm{PR}}\right)$, planet mass, or planet semimajor axis.} \label{tab:sims}
\end{table}

Simulations continue until all particles have been removed.

\subsection{Suite of Simulations: Results}
\label{section:results}
\subsubsection{Trapping probability}
\label{subsec:trapprob}
To automatedly identify resonance capture, we look for dust grains that are continuously between $a_{j:j+k}+2\Delta a_{\rm{max}}$~and $a_{j:j+k}-2\Delta a_{\rm{max}}$~for at least twice the time predicted by dividing the crossing distance $4 \Delta a_{\rm{max}}$~by the instantaneous migration speed when the grain first encounters the resonance, given by equation \ref{eq:dadtpr}.  We employ the factor of $2$~in the expected width to allow for deviations from our expression for the width owing to higher order terms.  We also require the dust grain to be at that semimajor axis for at least twice as long to avoid falsely identifying dust grains as captured either when they are slowed by the resonant perturbation but not captured, or when oscillations (as equation \ref{eq:kevolves})~cause variations in $a$~that result in dust grains staying slightly longer at the resonance than equation \ref{eq:dadtpr} predicts, or when close encounters with the planet cause $a$~to evolve other than as predicted by equation \ref{eq:dadtpr}, using the eccentricity of the dust grain when it first encounters the resonance.  A spot check of 80 particles found that this approach correctly identified 57 of 58 times when a particle crossing a $j+1:j$~resonance was captured, and 359 of 363 times when a particle crossing a $j+1:j$~resonance was not captured, considering only crossings up to the first reported capture.  The trapping lifetime (or resonance escape time) is recorded as the total time the dust grain spends between $a_{j:j+k}+2\Delta a_{\rm{max}}$~and $a_{j:j+k}-2\Delta a_{\rm{max}}$~less the time predicted by integrating equation \ref{eq:dadtpr}~across that change in semimajor axis.

\begin{figure}
  \centering
  {\includegraphics[width=0.44\textwidth,trim = 50 50 0 50, clip]{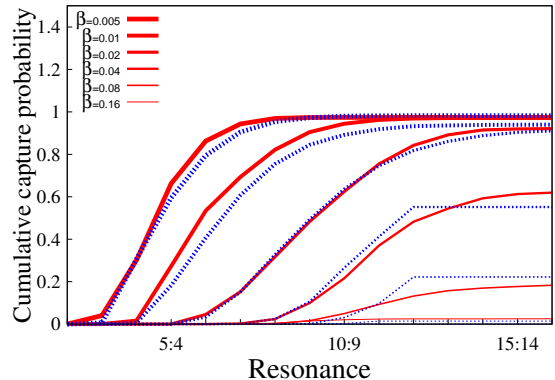}}
  \caption{Cumulative capture probability for particles with $e_0 = 0.01$, $i_0 = 2\pi \times 10^{-2}$, $a_p = 1 \rm{au}$, $m_p = m_{\oplus}$, for various $\beta_{\rm{PR}}$ (simulations A-G).  Comparing the outcomes of our simulations (red solid lines) with the calculations in \textsection \ref{sec:rescap} (blue dotted lines).  The fit works well overall, although the critical $j$~cutoff is perhaps smoother in the N-body simulations than in our Hamiltonian approach.}
\label{fig:capturechance}
\end{figure}

\begin{figure}
  \centering
  {\includegraphics[width=0.44\textwidth,trim = 50 50 0 50, clip]{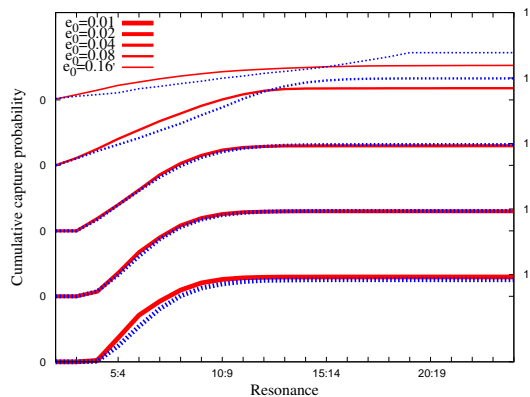}}
  \caption{Cumulative capture probability for particles with $\beta_{\rm{PR}} = 0.01$, $i_0 = 2\pi \times 10^{-2}$, $a_p = 1 \rm{au}$, $m_p = m_{\oplus}$, for various $e_0$~(simulations B, H-M), comparing the outcomes of our simulations (red solid lines) with the calculations in \textsection \ref{sec:rescap}~for different initial eccentricities (dotted blue lines).  Each initial eccentricity is offset vertically by 0.75 for clarity.  The $e_0 = 0.32$~and $e_0 = 0.64$~case both have $<1\%$~capture in both N-body simulations and the predictions of \textsection \ref{sec:rescap}, and are not plotted here.  At low eccentricity the agreement is very good, and it remains reasonably good at all eccentricities.}
\label{fig:capturechanceecc}
\end{figure}

We compare our observed capture probabilities to those calculated in \textsection \ref{sec:rescap}, and plot a comparison of the cumulative capture probability as a particle crosses for different values of $\beta_{\rm{PR}}$~(figure \ref{fig:capturechance} - simulations A-G) and for different values of $e_0$ (figure \ref{fig:capturechanceecc} - simulations B, H-M).  This comparison considers only the first capture of particles in the N-body simulation; any subsequent captures are neglected as eccentricity growth while in resonance (\textsection \ref{sec:resevolve}) makes such comparisons unsuitable for evaluating particles with the initial conditions of the experiment.  This comparison accounts for the eccentricity evolution as the semimajor axis decreases, and the eccentricity kicks due to resonance crossings (figure \ref{fig:singledust}); the quoted eccentricities are the initial values.  The calculated capture probabilities match the model well, although slight differences exist.

The 2:1 resonance presents a particular challenge as asymmetric libration results in capture into both the 2:1(l) and 2:1(u) resonance, with differing probabilities \citep{1958AJ.....63..443M,2002AJ....124.3430C,2005ApJ...619..623M}.  We emulate the approach of \citet{2003ApJ...598.1321W}~and calibrate the relative capture probabilities against our simulations in the form 
\begin{equation}
 \label{eq:twooneprobability}
 P_{2:1}\left(l\right) = 0.5 - a\theta^b \mu^c ,
\end{equation}
where $\theta = \left(\dot{a}/\rm{1 au/Myr}\right)/\sqrt{\left(a/\rm{1 au}\right)\left(M_\odot/m_*\right)}$.
We performed limited extra simulations of capture where capture into both the (u) and (l) resonance occurred, with 1000 particles; sampling planet mass every $25 m_\oplus$~from $150-400 m_\oplus$~(with $\beta = 0.01$)~and migration rate with factors of 2 in $\beta$~from $\beta = 0.0025$~to $\beta = 0.02$~with $m_p = 256 m_{\oplus}$.  Our best fit parameters were $a = 0.01$, $b = 0.25$, and $c = -0.4$, which we employ in \textsection \ref{section:makedisk}~to predict the relative rates of capture into the 2:1(l) and 2:1(u) resonances.

\subsubsection{Maximum j cutoff}
\label{sec:maxjfit}
\begin{figure}
  \centering
  {\includegraphics[width=0.44\textwidth,trim = 50 50 0 50, clip]{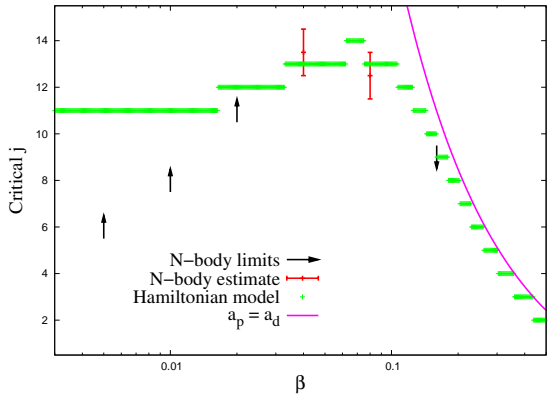}}
  \caption{The critical $j$~above which capture cannot occur.  Lower bounds come from simulations A, B, and C, where the N-body simulations capture $\sim 100\%$~of particles, so capture must be able to occur until the $j$~where the cumulative capture probability is $\sim 100\%$.  The upper bound is from simulation F, where the lack of any captures constrains the maximum $j$~to be less than the lowest $j$~for which capture is predicted in the Hamiltonian model.  The N-body estimates are from simulations D and E, with the critical $j$~taken from where the Hamiltonian model and N-body simulations predictions for capture probability diverge.  The green points show the $j$~given by equation \ref{eq:critjfinal}, with $\Delta \equiv 2.3$~chosen to bring the Hamiltonian model into agreement with the N-body results.  The solid pink shows where the dust grain would orbit at the same semi-major axis as the planet; here the impact parameter goes to zero, and hence the calculated kick becomes infinitely large.}
\label{fig:critj}
\end{figure}

Our derivation of the highest $j$~at which capture can occur has a numerical coefficient $\Delta$~that was not specified in \textsection \ref{sec:maxjguess}.  Comparison of our Hamiltonian capture model with our N-body simulations favours a coefficient of $\Delta \approx 2.3$~to bring the two into agreement (figure \ref{fig:critj}).  This is slightly higher than values previously reported (see section \ref{sec:maxjguess}).

\subsubsection{Libration width at capture}
\label{sec:widthatcapture}
\begin{figure}
  \centering
  {\includegraphics[width=0.44\textwidth,trim = 50 50 0 50, clip]{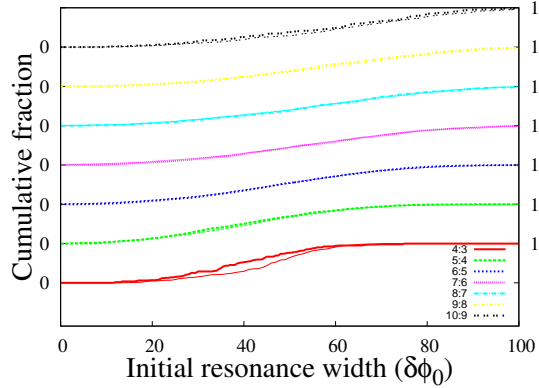}}
  \caption{Cumulative distribution of the resonant angle libration width at capture ($\delta \varphi_0$).  Each resonance is offset by $1$~vertically from one another for clarity.  Width is measured using the maximum and minimum libration angles ($\varphi$) while $0.04 < e < 0.06$.  Thick lines are simulations, thin lines are predictions from section \ref{sec:rescap} (for the most part, the two lines overlap too much to be distinguished).  These are all taken from the default case of $\beta_{\rm{PR}}=0.01$, $e_0 = 0.01$, $m_p = m_{\oplus}$, $a_p = 1 \rm{au}$~(Simulation B).  The overall agreement is excellent, except in the $4:3$~case, where the simulation had only $55$~captures. 
  }
\label{fig:varphidis}
\end{figure}

The model discussed in section \ref{sec:rescap} makes a prediction for what the distribution of libration widths should be at capture $\left(\delta \varphi_0\right)$.  In figure \ref{fig:varphidis}~we compare that prediction to our simulation results from simulation B, the canonical case,  using the maximum and minimum libration angles ($\varphi$) while $0.04 < e < 0.06$.  This is chosen to be confident the particle is in resonance, but has not experienced significant libration width growth.  The agreement is generally excellent, except in the lowest $j$~case, where the predicted libration widths are slightly too large.  We will see that it is important this distribution is well characterised, as the initial libration width sets the length of time the particle remains in resonance after capture.

\subsubsection{Eccentricity evolution}
\label{sec:eccentricityevolution}

In section \ref{sec:resevolve}, we presented an expression for the eccentricity evolution of a particle caught in resonance (equation \ref{eq:wandsgrowth}).  Here, we evaluate how closely that expression matches what we find in N-body simulations.  

\begin{figure}
  \centering
  {\includegraphics[width=0.44\textwidth,trim = 50 50 0 50, clip]{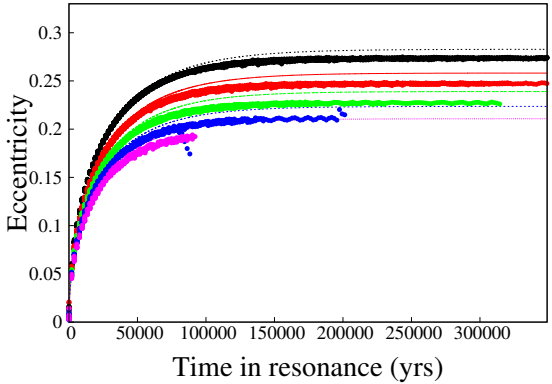}}
  \caption{A comparison between the expected evolution of the eccentricity $e_{\rm{predicted}}$~predicted from equation \ref{eq:wandsgrowth} (solid lines), and that found in the simulations $e_{\rm{sim}}$ (dots).  All particles plotted here are from the canonical case of $\beta_{\rm{PR}}=0.01$, $e_0 = 0.01$, $m_p = m_\oplus$, and $a_p = 1 \rm{au}$~(simulation B).  Plotted are 1000 random particles caught in 5:4 (black), 6:5 (red), 7:6 (green), 8:7 (blue), and 9:8 (magenta)~mean motion resonances for at least $5 \times 10^4$~years.  
  }
\label{fig:ematch}
\end{figure}

In figure \ref{fig:ematch}, we compare the evolution of eccentricity while in resonance for particles from simulation B to equation \ref{eq:wandsgrowth}, plotting all particles caught in 5:4, 6:5, 7:6, 8:7, 9:8, and 10:9~mean motion resonances for at least $5 \times 10^4$~years.  The evolution timescale is in excellent agreement, and the magnitude is in reasonable agreement, with only a $\sim 4\%$~deviation of the simulation from the analytic solution.  The analytic solution assumes particles are captured at zero eccentricity, but as the growth is an exponential decay away from 0, this difference is not significant.  

\subsubsection{Centre of Libration}
\label{subsec:librationcenter}

As seen in figure \ref{fig:singledust}, after capture, the eccentricity rises, and the predicted centre of libration moves quickly towards a value set by the final eccentricity.  The centre of libration found in simulation lags the predicted value, however, as seen in the bottom left panel of figure \ref{fig:singledust}.  The same general behaviour of an initial offset with a linear relaxation towards the equilibrium value is also found for the other particles in the simulations.  Thus, we set out to characterise this behaviour with a phenomenological model of the form
\begin{equation}
 \varphi_0^{sim} = \varphi_0^{cal} + \Delta \varphi_o \left(1-\frac{t}{\tau_{relax}}\right).
\end{equation}
Here the maximum offset in the centre of libration, $\Delta \varphi_0$~and the relaxation time, $\tau$,~are parameters that we will fit phenomenologically, and $t$~is the time after resonance capture.

\begin{figure}
 \centering
{\includegraphics[width=0.44\textwidth,trim = 50 50 0 50, clip]{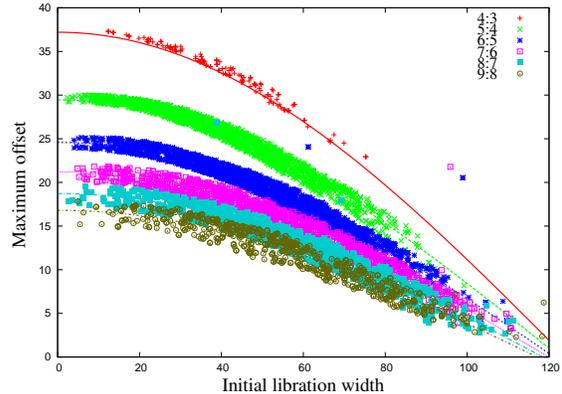}}
 \caption{Maximum offset between the predicted centre of libration and that found in simulations, compared to the initial libration width, for particles from simulation B.  The solid lines are a fit to all simulations, given by equations \ref{eq:c1andc2}, \ref{eq:c1}, and \ref{eq:c2}.}
 \label{fig:offsetvswidth}
\end{figure}

We begin with a fit of the maximum offset $\left( \Delta \varphi_0\right)$~between the equilibrium and actual centre of libration, from simulation B, plotted against the initial libration width, for several resonances (figure \ref{fig:offsetvswidth}).  Here, the maximum offset is found to depend only on two parameters, the $j$~of the resonance, and the initial libration width of the captured particle, $\delta \varphi_0$~(\textsection \ref{sec:widthatcapture}).  The best fit is of the form 
\begin{equation}
 \Delta \varphi_0 = C_1\cos{\left(\delta \varphi_0/C_2\right)}.
 \label{eq:c1andc2}
\end{equation}
Fitting the same for all simulations, we find
\begin{equation}
 C_1 = 4475 \beta_{\rm{PR}}^{0.847}j^{-0.81}\left(\frac{m_p}{m_{\oplus}}\right)^{-0.864}\left(\frac{a_p}{\rm{1 au}}\right)^{-0.423}
 \label{eq:c1}
\end{equation}
and 
\begin{equation}
 C_2 = 163.9 -1.76j -1.4\left(\frac{m_p}{m_{\oplus}}\right)+0.73\left(\frac{a}{1 \rm{au}}\right).
\label{eq:c2}
\end{equation}
We neglect the $\beta_{\rm{PR}}$ dependence of $C_2$~as the best fit coefficient of $-1.93 \pm 1.76$~is consistent with zero.

\begin{figure}
 \centering
{\includegraphics[width=0.44\textwidth,trim = 50 50 0 50, clip]{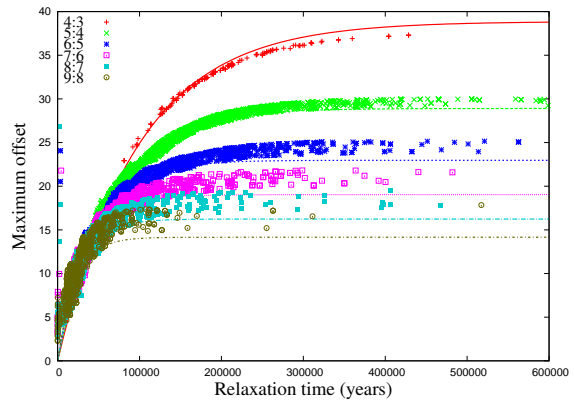}}
 \caption{The time it takes for the libration centre to relax to the equilibrium value of equation \ref{eq:varphi0}, compared to the initial libration width, for particles from simulation B.  The solid lines are a joint fit to all the simulations of the form $\Delta \varphi_0 = C_3\left(1-e^{-\tau_{relax}/C_4}\right)$.}
 \label{fig:relaxvswidth}
\end{figure}

Plotting the maximum offset against the relaxation time, we find the two have an approximately functional correspondence (figure \ref{fig:relaxvswidth}).  We fit the maximum offset between the predicted centre of libration and that found in simulations against time it takes for the centre of libration to to relax the equilibrium value of equation \ref{eq:varphi0}.  The best fit value is 
\begin{equation}
 \tau_{relax} = C_4\ln{\left(1-\frac{\Delta \varphi_0}{C_3}\right)}
  \label{eq:c3andc4}
\end{equation}
with 
\begin{equation}
  C_3 = 7605 j^{-1.03}\beta_{\rm{PR}}^{0.9}m_p^{-0.94}a_p^{-0.45}
\end{equation}
and 
\begin{equation}
C_4 = 13949 j^{-1.54}\beta_{\rm{PR}}^{-0.79}m^{-0.385}a_p^{1.73}.
\end{equation}
An example of this fit for simulation B is plotted in figure \ref{fig:relaxvswidth}.  Here, of course, the combination of equations \ref{eq:c1andc2} and \ref{eq:c3andc4} allows one to express the relaxation time as a function of the initial libration width in the form
\begin{equation}
 \tau_{\rm{relax}} = C_4 \ln{\left(1-\frac{C_1}{C_3}\cos{\left(\frac{\delta \varphi_0}{C_2}\right)}\right)},
\end{equation}
and thus for particles with the same system parameters caught in the same resonance, their evolution is dictated entirely by their initial libration width.

\subsubsection{Libration width evolution}

\begin{figure}
  \centering
  {\includegraphics[width=0.44\textwidth,trim = 50 50 0 50, clip]{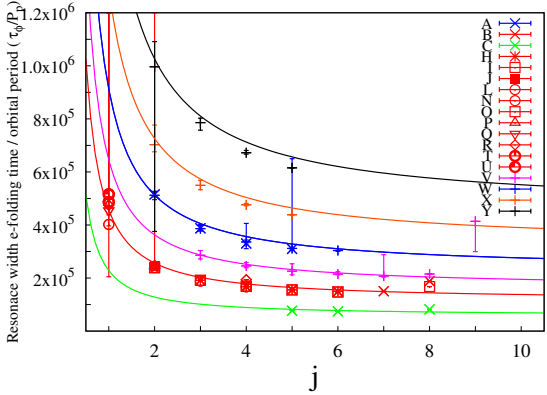}}
  \caption{The $e$-folding time of the resonance width for dust grains in all simulations.   We fit the resonance width as exponential growth between $160 (0.01/\beta_{\rm{PR}})\left(a_p/1 \rm{au}\right)^2$ kyrs and $320 (0.01/\beta_{\rm{PR}})\left(a_p/1 \rm{au}\right)^2$ kyrs for those particles which are trapped in resonance for at least 480 $(0.01/\beta_{\rm{PR}})\left(a_p/1 \rm{au}\right)^2$ kyrs (all simulations not plotted have zero particles meeting this criterion).  Points are values from the N-body simulations, the solid lines are equation \ref{eq:tauvarphi}.   Here the error bars cover the smallest to largest values, with the point centred on the median value.  Rare catastrophic failures of the automatic fitting produce extremely large values, so we employ the median value, rather than the mean value.}
\label{fig:varphiev}
\end{figure}

The resonance width $\left(\delta \varphi\right)$~was found to grow exponentially with time by \citet{1994A&A...289..972S}, but they did not quantify this behaviour.  We also observed exponential growth with time (Figure \ref{fig:singledust}).  This growth is of the form
\begin{equation}
 \label{eq:widthexpgrowth}
 \delta \varphi = \delta \varphi_0 e^{t/\tau_{\varphi}}
\end{equation}
 where $\delta \varphi_0$~is characterised in \textsection \ref{sec:widthatcapture}.  To fit the resonance width of a particle in resonance, we bin all outputs of $\varphi$~within a 4000 $\left(0.01/\beta_{\rm{PR}}\right)$ $\left(a/1~\rm{au}\right)^{1.5}$~year period (as in \textsection \ref{subsec:librationcenter}), taking the largest difference between two values as the resonance width at that time.  For those particles that persist in resonance for more than $4.8 \times 10^5 \left(0.01/\beta_{\rm{PR}}\right)$ $\left(a/1~\rm{au}\right)^{1.5}$~years, we fit an exponential curve to the growth between $1.6 \times 10^5 \left(0.01/\beta_{\rm{PR}}\right)$ $\left(a/1~\rm{au}\right)^{1.5}$~and $3.2 \times 10^5 \left(0.01/\beta_{\rm{PR}}\right)$ $\left(a/1~\rm{au}\right)^{1.5}$~years~to assign a $\tau_{\varphi}$~to that particle's evolution.  The $\tau_{\varphi}$~of different particles with the same $\beta_{\rm{PR}}$~caught in the same resonance cluster tightly (Figure \ref{fig:varphiev}). 

By fitting all of the values derived this way for difference resonances and simulations, we find

\begin{equation}
 \label{eq:tauvarphi}
 \tau_{\varphi} \approx 1.14 \times 10^5 \left(\frac{0.01}{\beta_{\rm{PR}}}\right) \left(\frac{j+1}{j}\right)^2 \left(\frac{a_p}{1 \rm{au}}\right)^2 \rm{years},
\end{equation}

a fit of which can be seen in plot \ref{fig:varphiev}.  Fixing the constants of equation \ref{eq:tauvarphi}, varying one while keeping the rest fixed, a non-linear least-squares Marquardt-Levenberge fit, weighting each point by $\sqrt{n}$~where $n$~is the number of particle-histories represented by a point, gives the normalisation as $1.14 \pm 0.01 \times 10^5$, the exponent of the $\left(j+1\right)/j$~term as $2.02 \pm 0.01$, the exponent of the $\beta_{\rm{PR}}$~term is $-0.94 \pm 0.05$, the exponent of the planet mass term would be $-0.006 \pm 0.002$ if we included a power law term dependent on the planet's mass, and the exponent of the semimajor axis is $1.99 \pm 0.02$.  

Note that the scalings in equation \ref{eq:tauvarphi} are very similar to the PR drag lifetime in equation \ref{eq:lifetimepredict}.  They are related to one another as
\begin{equation}
 \tau_{\varphi} \approx 3 \left(\frac{j+1}{j}\right)^{\frac{2}{3}}\left(1-\beta_{\rm{PR}}\right)^{-\frac{2}{3}} \tau_{\rm{PR}}
\end{equation}
 and here the $j$~and $\beta_{\rm{PR}}$~terms will be close to one, and thus the e-folding time will always be larger than the PR drag time by a factor of a few.  Because any resonant object must have $\delta \varphi < \pi$, unless dust grains are captures at very small libration widths (which is not found by \citet{2011MNRAS.413..554M}~for any parameter space they investigate), they can only undergo a few e-folds of their libration width before they escape the resonance.  Therefore, resonantly captured rings should never exceed the background disk in brightness by more than a factor of $\mathcal{O}\left(10\right)$.

\subsubsection{Escape from resonance}
\label{sec:escape}

\begin{figure}
  \centering
  {\includegraphics[width=0.48\textwidth,trim = 50 50 50 50, clip]{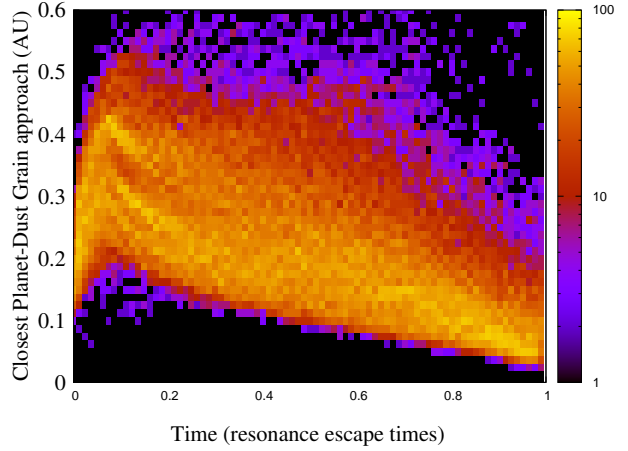}}
  \caption{Closest approach between the planet and the dust particle in simulation B.  Dust grains are binned with all grains trapped in the same resonance in resonance escape time bins~with width $10^4$~yrs.  Closest approach is sampled across 1000 years for each bin.  Once the particles come within $\sim 0.04~\rm{au}$, they are ejected from the resonance.  We measure this escape radius across all simulations using the second last time bin, where 10\% of measurements are below the escape radius.  Only those particles caught for more than $10^4$~years are included.}
\label{fig:closeapproachevolution}
\end{figure}

At early times, dust grains undergo close encounters with the planet as they are pushed into the resonance.  The resonance then protects the dust grain from close encounters, but as the eccentricity and libration width increase, encounters become increasingly closer, until the particle is lost from the resonance.  To analyse if and when close encounters cause a dust grain to be lost from resonance, we bin together all dust grains from the same simulation which are caught in the same resonance and which are trapped in that resonance for the same amount of time (rounded to the nearest $10^4 \left(a/1~\rm{au}\right)^2\left(m_p/m_{\oplus}\right)^{0.5}$~years).  We define the closest approach for each bin of dust grains by the smallest planet-dust grain separation output by the simulation over each $10^3\left(a/1~\rm{au}\right)^2\left(m_p/m_{\oplus}\right)^{0.5}$ year interval (Figure \ref{fig:closeapproachevolution}).  Because the planet-dust grain closest approaches decrease during the evolution, and escape occurs for all particles at a fixed encounter distance, we conclude escape from resonance occurs due to a close encounter with the planet, which provides an impulse sufficient to move a particle out of resonance.

\begin{figure*}
  \centering
  \subfigure{\includegraphics[width=0.48\textwidth,trim = 50 50 0 50, clip]{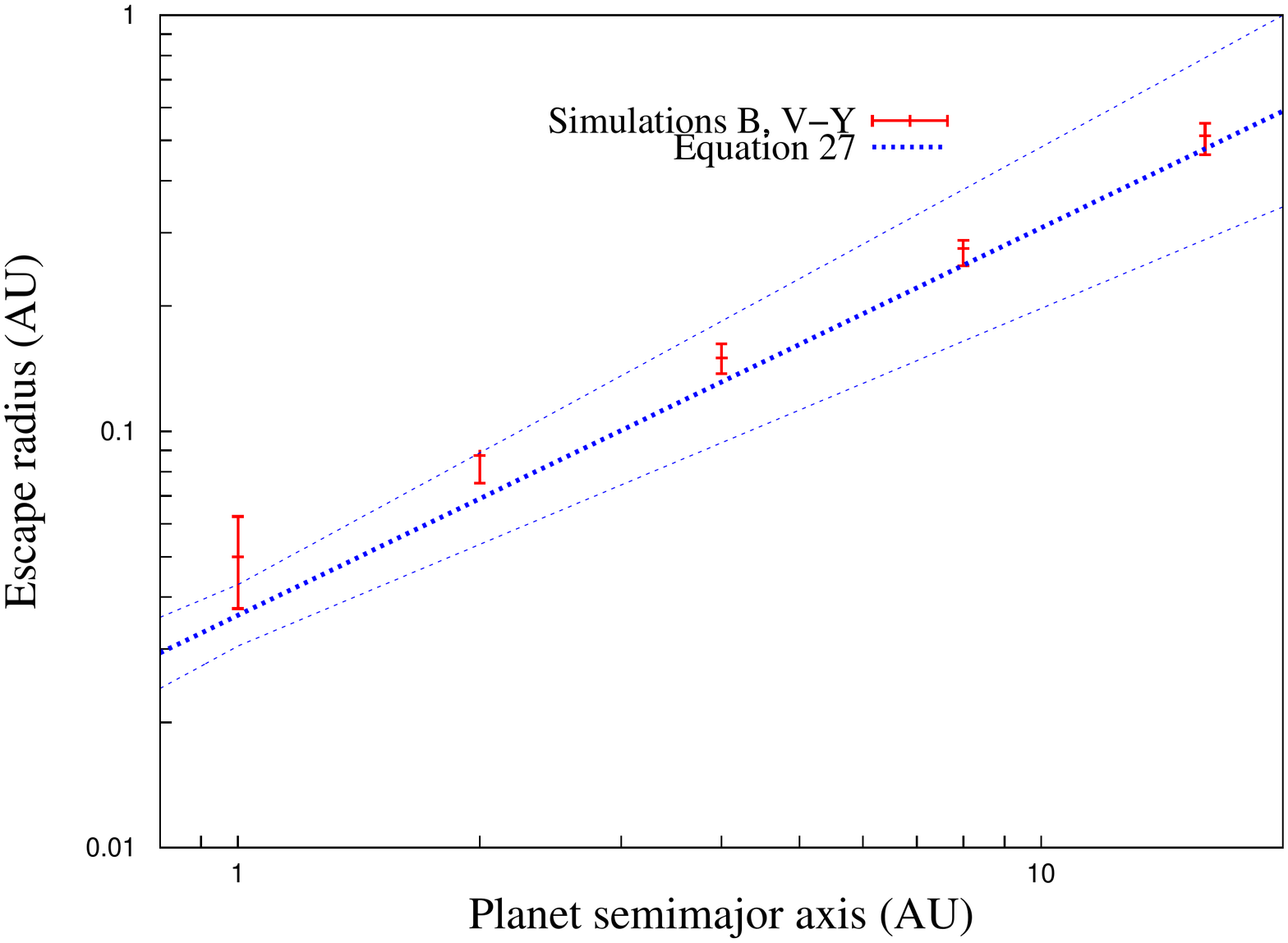}}
  \subfigure{\includegraphics[width=0.48\textwidth,trim = 50 50 0 50, clip]{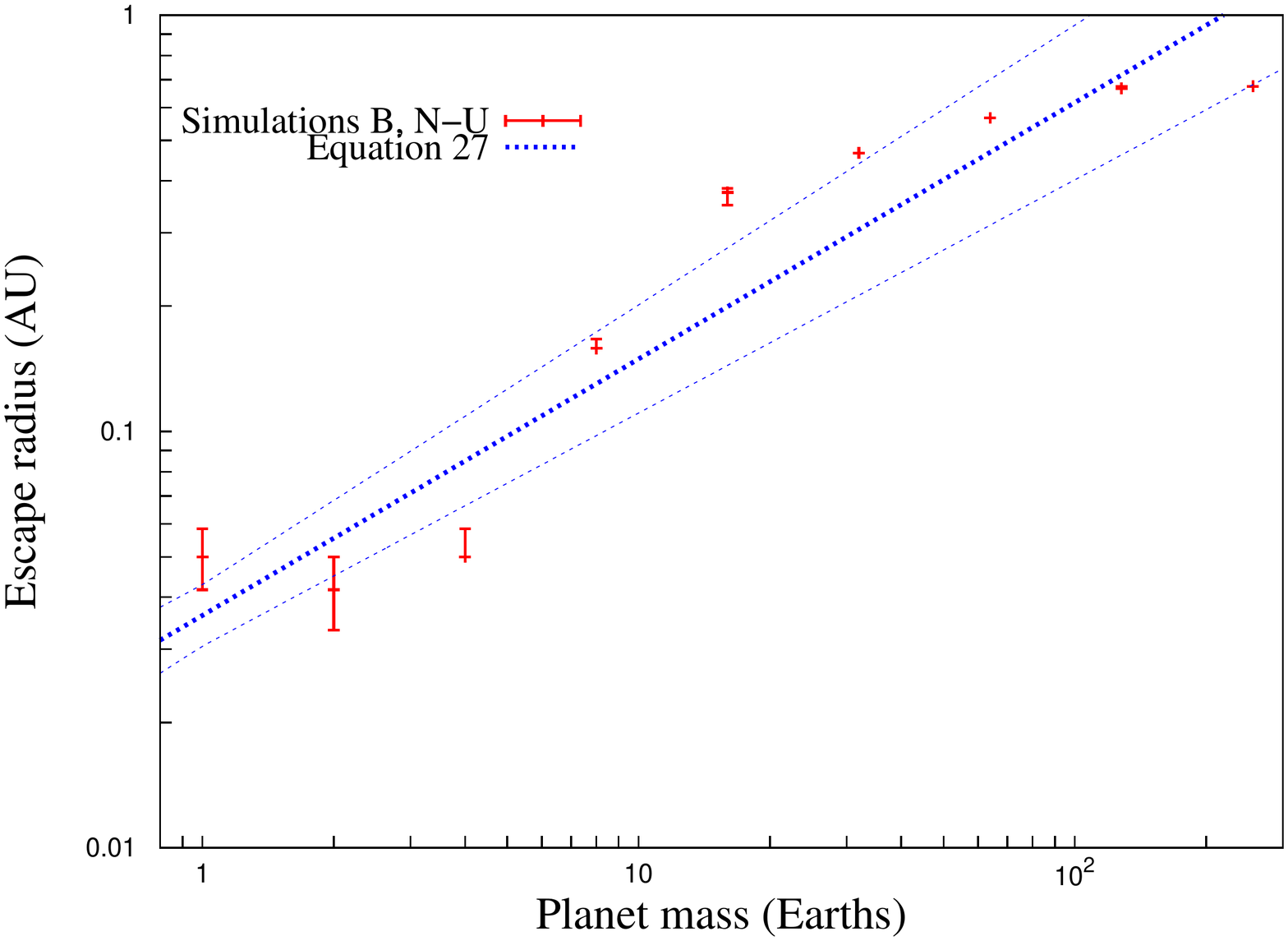}} \\
  \subfigure{\includegraphics[width=0.48\textwidth,trim = 50 50 0 50, clip]{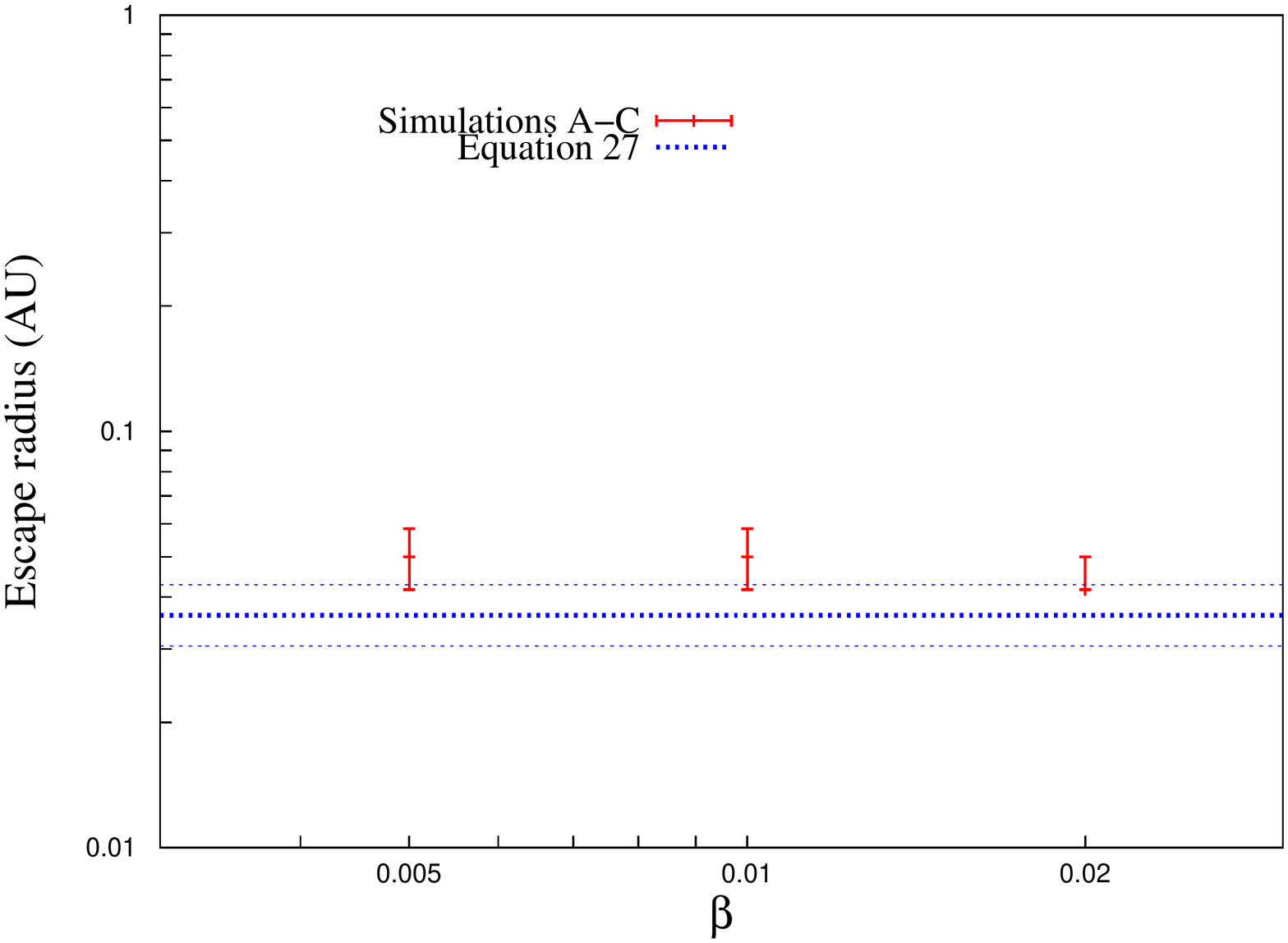}}
  \subfigure{\includegraphics[width=0.48\textwidth,trim = 50 50 0 50, clip]{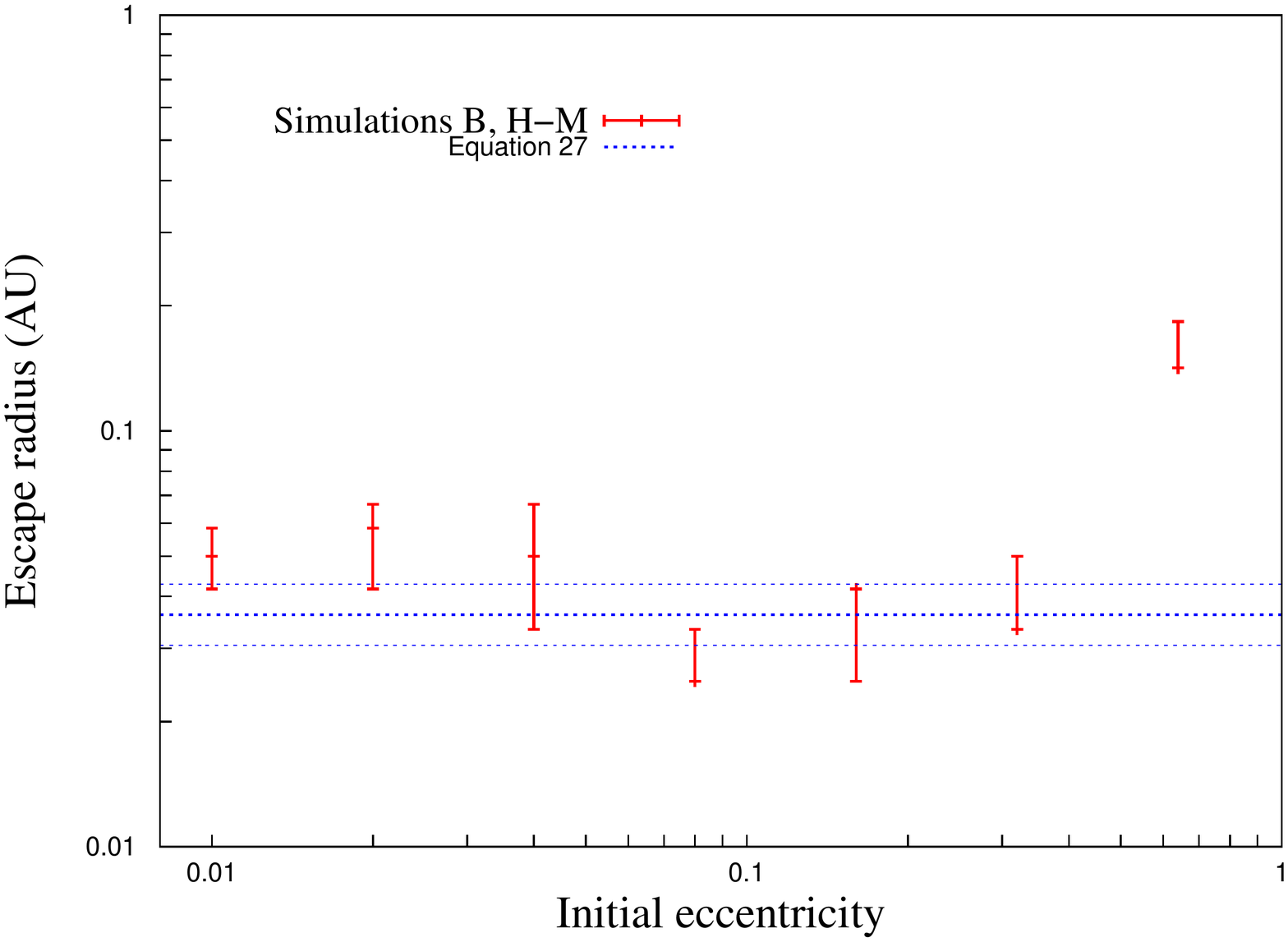}}
  \caption{Escape distance vs. planet semimajor axis in simulations B, V, W, X, Y (top left), escape radius vs planet mass, from simulations B, N-U (top right), escape radius vs $\beta_{\rm{PR}}$, from simulations $A-C$ (bottom left), and escape radius vs initial eccentricity, featuring simulations B, and H through M (bottom right).  Error bars are the $10^\text{th}$, $20^\text{th}$, and $30^\text{th}$~percentile fits to the second last time column in plots like figure \ref{fig:closeapproachevolution} for the relevant simulations.  The fit (blue dotted line) is a joint fit to mass and semimajor axis simulations (B, N-Y).}
\label{fig:rescfits}
\end{figure*}

To measure this distance, we consider the 10$^\text{th}$ percentile closest approach in the second to last time bin in the separation (to allow for some error in the automated fitting, sampling time, and identification of when escape from resonance occurred).  Because eccentricity evolves to a fixed value (\textsection \ref{sec:eccentricityevolution})~we do not expect eccentricity at capture to affect the resonance escape distance, and our simulations are compatible with no $e$~dependence (figure \ref{fig:rescfits}).  Additionally, no $\beta_{\rm{PR}}$, nor $j$~dependence is observed.  The escape radius is observed to depend most strongly on the planet mass and semimajor axis.   A non-linear least-squares Marquardt-Levenberge fit of the escape radius in simulations B and N-Y gives the escape radius as
\begin{equation}
 \label{eq:resc}
 R_{\rm{e}} \approx 0.036^{+0.0067}_{-0.0057} \left(\frac{m_{p}}{m_{\oplus}}\right)^{0.616 \pm 0.056}\left(\frac{a_p}{1 \rm{au}}\right)^{0.931 \pm 0.012}~\rm{au}
\end{equation}
of the planet.  

\subsubsection{Resonance lifetimes}

\begin{figure}
  \centering
  {\includegraphics[width=0.48\textwidth,trim = 50 50 0 50, clip]{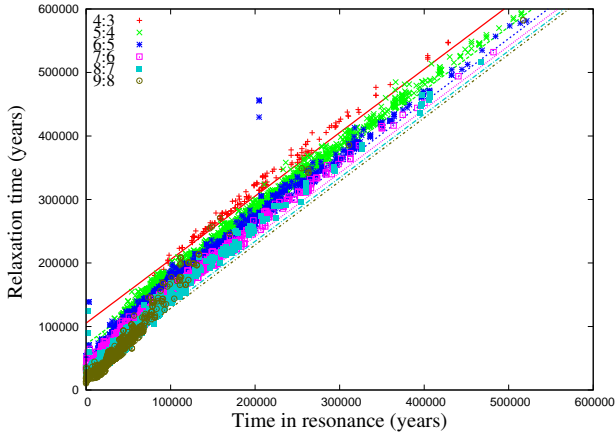}}
  \caption{Lifetime of particle in resonance plotted against the time it takes the centre of libration to relax to the equilibrium value.  Points are from simulation B.  A fit to all simulations of the form  $\tau_{res} = \tau_{relax} + C$ is plotted for the appropriate resonances.}
\label{fig:lifetimevsrelax}
\end{figure}

Comparing the relaxation time to the total time spent in resonance shows that the two show a functional relationship, which can be fitted as a linear relationship with a constant offset that depends on the resonance and other properties of the simulation.  We perform a joint fit across all simulations of 
\begin{equation}
 \tau_{res} = \tau_{relax} + C_5 ,
 \label{eq:c5}
\end{equation}
and get a best fit of 
\begin{equation}
 C_5 = 377.5 j^{-1.33}\beta_{\rm{PR}}^{-1.54}m_p^{0.21}a_p^{2.2} .
\end{equation}
An example of this is plotted in figure \ref{fig:lifetimevsrelax} for simulation B.

\begin{figure}
  \centering
  {\includegraphics[width=0.49\textwidth,trim = 50 50 0 50, clip]{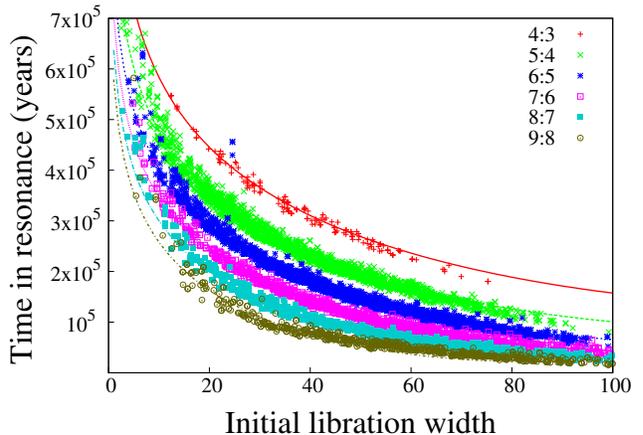}}
  \caption{Time in resonance versus the initial libration width in simulation B (points). The resonant lifetime is fixed for a given capture by the initial libration width and $j$, as the libration width grows predictably in that case (equation \ref{eq:tauvarphi}), until the particle has a close encounter with the planet (equation \ref{eq:resc}), causing it to be ejected from resonance.  This results in a predictable resonance lifetime (the lines are equation \ref{eq:lifetime} for the corresponding resonances).}
\label{fig:lifetimevsphi0}
\end{figure}

Knowing how dust grains escape from resonance (equation \ref{eq:resc}), and how they evolve in resonance (equations \ref{eq:wandsgrowth} and \ref{eq:tauvarphi}), we can, in principle, predict the resonance lifetime of a trapped dust grain.  That analysis does not, however, lend itself to a neat analytic expression.  Instead, we derive a lifetime by combining equations \ref{eq:c1andc2}, \ref{eq:c3andc4}, and \ref{eq:c5}, which gives the time in resonance as $\tau_{res} = C_4\ln{\left(1-C_1\cos{\left(\delta \varphi_0/C_2\right)}/C_3\right)}+C_5$.  In practice, $\tau_{res}$~changes very rapidly at small $\varphi_0$, and the uncertainties accumulated in bootstrapping make the expression a poor fit.  Using the general expression, we refit with

\begin{equation}
 \label{eq:lifetime}
 \tau_{res} = C_A\ln{\left(1-\cos{\frac{\delta \varphi_0}{C_B}}\right)}+C_C .
\end{equation}
In equation \ref{eq:lifetime}, the best fit values across all simulations are 
\begin{equation}
 C_A = -1.27\times 10^3j^{-0.37}\beta_{\rm{PR}}^{-1.0}\left(\frac{m_p}{m_{\oplus}}\right)^{0.06}\left(\frac{a_p}{1\rm{au}}\right)^{2.1} \rm{years} ,
\end{equation}
\begin{equation}
 C_B = 66-4.4j-3.2\left(\frac{m_p}{m_{\oplus}}\right)-0.7\left(\frac{a_p}{1 \rm{au}}\right) ,
\end{equation}
and 
\begin{equation}
C_C = 3959j^{-1.04}\beta_{\rm{PR}}^{-1.1}\left(\frac{m_p}{m_{\oplus}}\right)^{0.01}\left(\frac{a_p}{1 \rm{au}}\right)^{2.0}
\end{equation}
 years.  A comparison of this fit to the results of simulation B is plotted in figure \ref{fig:lifetimevsphi0}.

\section{Application: Modelling a whole disk}
\label{section:makedisk}

We are now in a position to predict the dynamical evolution of a population of dust grains as they evolve past a planet.  Disk images are generated by the following step-by-step approach:
\newpage
\begin{enumerate}
 \item[{\bf 1.}] Initial conditions are specified by assigning a mass to the star, mass and semimajor axis to the planet, and semimajor axis, eccentricity, and inclination to the parent body of the dust grains, and $\beta_{\rm{PR}}$~to the dust grains.   To consider a population of parent bodies (as we will in \textsection \ref{sec:earthicandisk}), the parent body for each grain is chosen randomly from the population.  
 \item[{\bf 2.}] Dust grains are placed on orbits that correspond to their parent body orbits (modified for their $\beta_{\rm{PR}}$, as outlined in \textsection \ref{subsec:collisionalproduction})   To consider a population of grains with different $\beta_{\rm{PR}}$~is somewhat complicated, as the number of grains may be a steep function of $\beta_{\rm{PR}}$.  This can result in under-sampling or oversampling, and we create images for single $\beta_{\rm{PR}}$~values, and weight and add the images afterwards to model a grain population.
 \item[{\bf 3.}] Dust grains evolve in $a_d$~and $e_d$~due to PR drag as calculated by \citet{1950ApJ...111..134W} (\textsection \ref{sec:prdrag}).
 \item[{\bf 4.}] Whenever a particle encounters a first order mean motion resonance with the planet (\textsection \ref{sec:res}), it may be captured into resonance with probability given in \textsection \ref{sec:rescap} (step {\bf 5.A})~or not (step {\bf 5.B}).
 \begin{enumerate}
  \item[{\bf 5.A.i}] If a particle is captured into resonance, the initial libration width is taken from the distribution given by the model of \citet{2011MNRAS.413..554M}, updated in \textsection \ref{sec:rescap}.  The initial centre of libration is offset from the equilibrium value of equation \ref{eq:varphi0} by an amount dependent on the resonant width at capture, as given by equation \ref{eq:c1andc2}.  
  \item[{\bf 5.A.ii}] While trapped in resonance, particles evolve in eccentricity as per equation \ref{eq:wandsgrowth} and libration width as per equation \ref{eq:widthexpgrowth}.  The offset from the equilibrium centre of libration relaxes linearly to the value given by equation \ref{eq:varphi0} in a relaxation time given by equation \ref{eq:c3andc4}.
  \item[{\bf 5.A.iii}] When the particle comes within a distance given by equation \ref{eq:resc} of the planet, it is removed from resonance, and begins evolving again under PR drag (i.e., return to step {\bf 3.}).
  \item[{\bf 5.B}] If a particle is not captured into resonance, an eccentricity jump is applied as found in \citet{2011MNRAS.413..554M}~(with the modifications found in the Appendix), and the particle continues evolving under PR drag (i.e., return to step {\bf 3.}).
 \end{enumerate}
 \item[{\bf 6.}] When the dust particle encounters the star it is removed from the simulation.  
\end{enumerate}

\begin{figure*}
  \centering
  \subfigure{\includegraphics[width=0.24\textwidth,trim = 100 100 100 100]{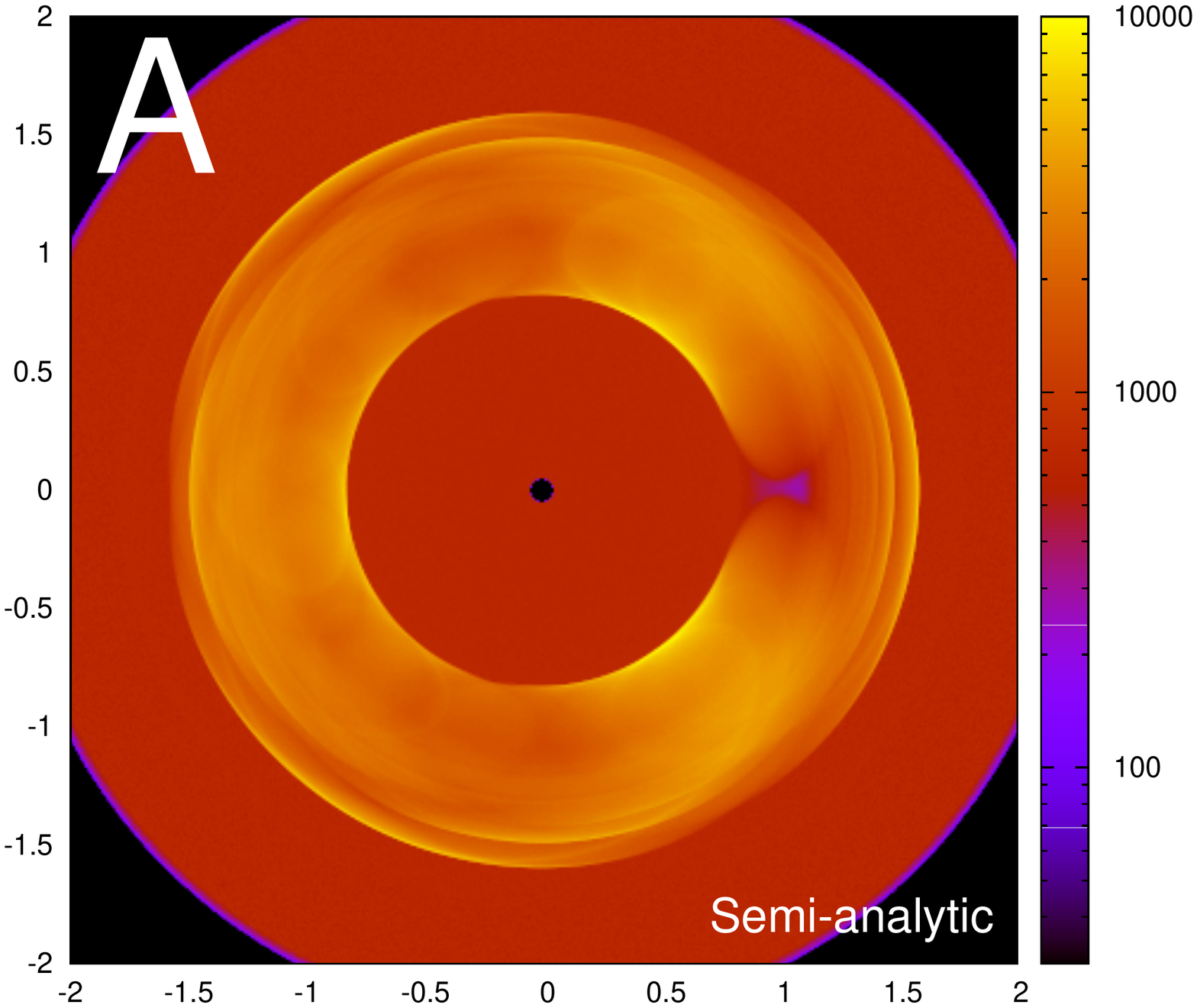}}
  \subfigure{\includegraphics[width=0.24\textwidth,trim = 100 100 100 100]{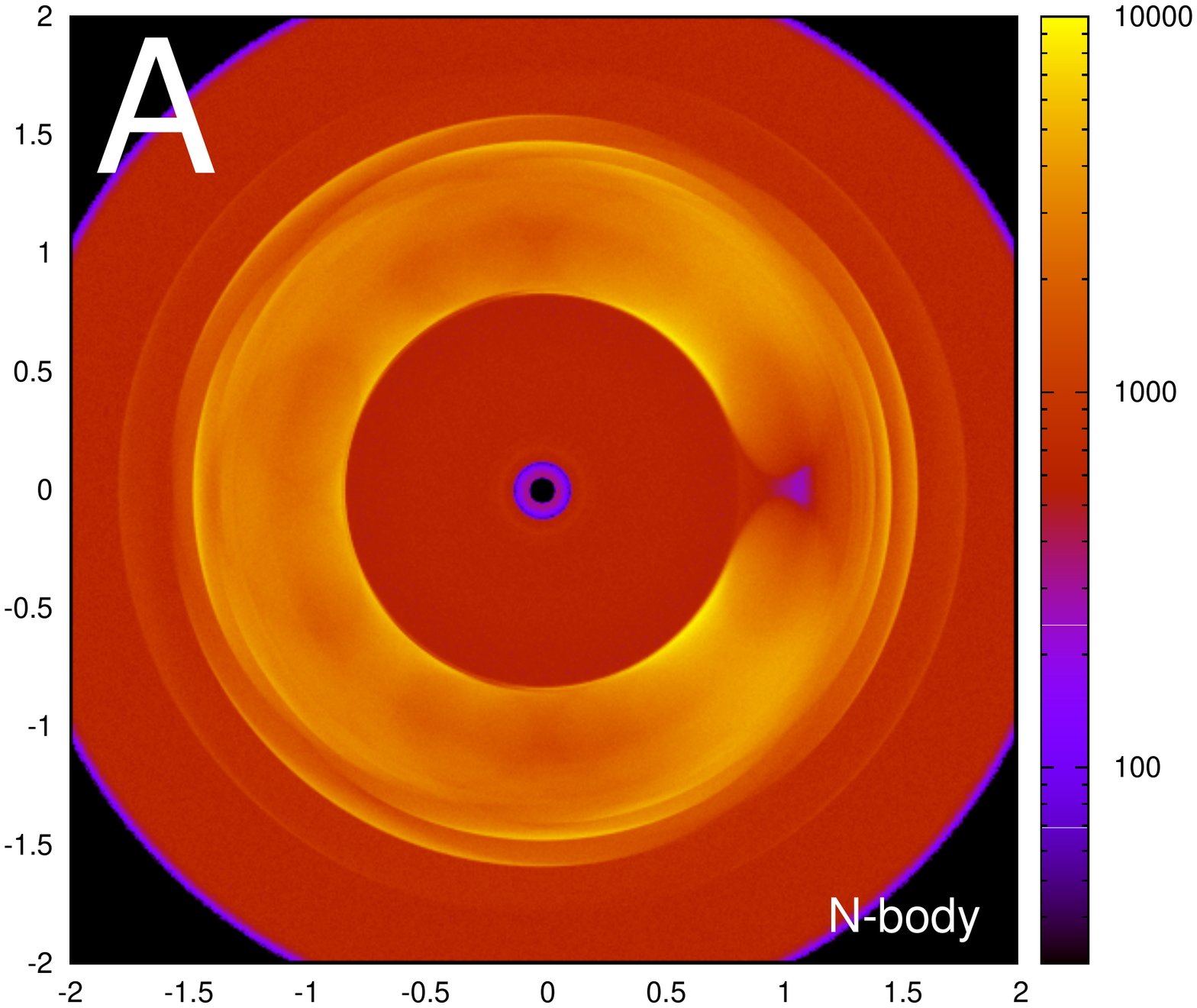}}
  \subfigure{\includegraphics[width=0.24\textwidth,trim = 100 100 100 100]{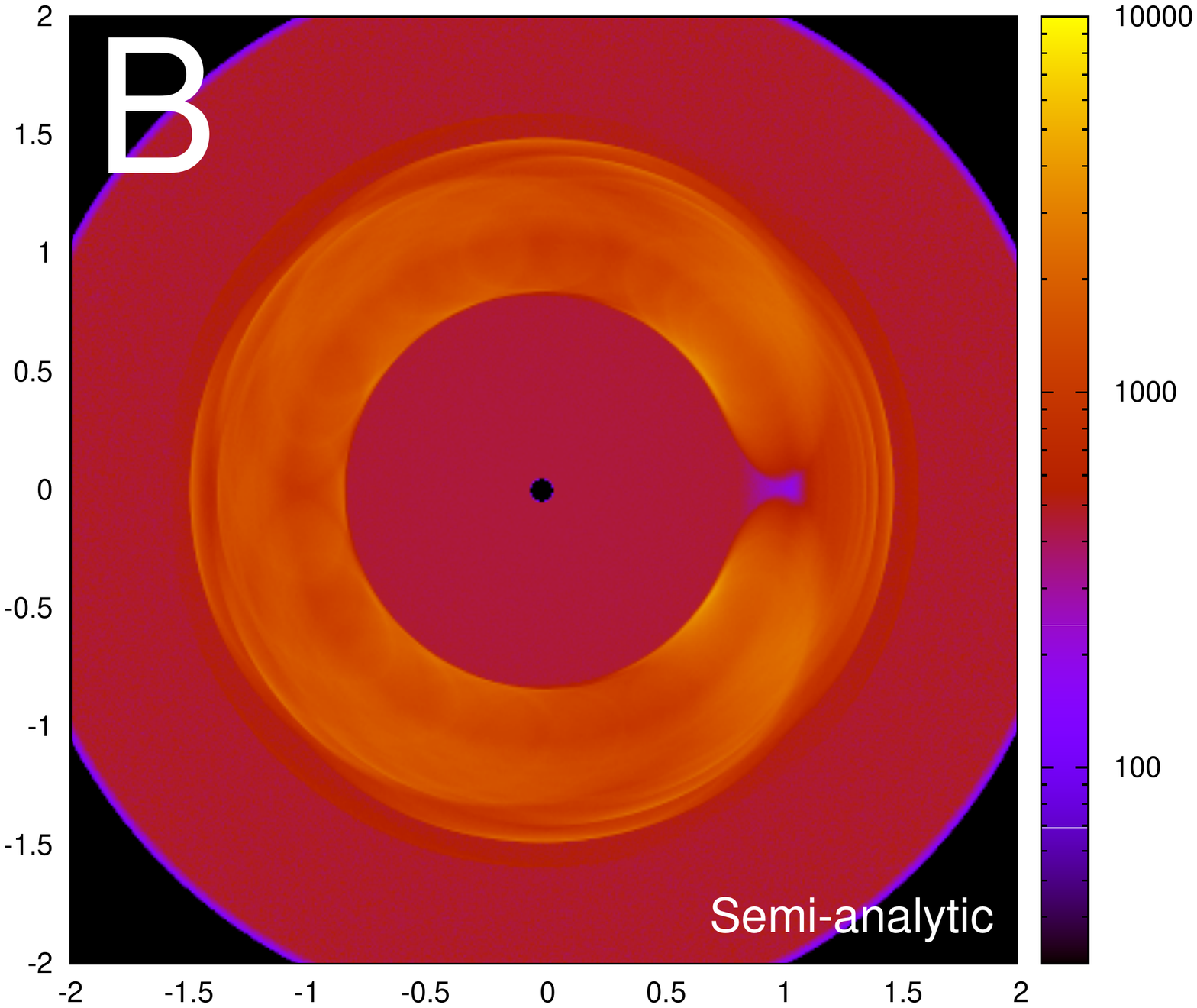}}
  \subfigure{\includegraphics[width=0.24\textwidth,trim = 100 100 100 100]{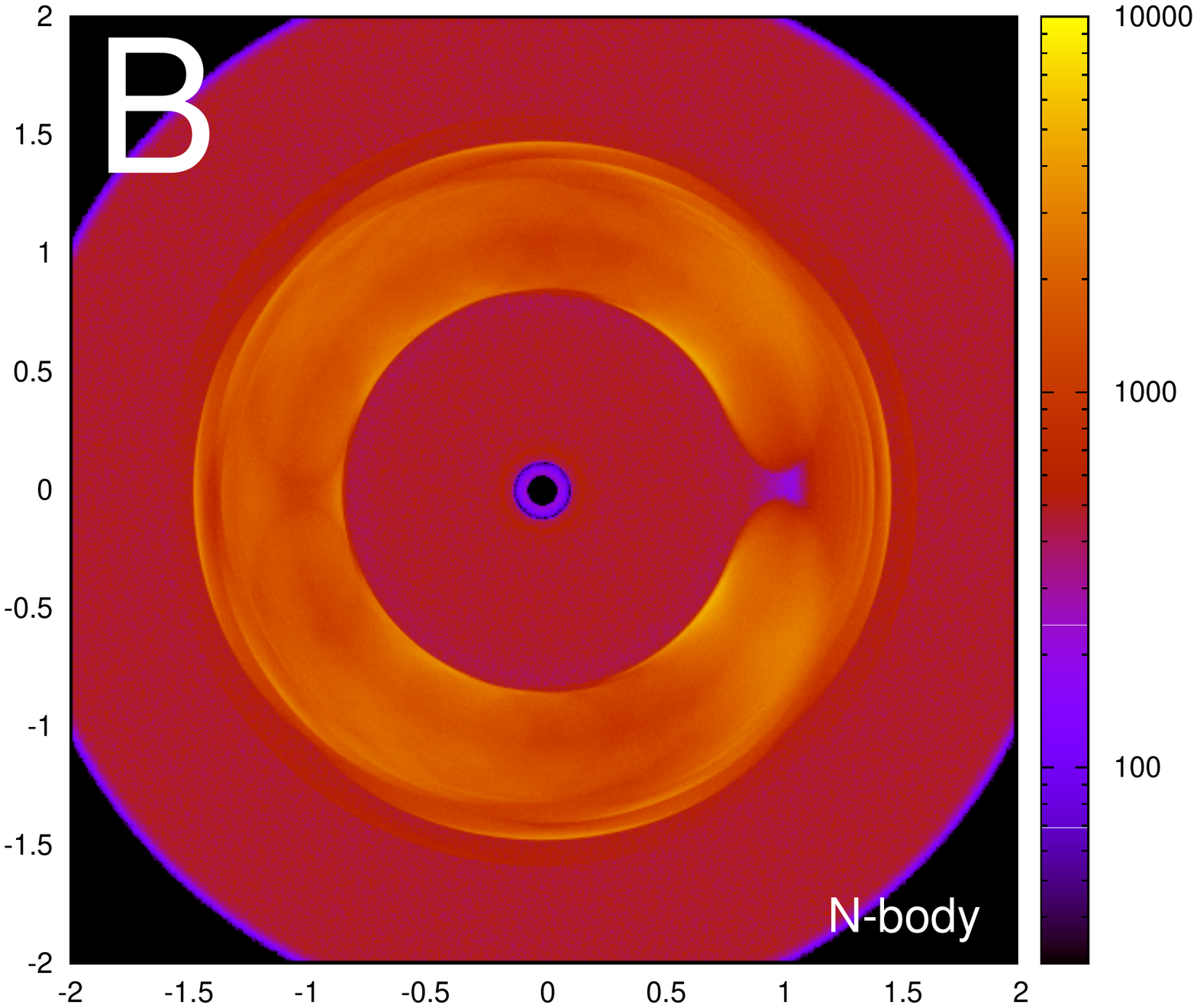}} \\
  \subfigure{\includegraphics[width=0.24\textwidth,trim = 100 100 100 100]{semianalyticbp01disk.pdf}}
  \subfigure{\includegraphics[width=0.24\textwidth,trim = 100 100 100 100]{nbodybp01disk.pdf}}
  \subfigure{\includegraphics[width=0.24\textwidth,trim = 100 100 100 100]{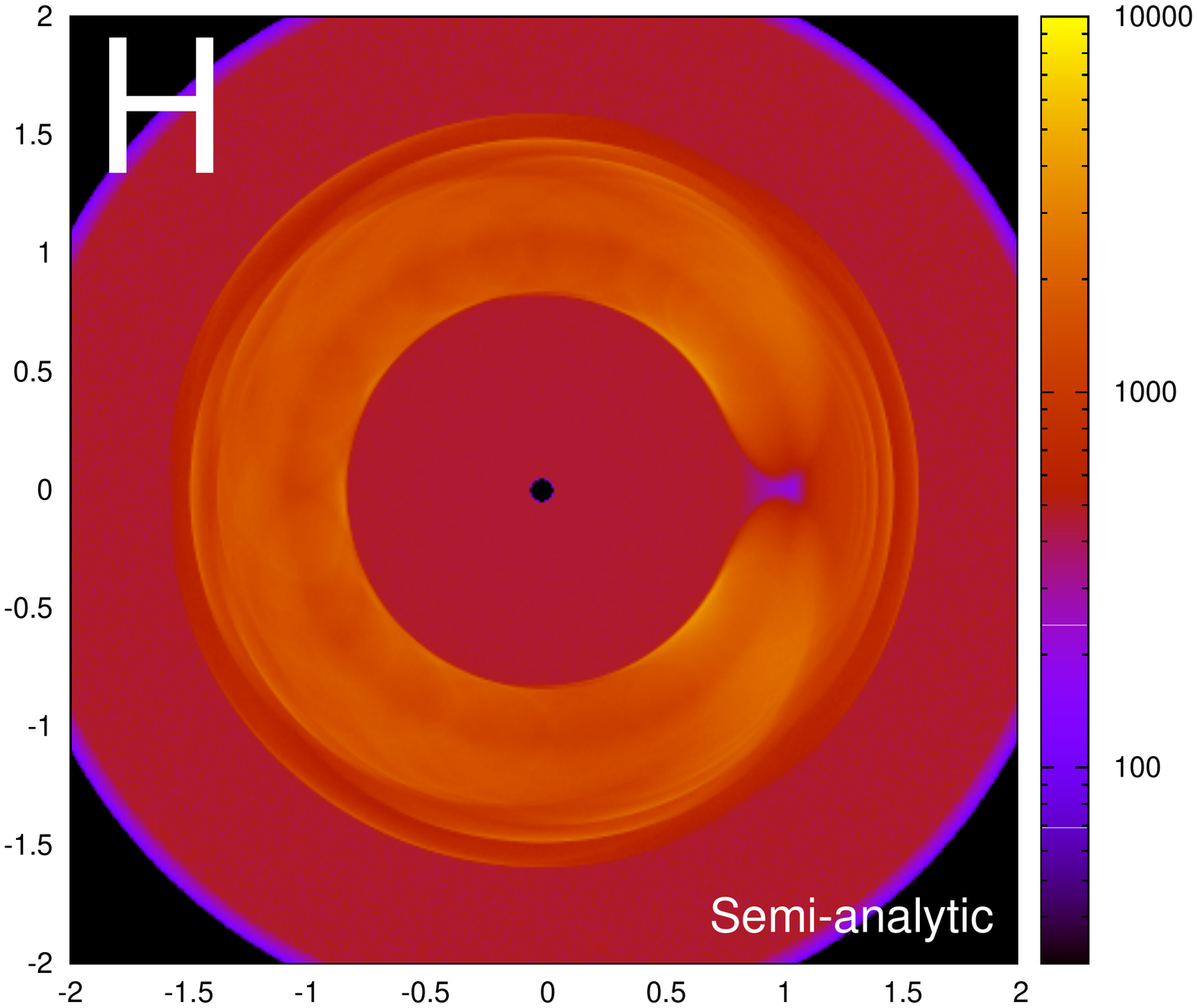}}
  \subfigure{\includegraphics[width=0.24\textwidth,trim = 100 100 100 100]{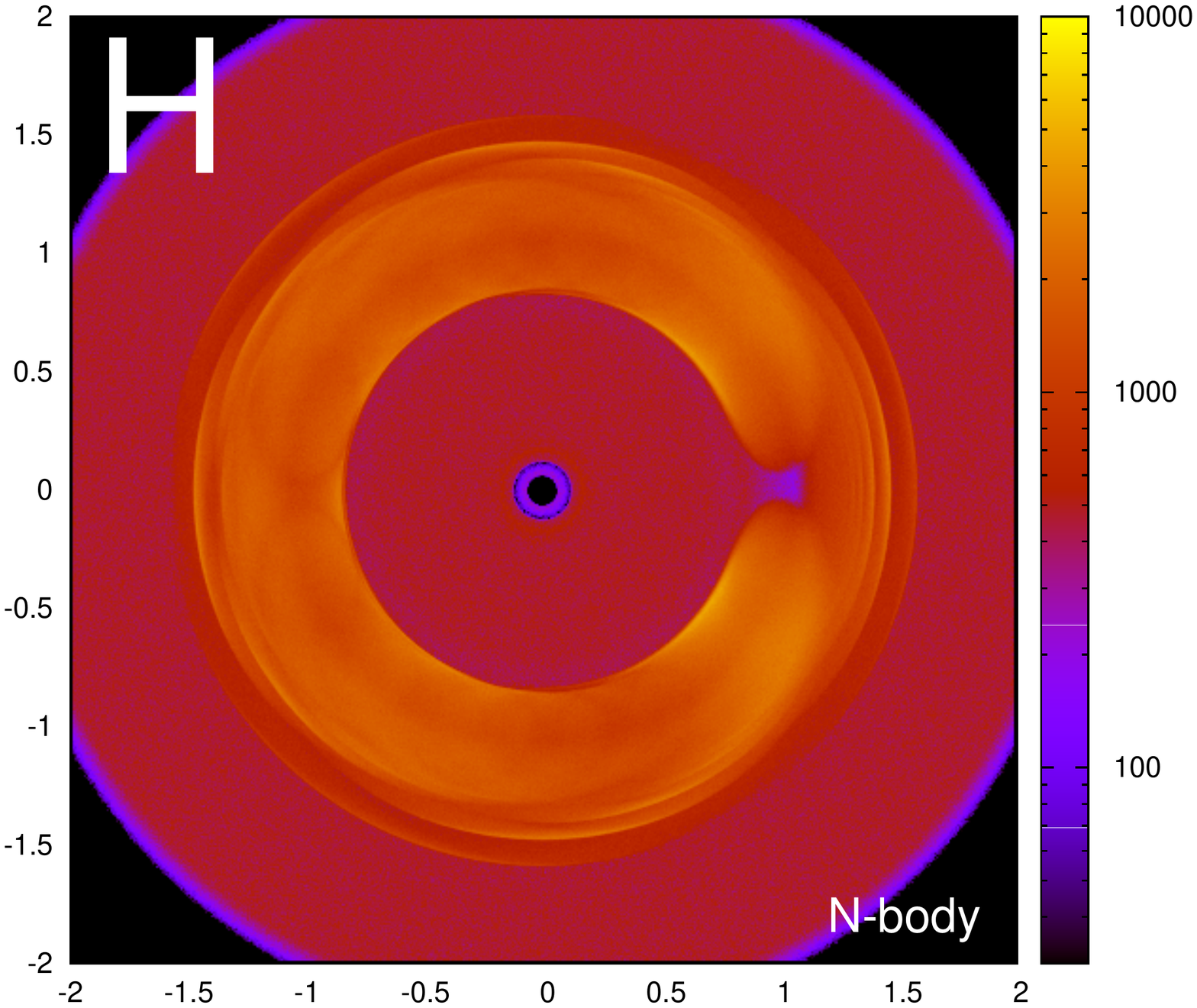}} \\
  \subfigure{\includegraphics[width=0.24\textwidth,trim = 100 100 100 100]{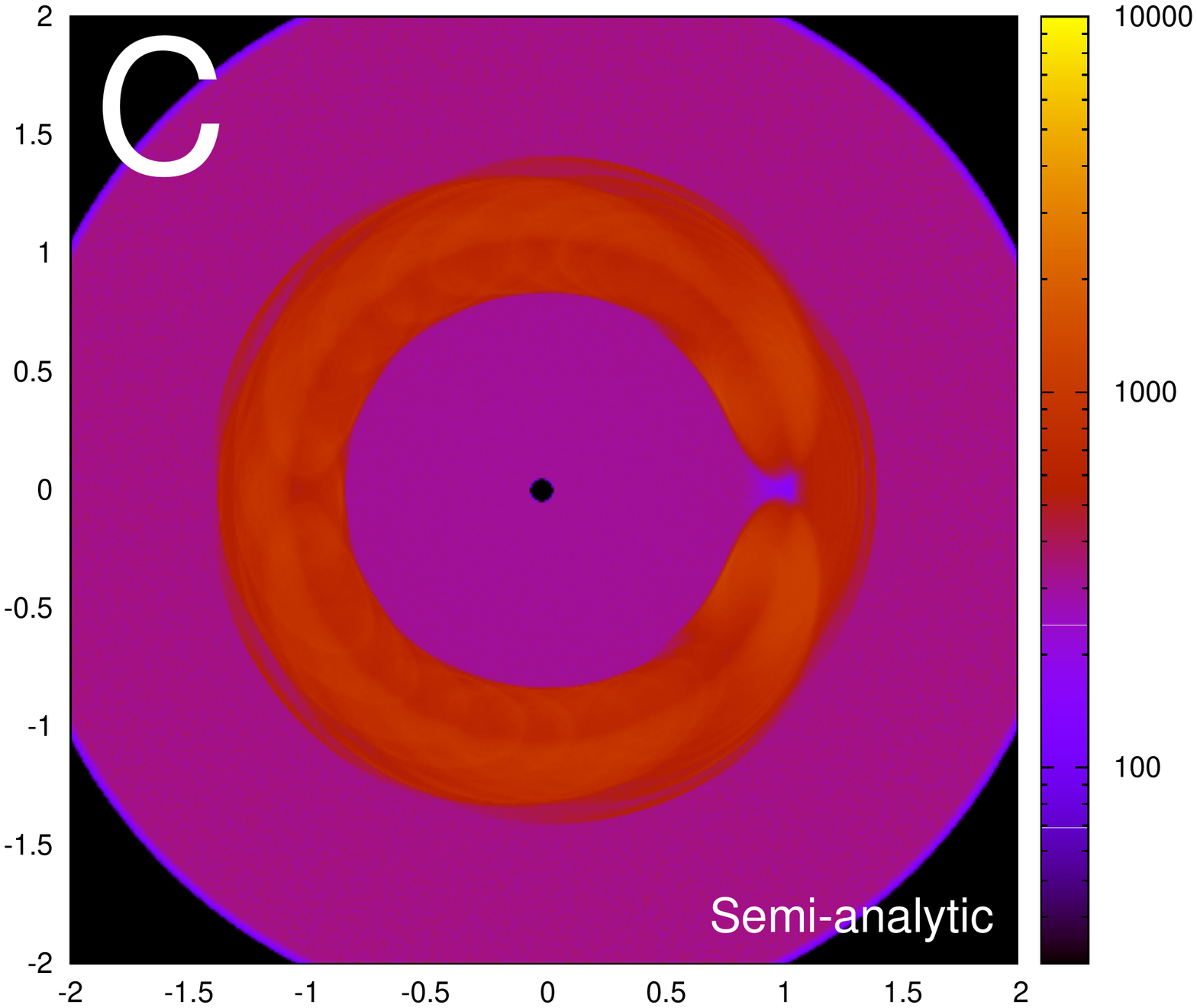}}
  \subfigure{\includegraphics[width=0.24\textwidth,trim = 100 100 100 100]{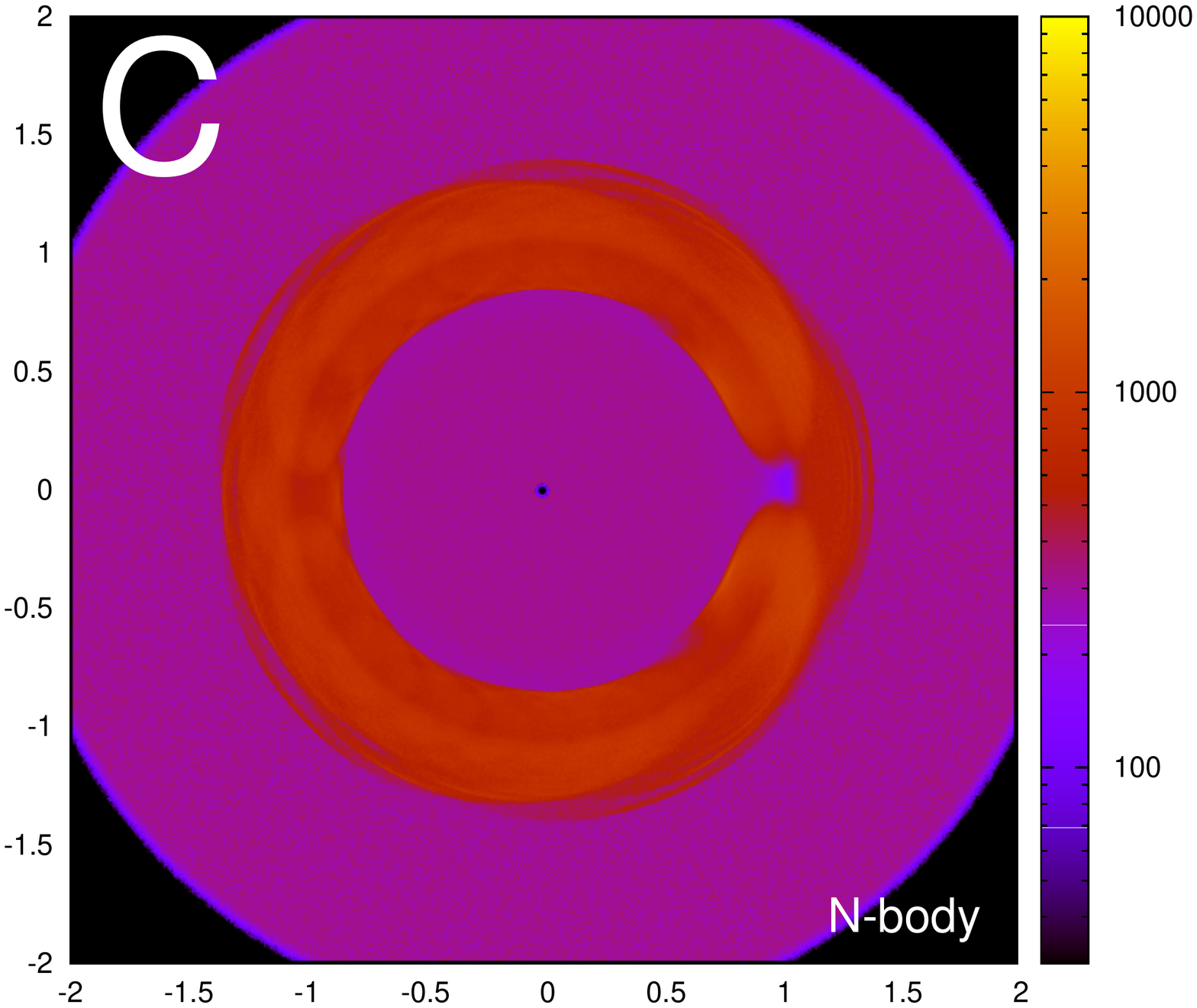}}
  \subfigure{\includegraphics[width=0.24\textwidth,trim = 100 100 100 100]{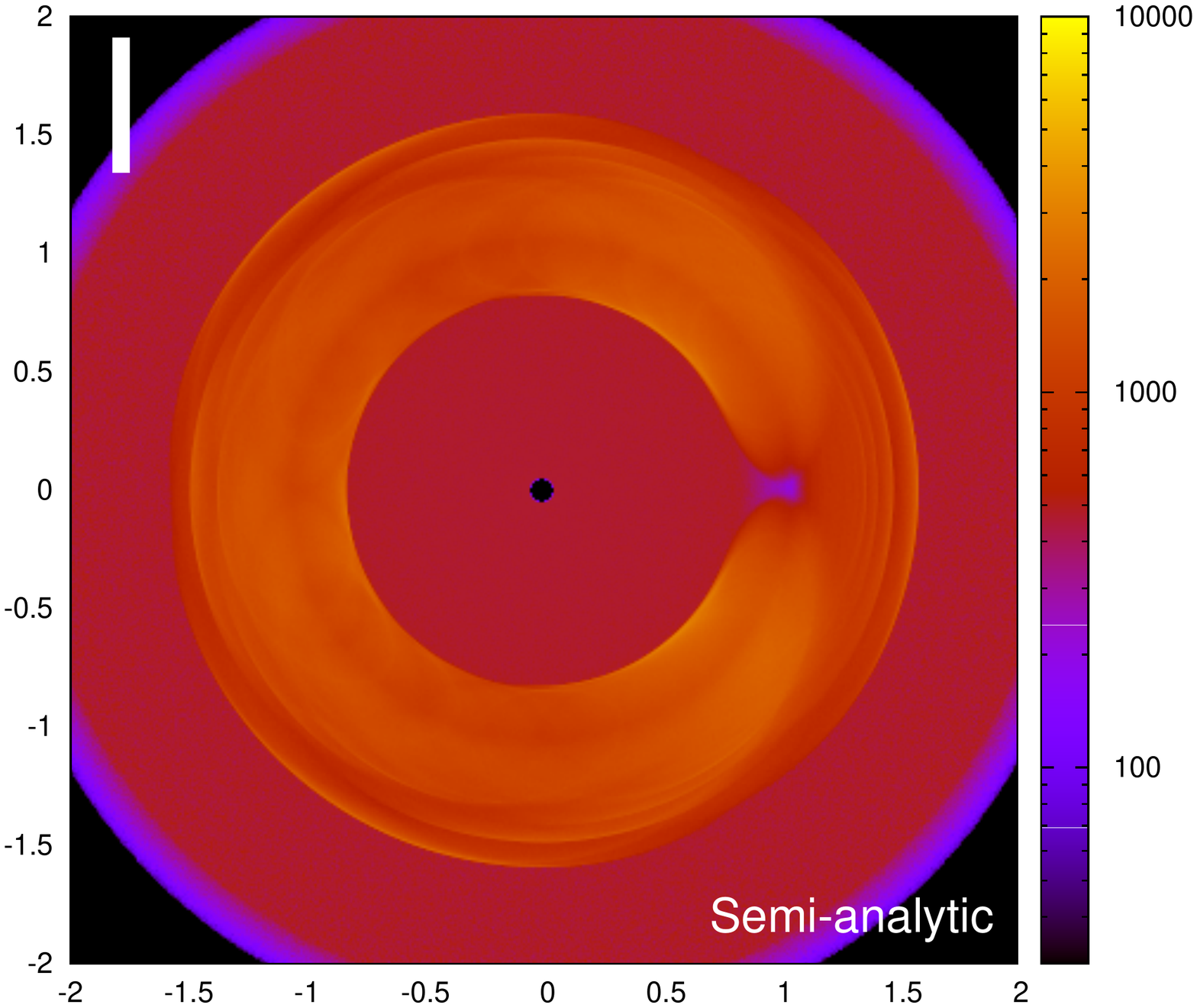}}
  \subfigure{\includegraphics[width=0.24\textwidth,trim = 100 100 100 100]{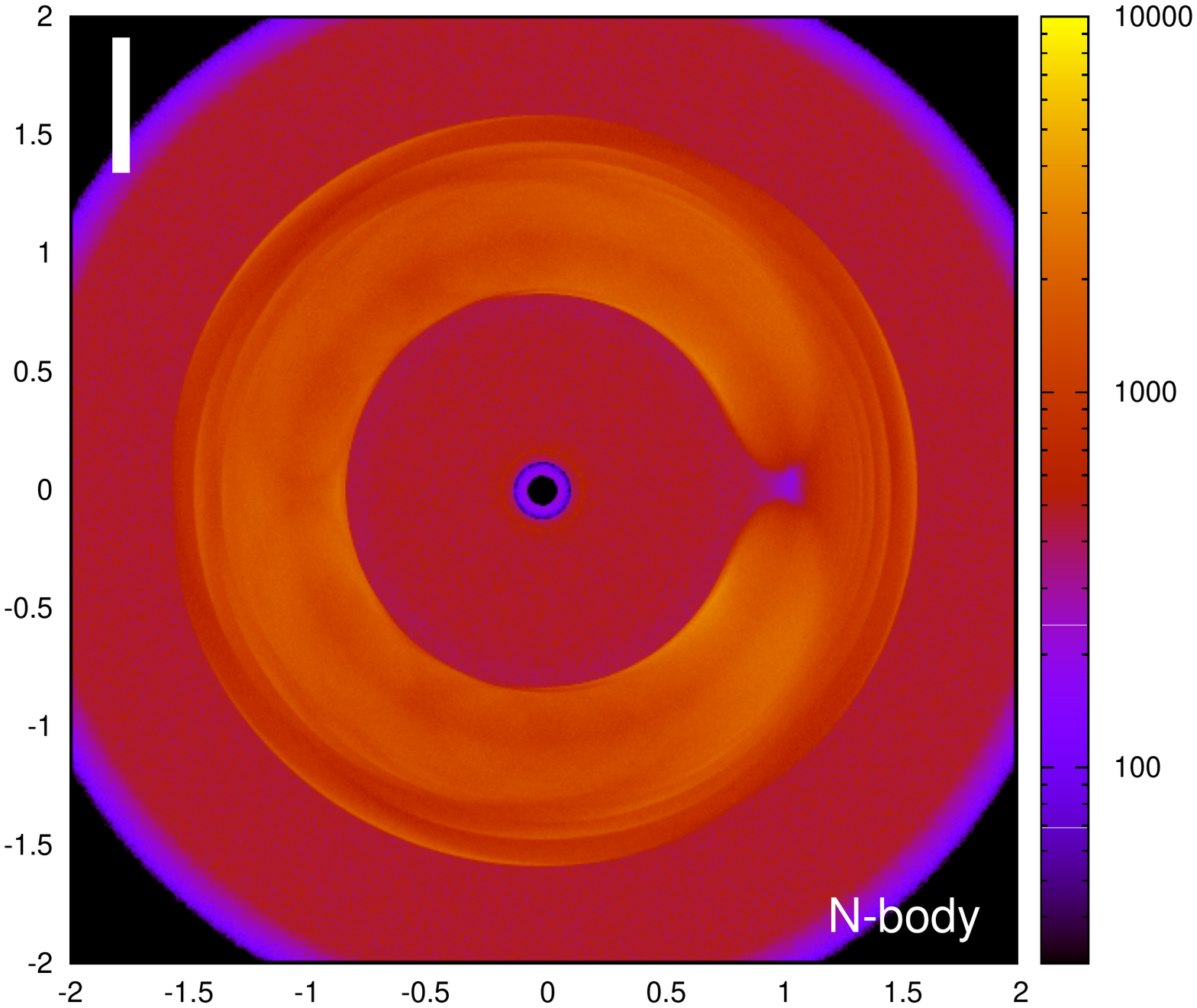}} \\
  \subfigure{\includegraphics[width=0.24\textwidth,trim = 100 100 100 100]{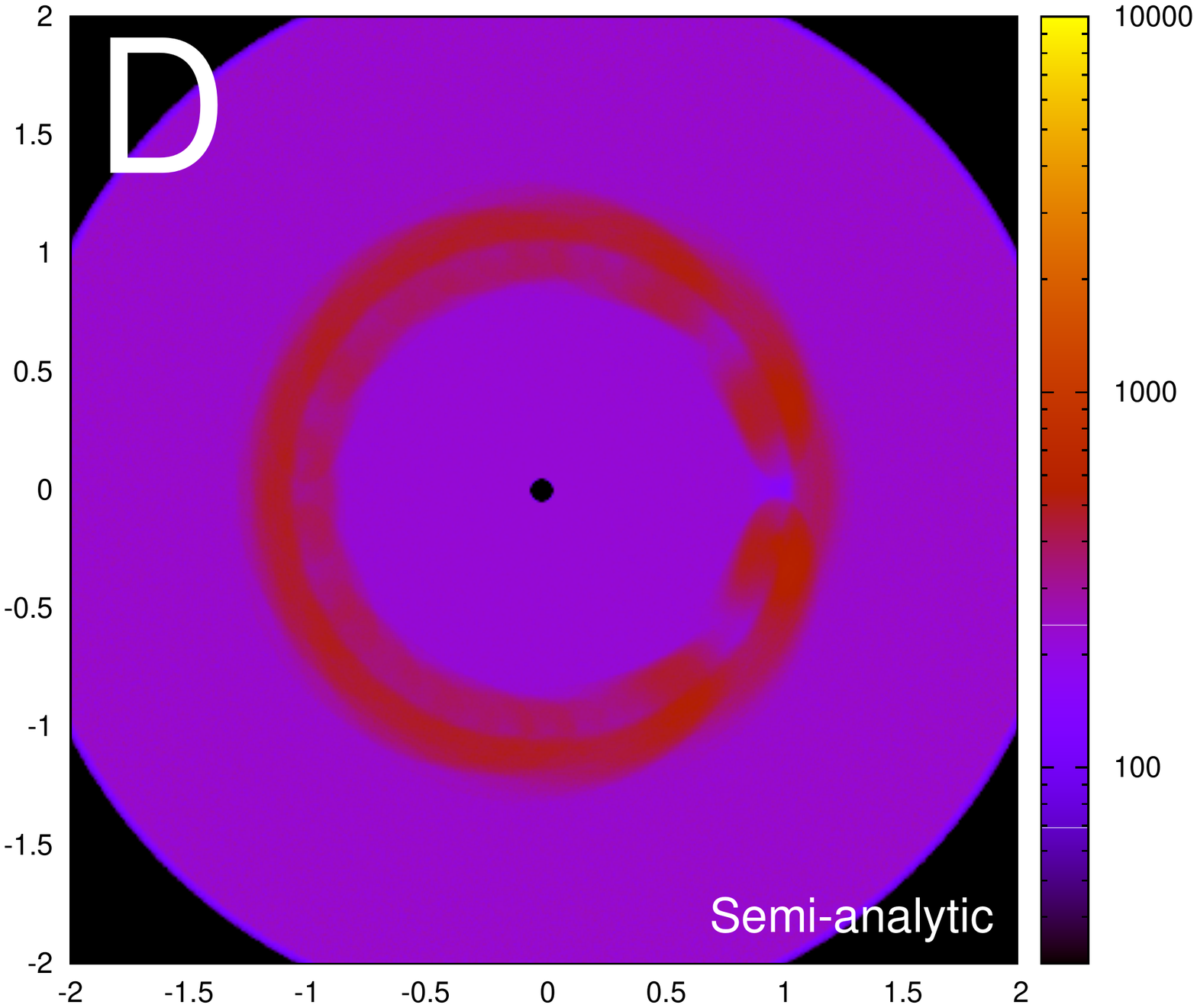}}
  \subfigure{\includegraphics[width=0.24\textwidth,trim = 100 100 100 100]{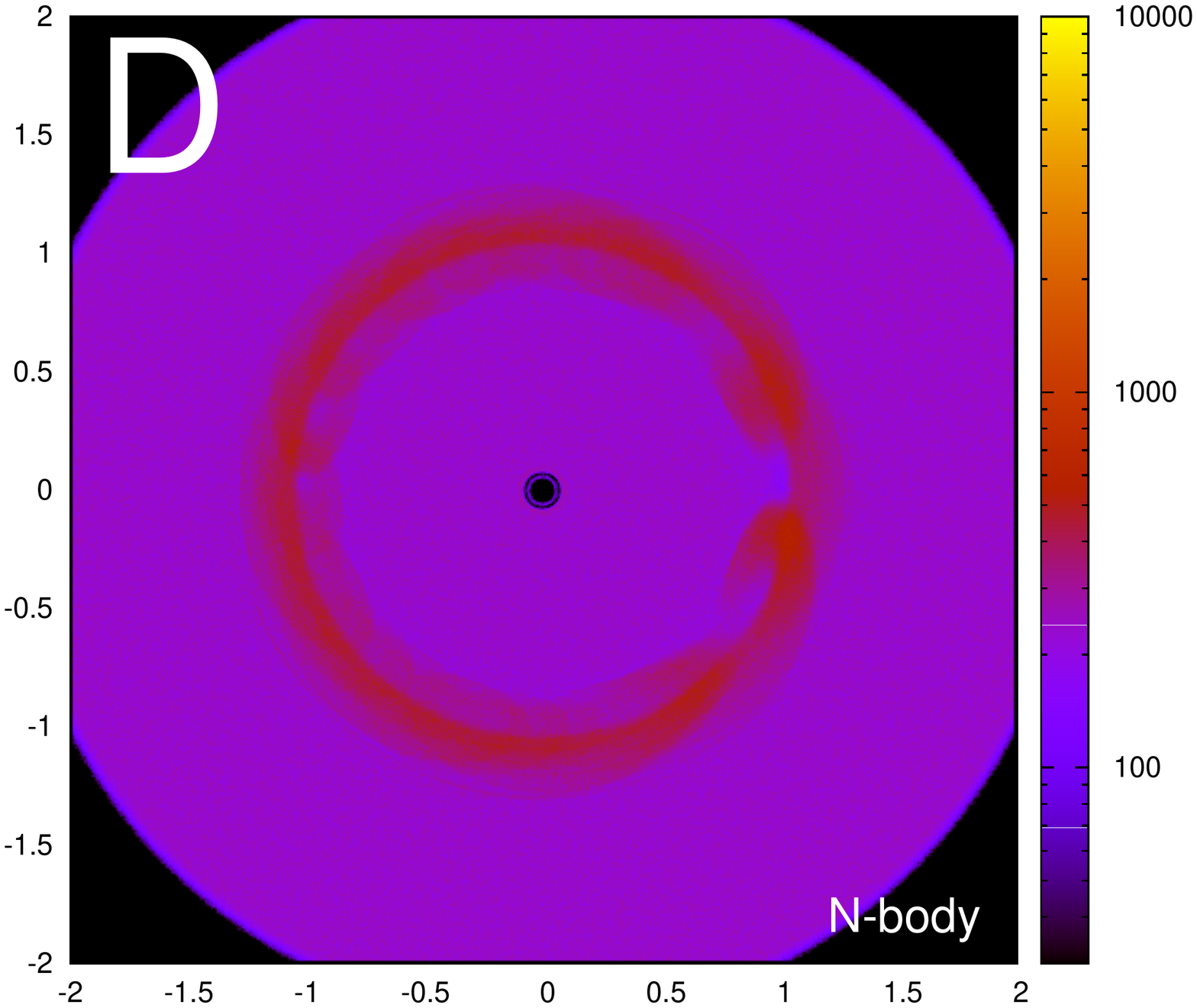}}
  \subfigure{\includegraphics[width=0.24\textwidth,trim = 100 100 100 100]{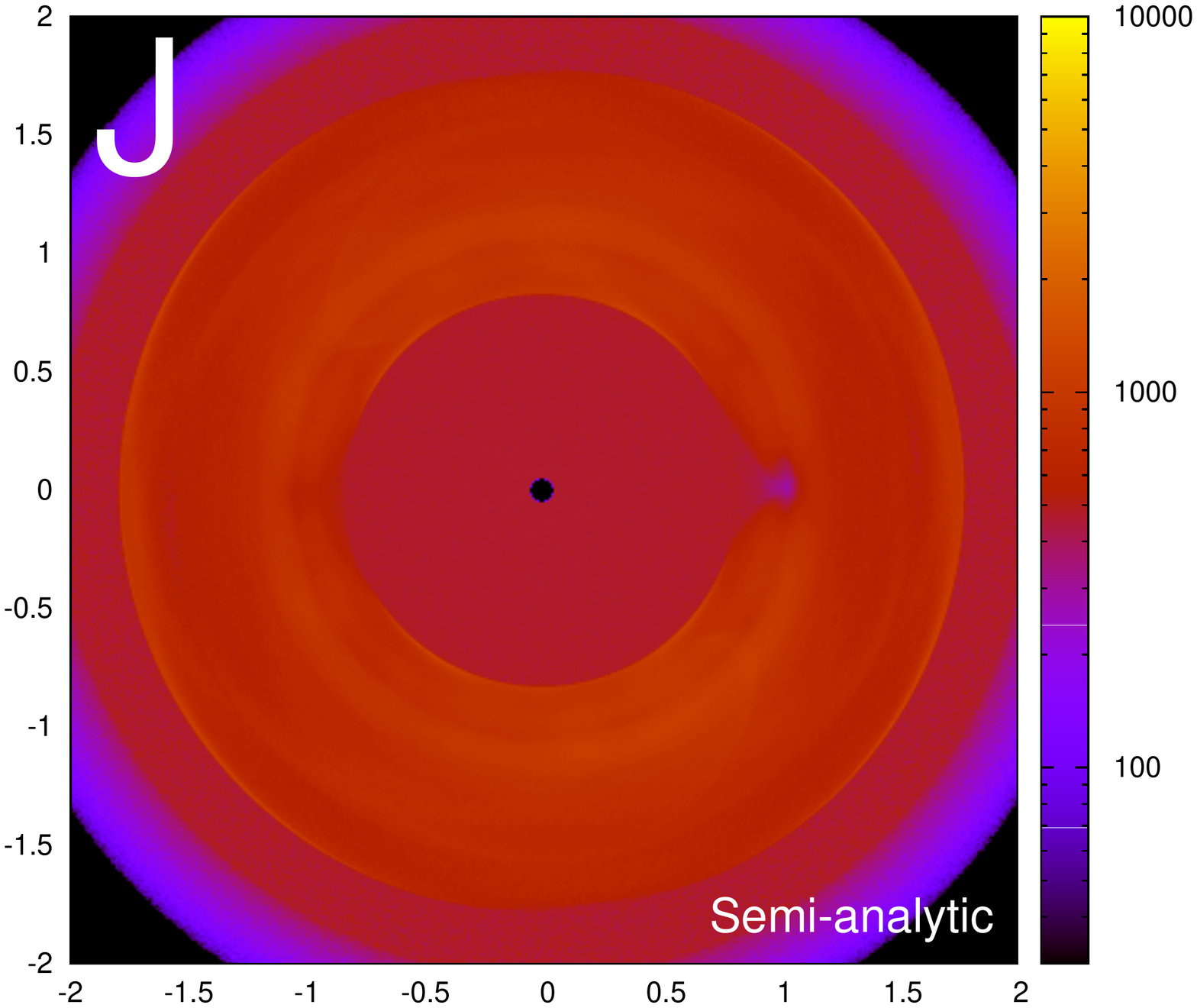}}
  \subfigure{\includegraphics[width=0.24\textwidth,trim = 100 100 100 100]{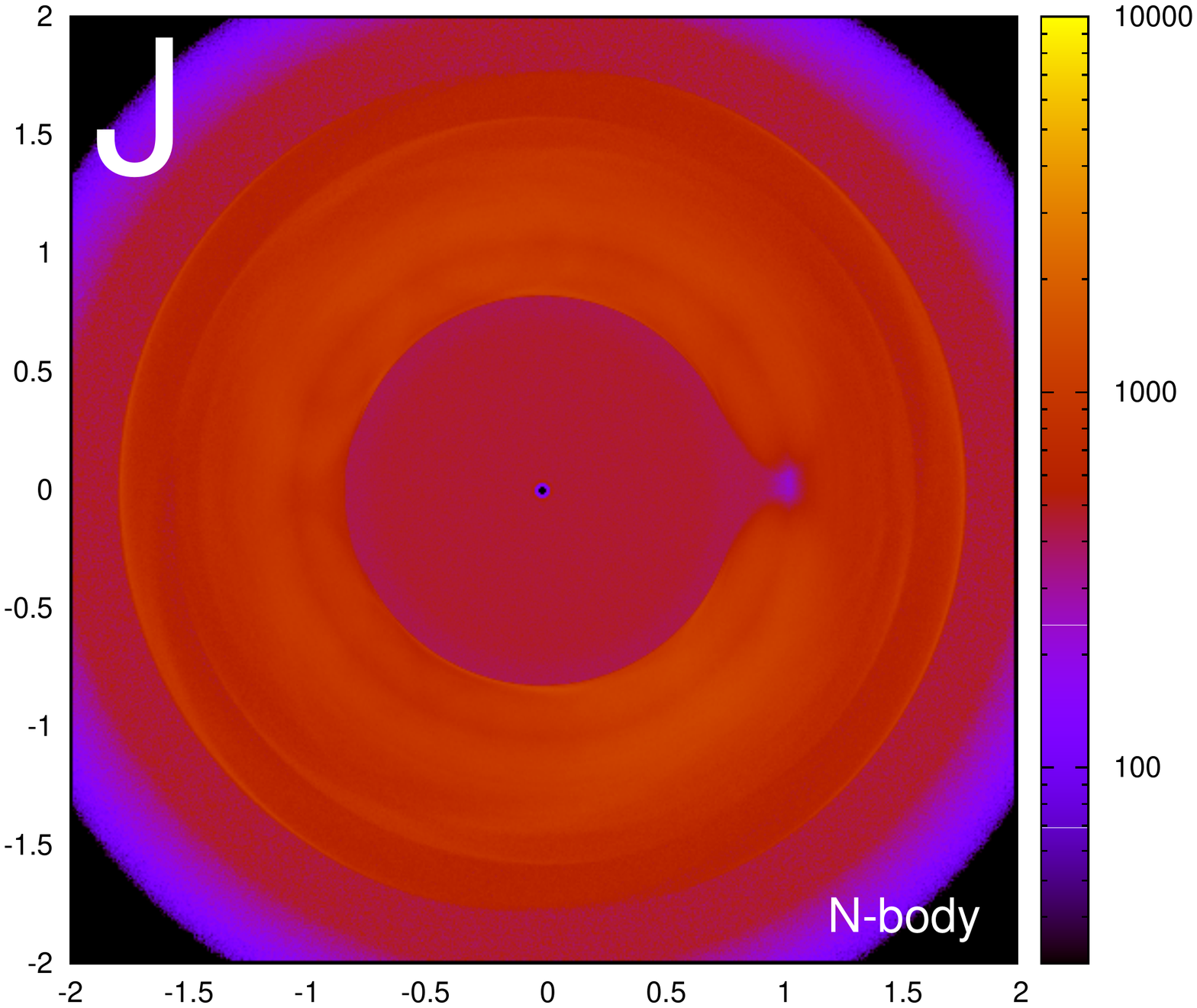}} \\
  \subfigure{\includegraphics[width=0.24\textwidth,trim = 100 100 100 100]{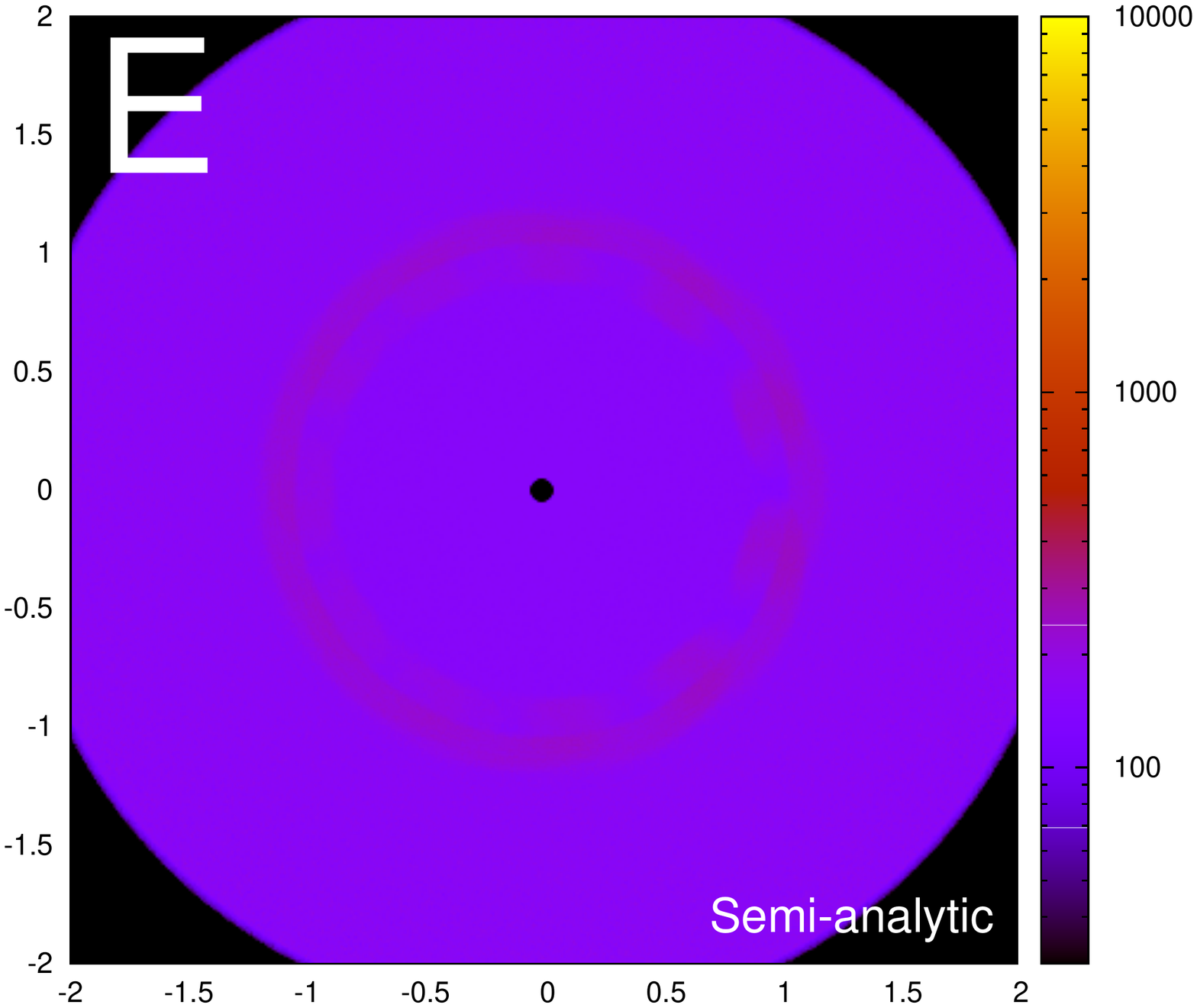}}
  \subfigure{\includegraphics[width=0.24\textwidth,trim = 100 100 100 100]{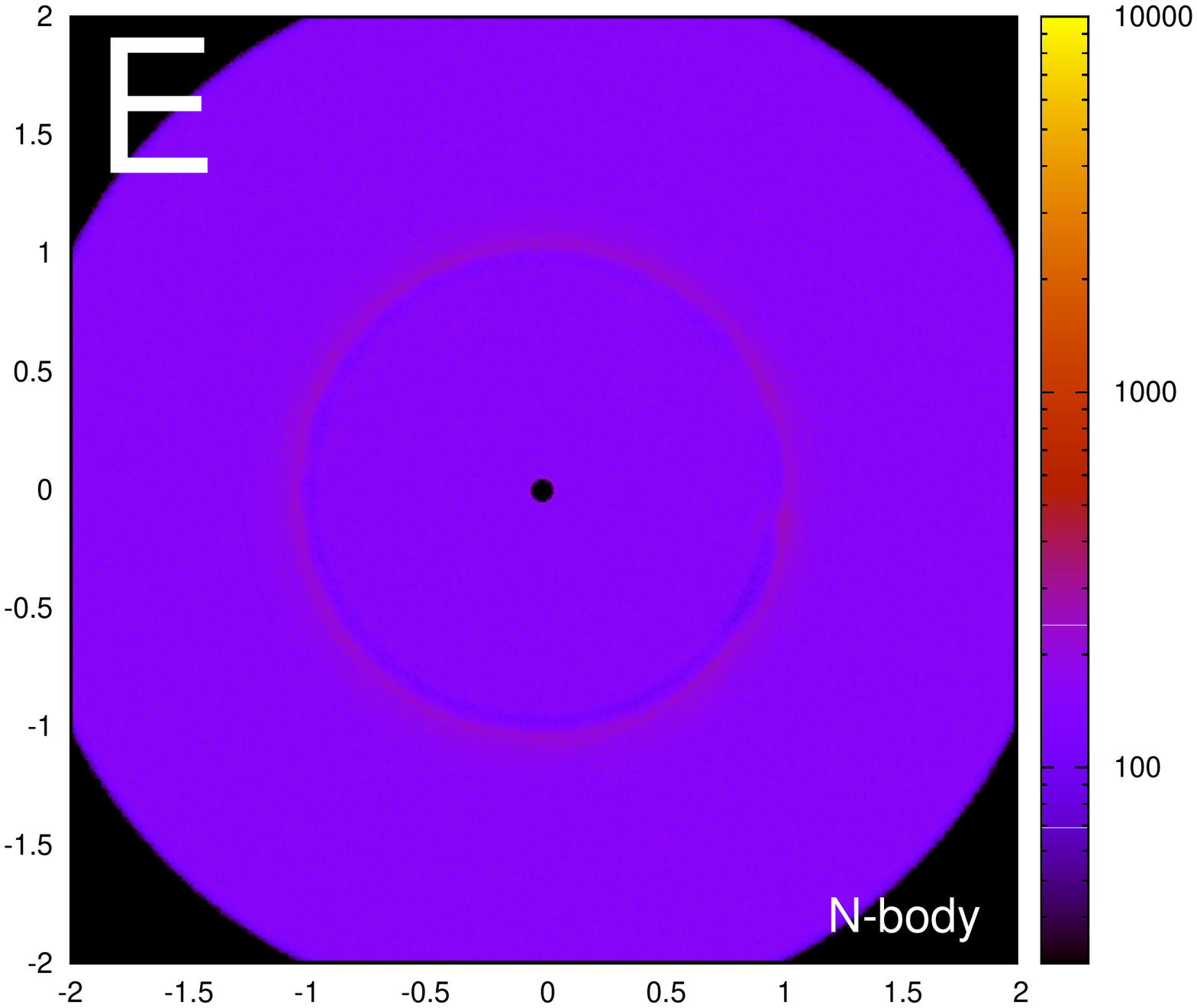}}
  \subfigure{\includegraphics[width=0.24\textwidth,trim = 100 100 100 100]{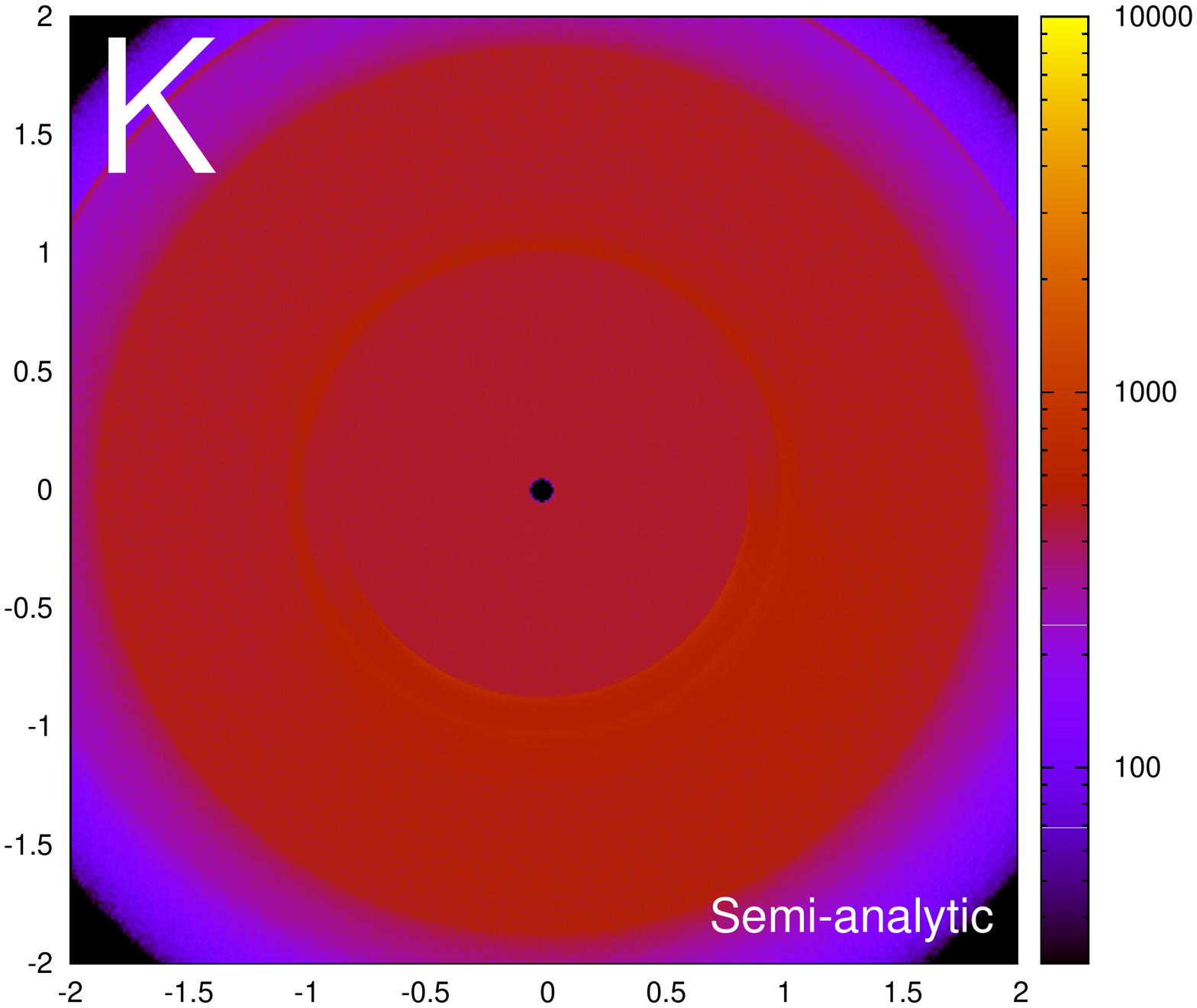}}
  \subfigure{\includegraphics[width=0.24\textwidth,trim = 100 100 100 100]{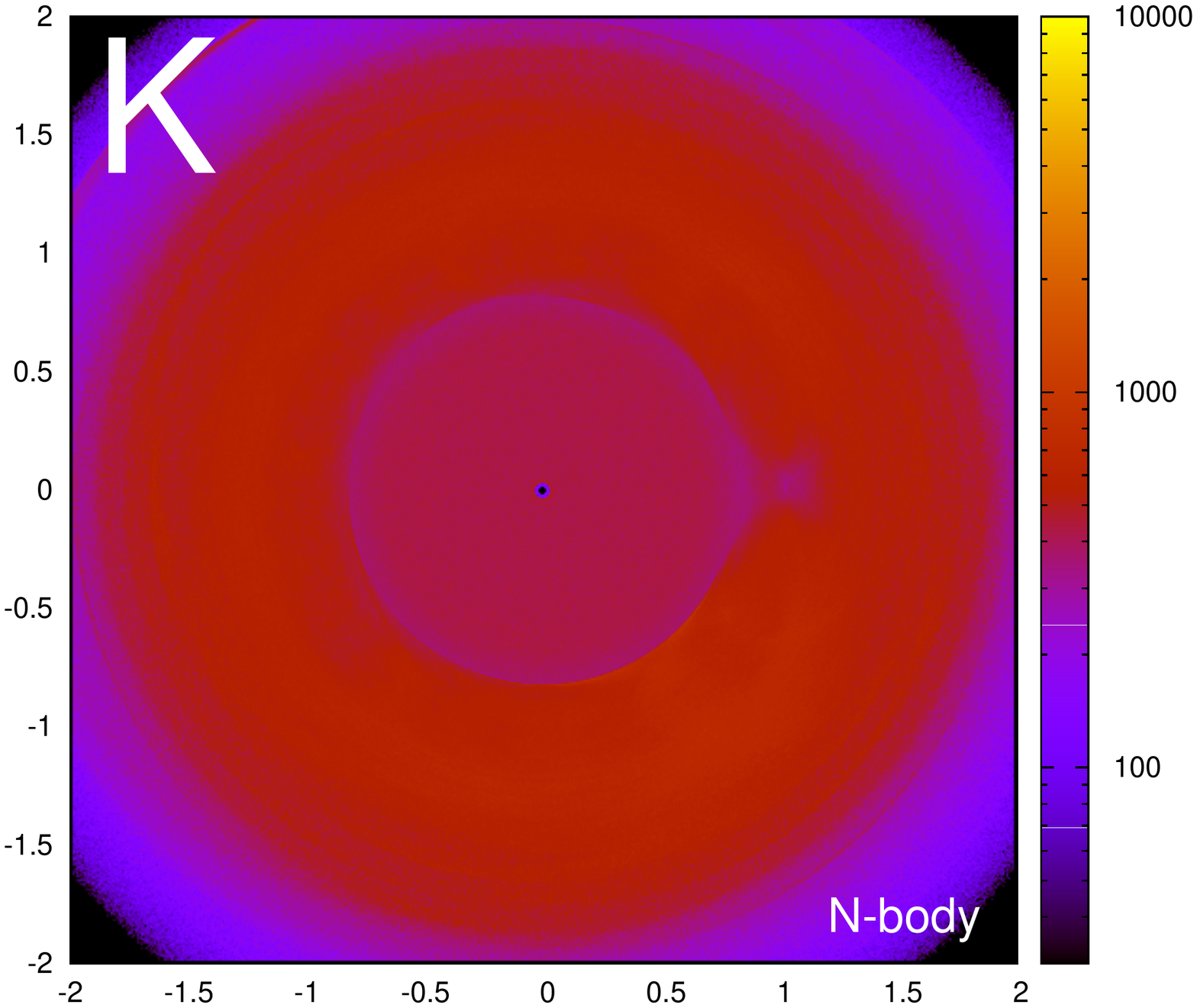}} \\
  \subfigure{\includegraphics[width=0.24\textwidth,trim = 100 100 100 100]{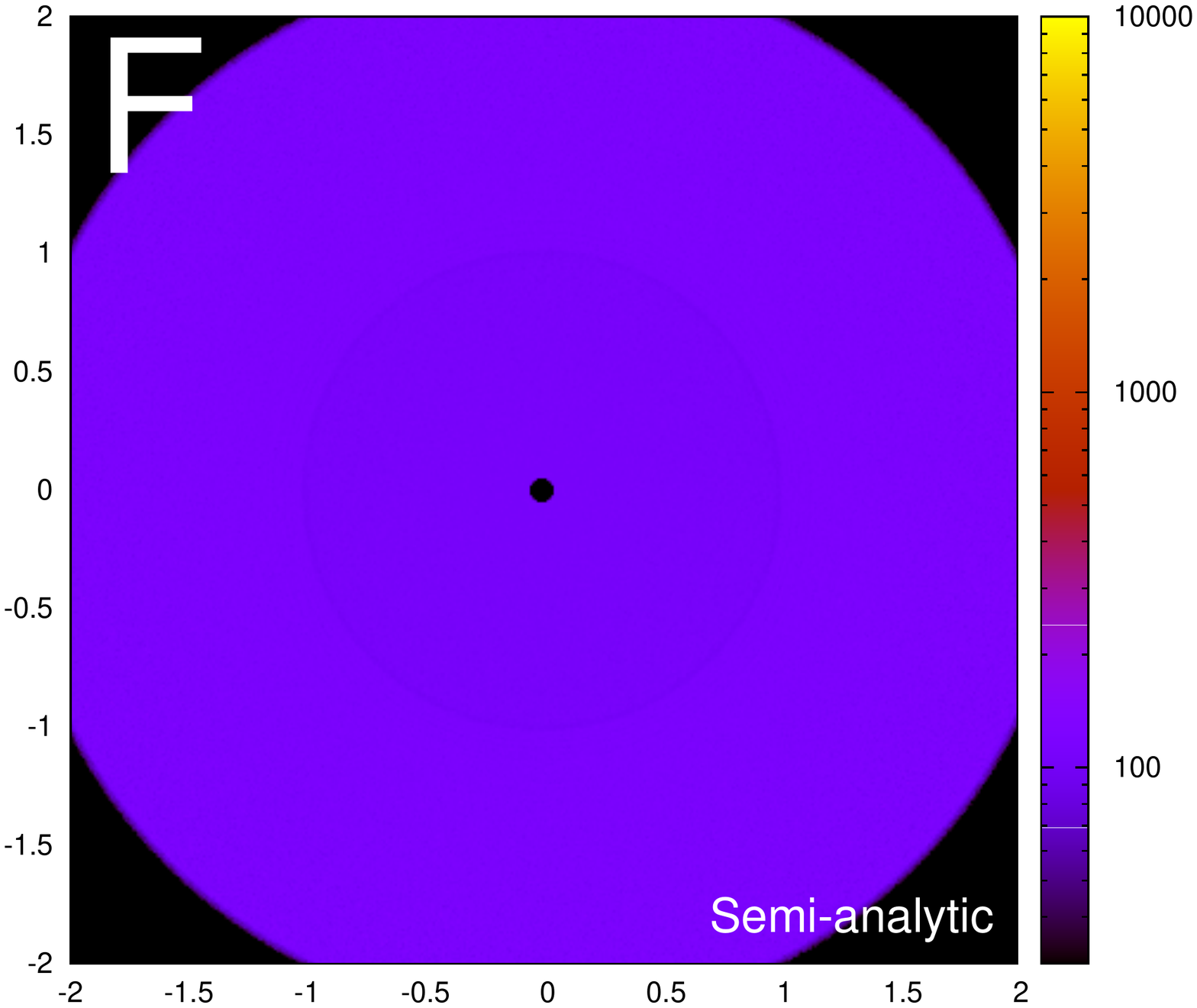}}
  \subfigure{\includegraphics[width=0.24\textwidth,trim = 100 100 100 100]{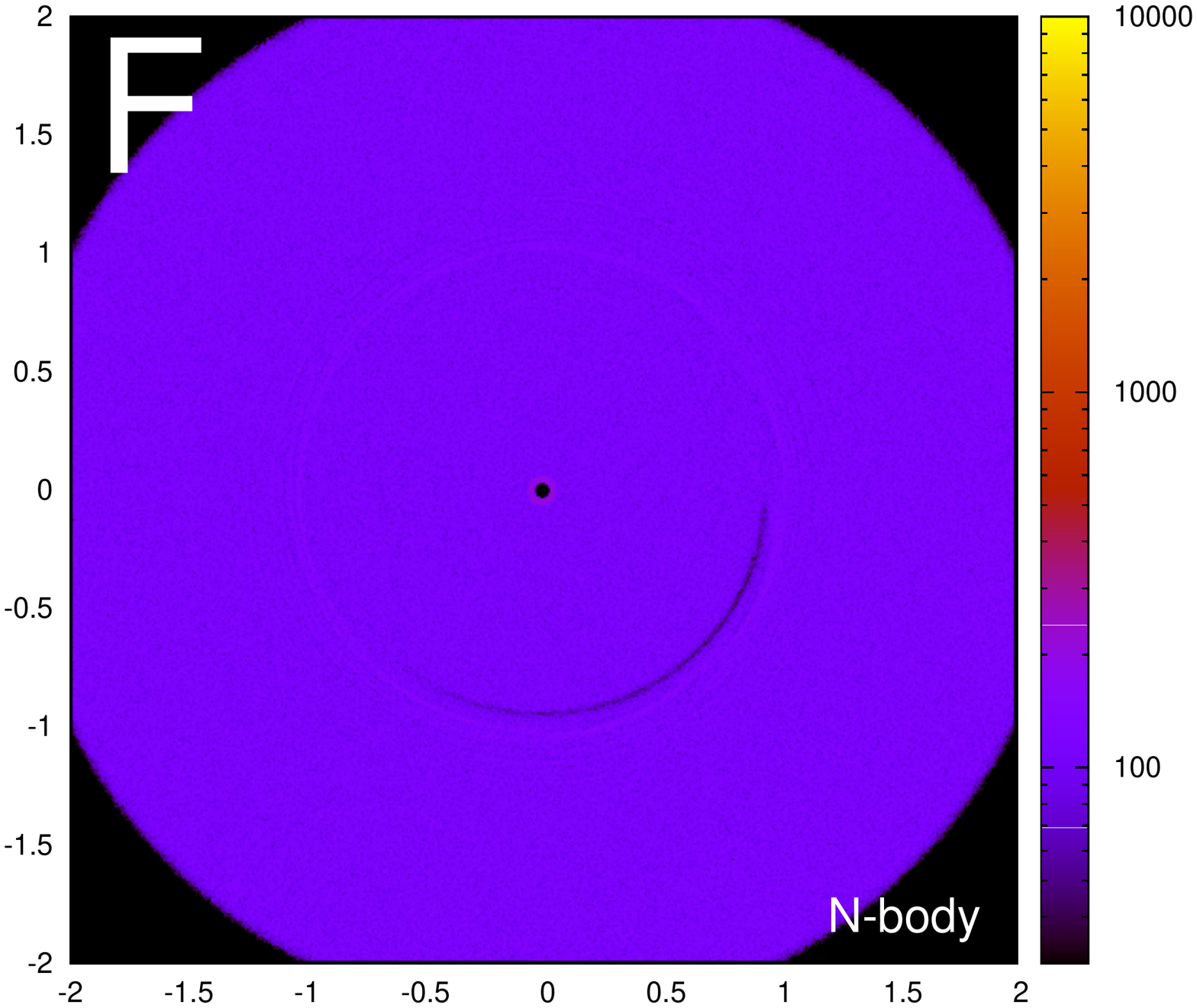}}
  \subfigure{\includegraphics[width=0.24\textwidth,trim = 100 100 100 100]{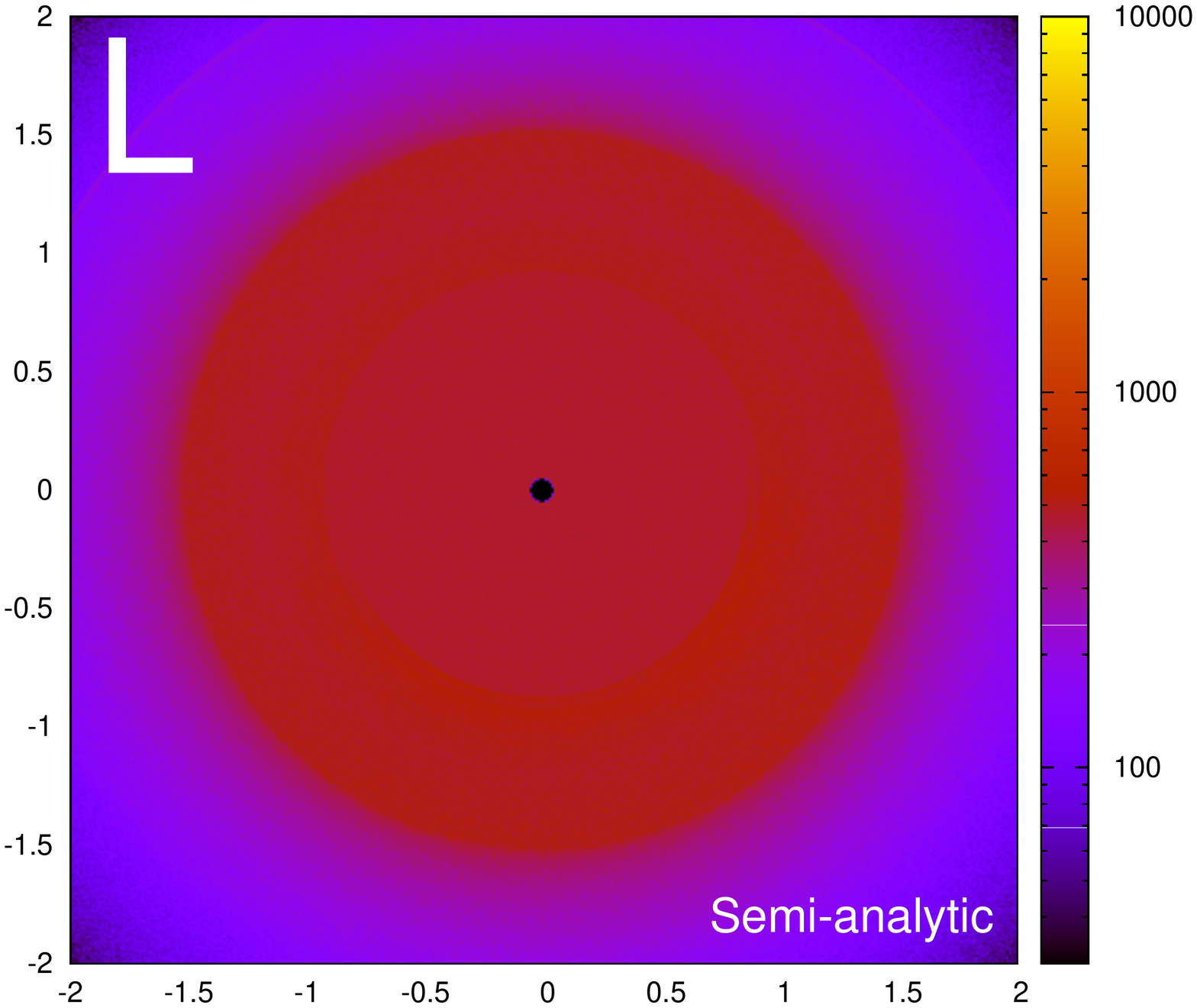}}
  \subfigure{\includegraphics[width=0.24\textwidth,trim = 100 100 100 100]{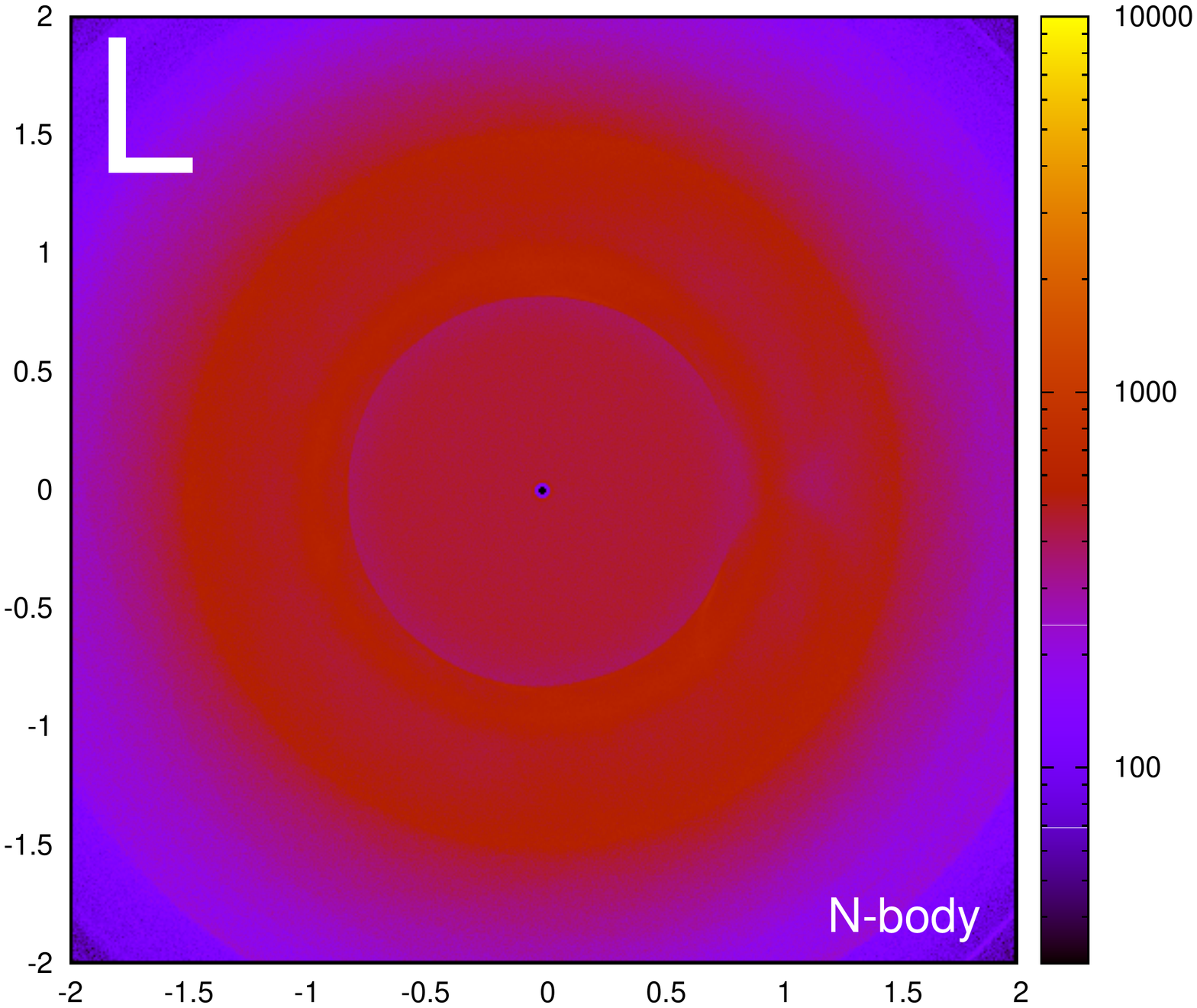}} \\
  \subfigure{\includegraphics[width=0.24\textwidth,trim = 100 100 100 100]{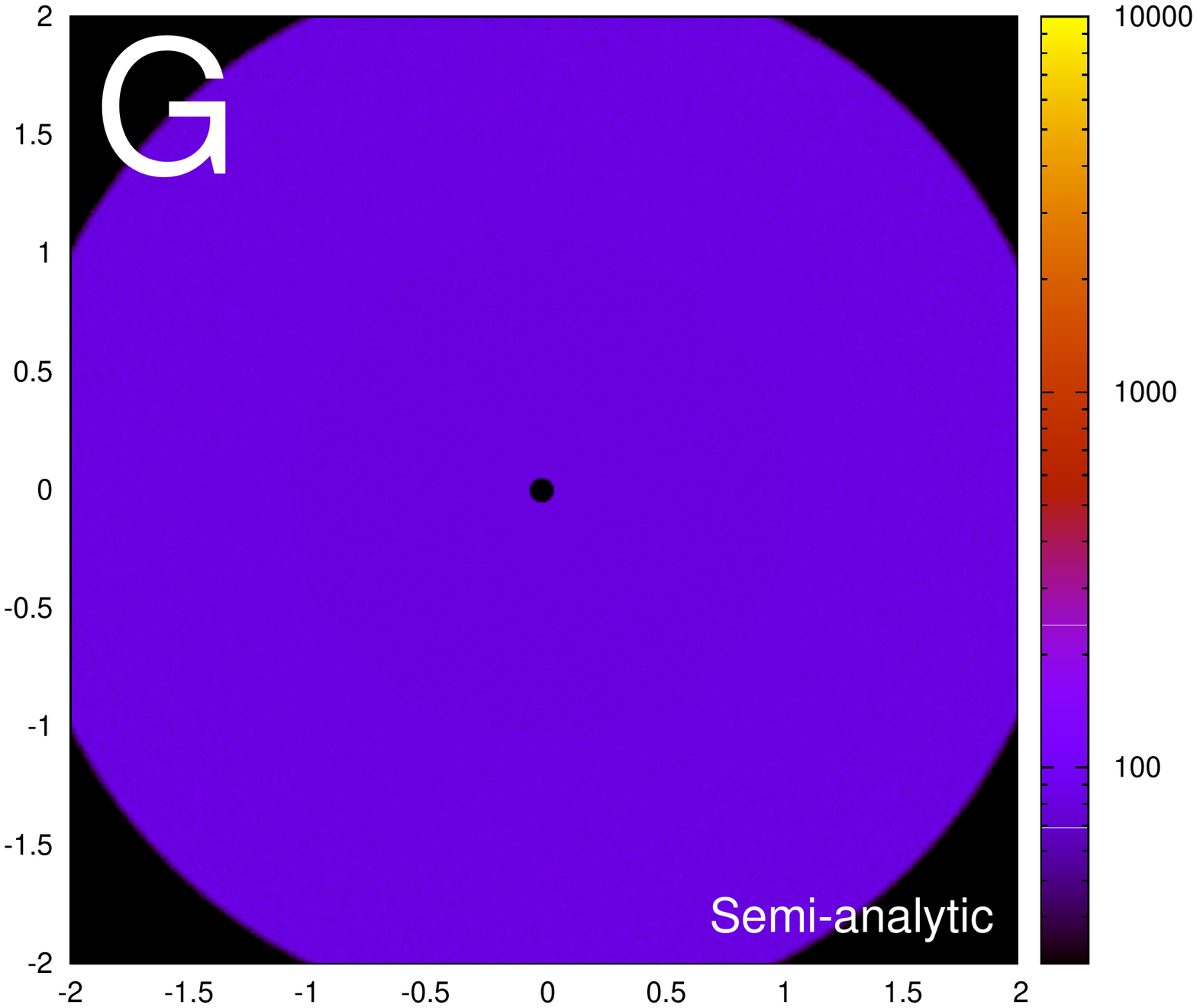}}
  \subfigure{\includegraphics[width=0.24\textwidth,trim = 100 100 100 100]{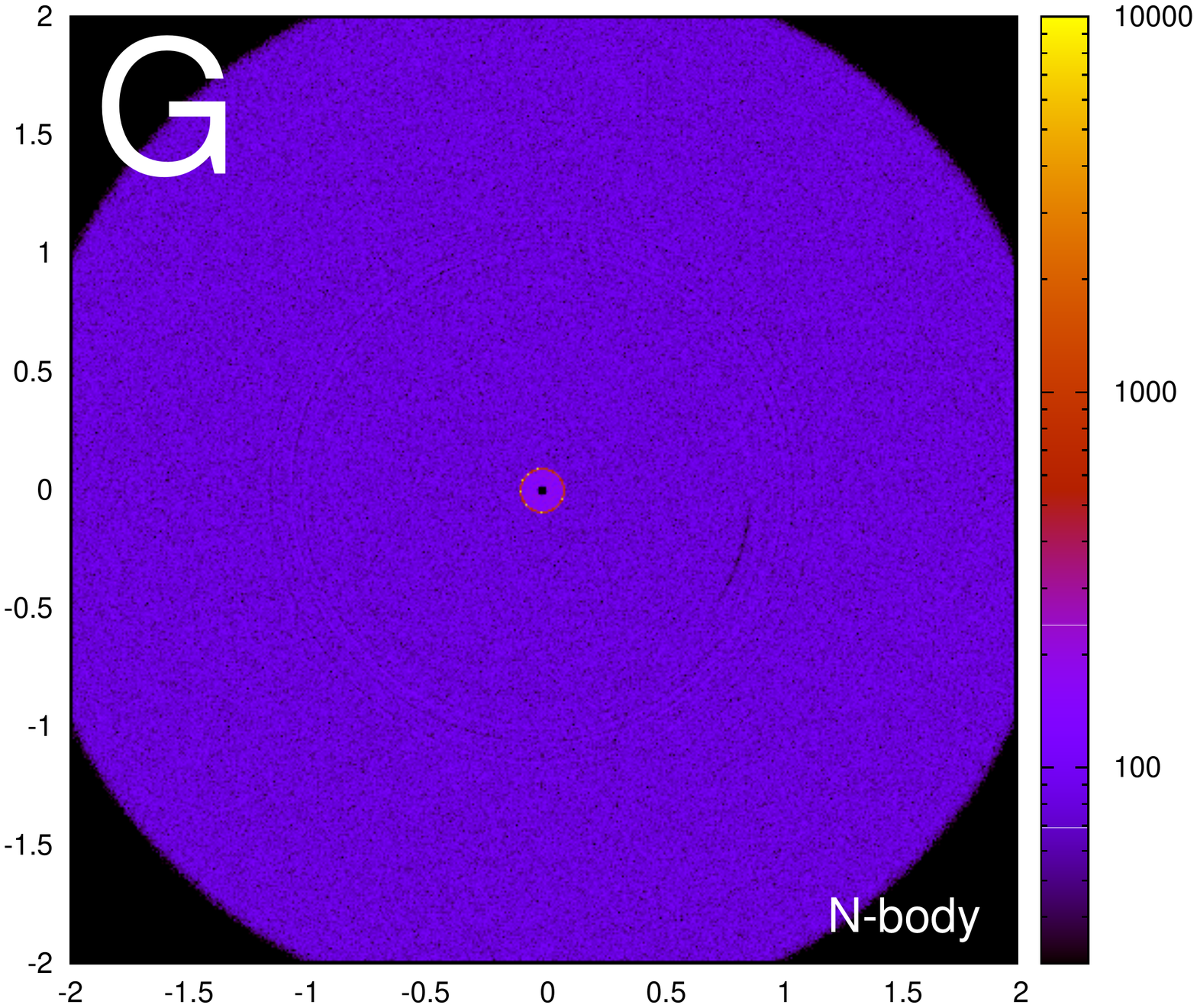}}
  \subfigure{\includegraphics[width=0.24\textwidth,trim = 100 100 100 100]{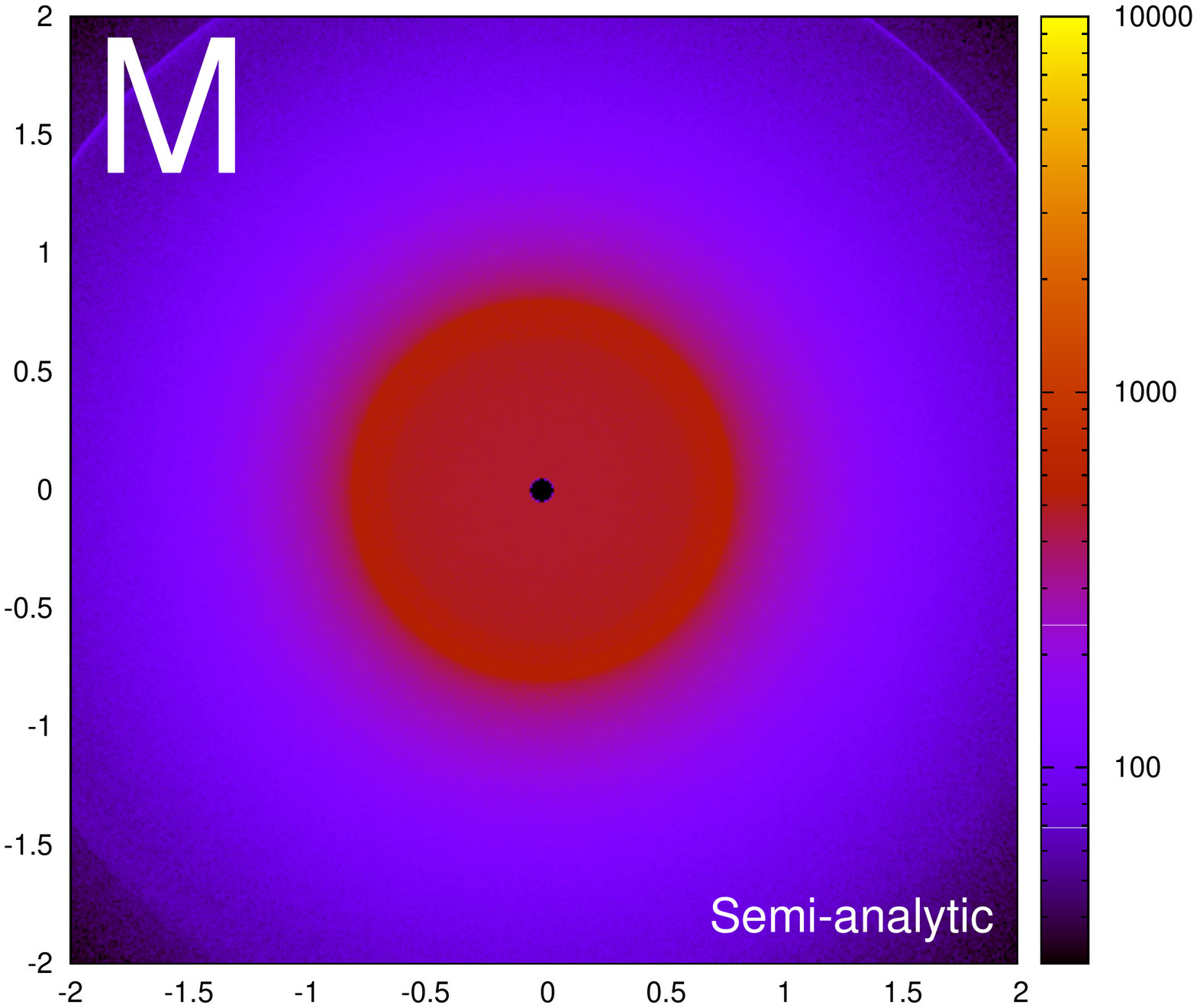}}
  \subfigure{\includegraphics[width=0.24\textwidth,trim = 100 100 100 100]{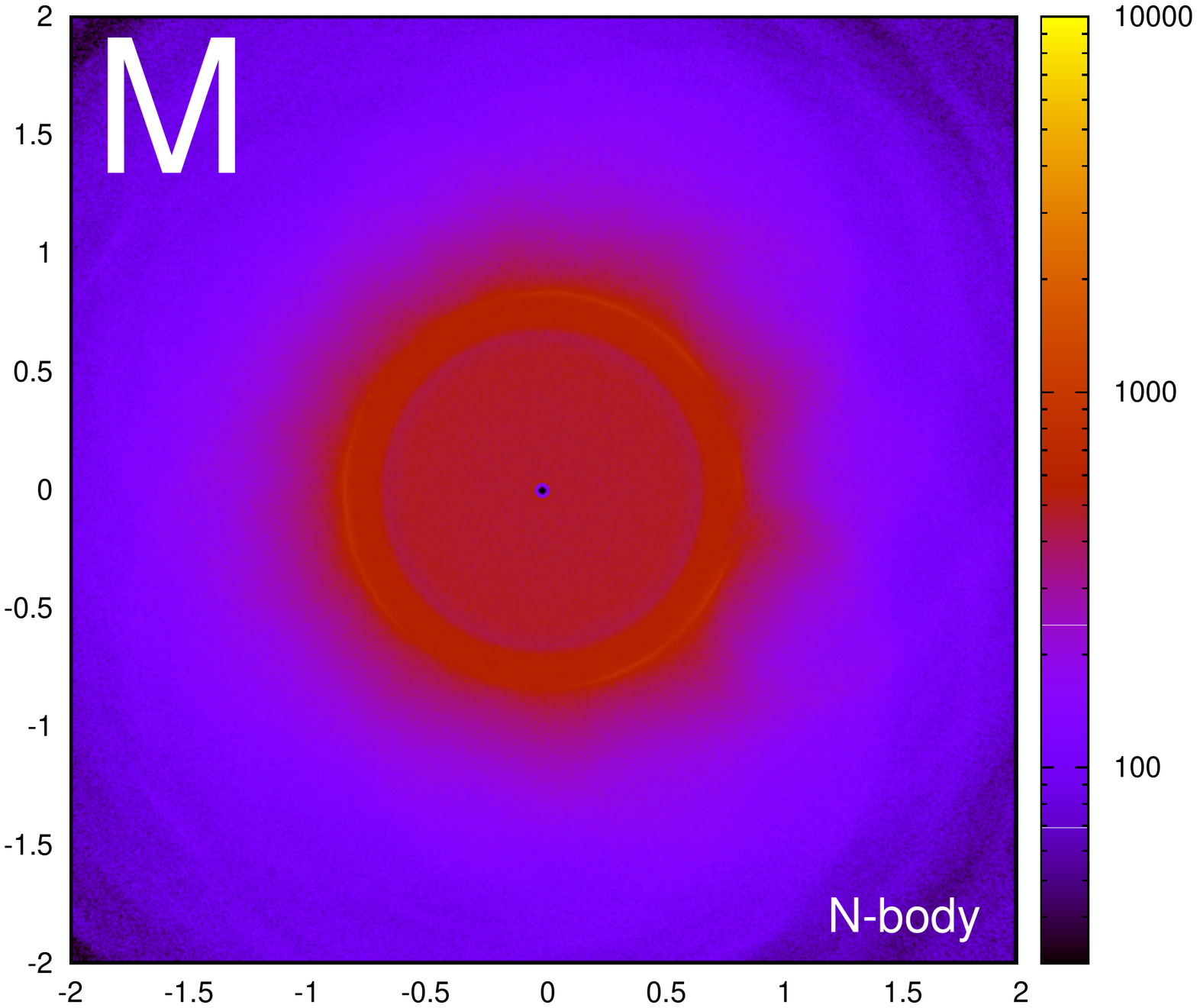}} \\
  \caption{Simulations A - G (variations of $\beta_{\rm{PR}}$)~left (1st column is semi-analytic approach, 2nd column is N-body approach), and Simulations B, H-M (variations of $e_0$)~right (3rd column is semi-analytic approach, 4th column is N-body approach).  In all plots, 10000 particles are each plotted once every $10^3\times\beta_{\rm{PR}}^{-0.5}a^{1.5}$~days, in a grid of 400 by 400 cells covering from $-2a_p$~to $2a_p$.  To produce the N-body plot, we rotate the dust grain around the star so the planet is located at (1,0).  All the subsequent disk images are produced in the same fashion.}
  \label{fig:allplots}
\end{figure*}

\begin{figure*}
  \centering
  \subfigure{\includegraphics[width=0.24\textwidth,trim = 100 100 100 100]{semianalyticbp01disk.pdf}}
  \subfigure{\includegraphics[width=0.24\textwidth,trim = 100 100 100 100]{nbodybp01disk.pdf}}
  \subfigure{\includegraphics[width=0.24\textwidth,trim = 100 100 100 100]{semianalyticbp01disk.pdf}}
  \subfigure{\includegraphics[width=0.24\textwidth,trim = 100 100 100 100]{nbodybp01disk.pdf}} \\
  \subfigure{\includegraphics[width=0.24\textwidth,trim = 100 100 100 100]{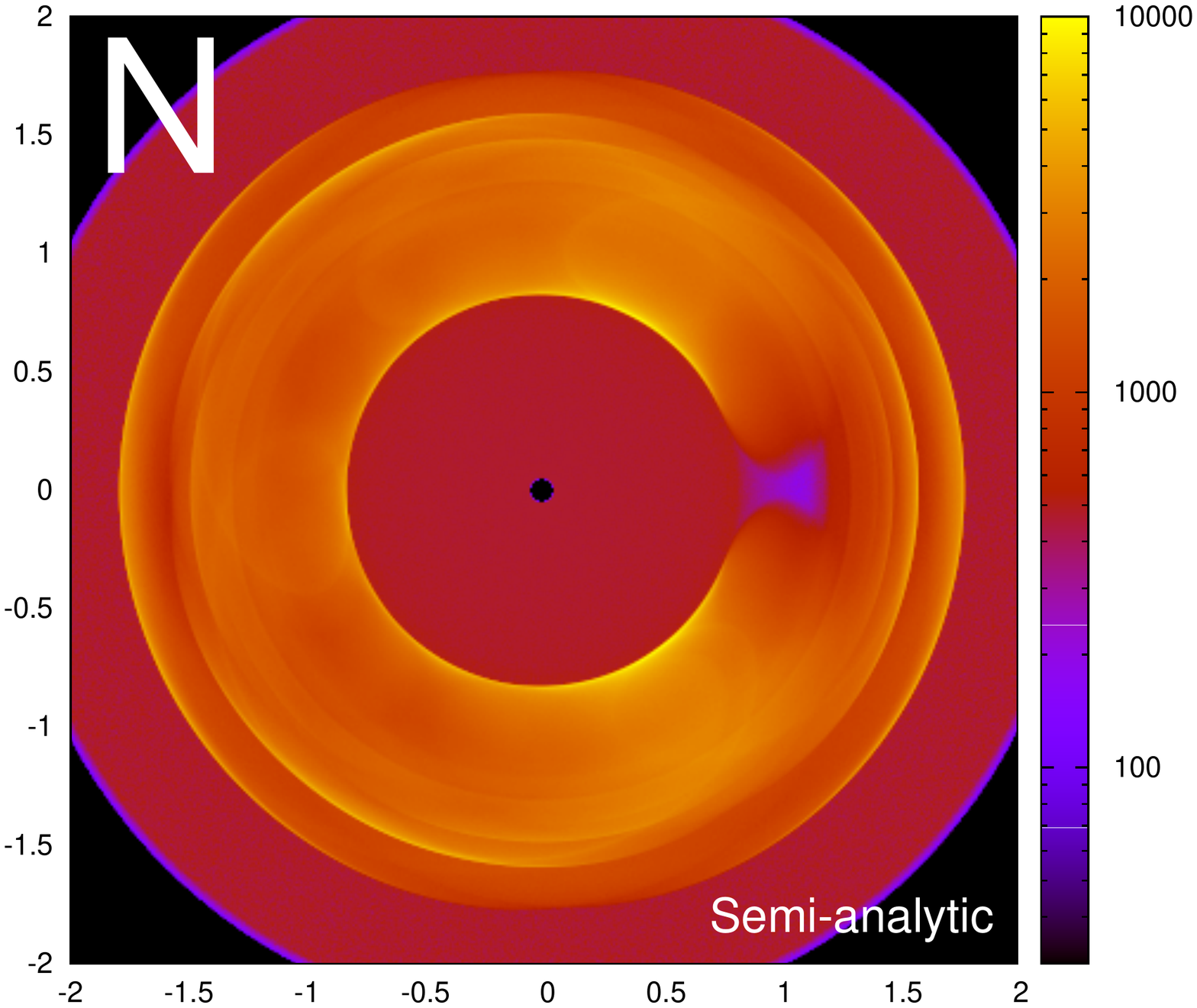}}
  \subfigure{\includegraphics[width=0.24\textwidth,trim = 100 100 100 100]{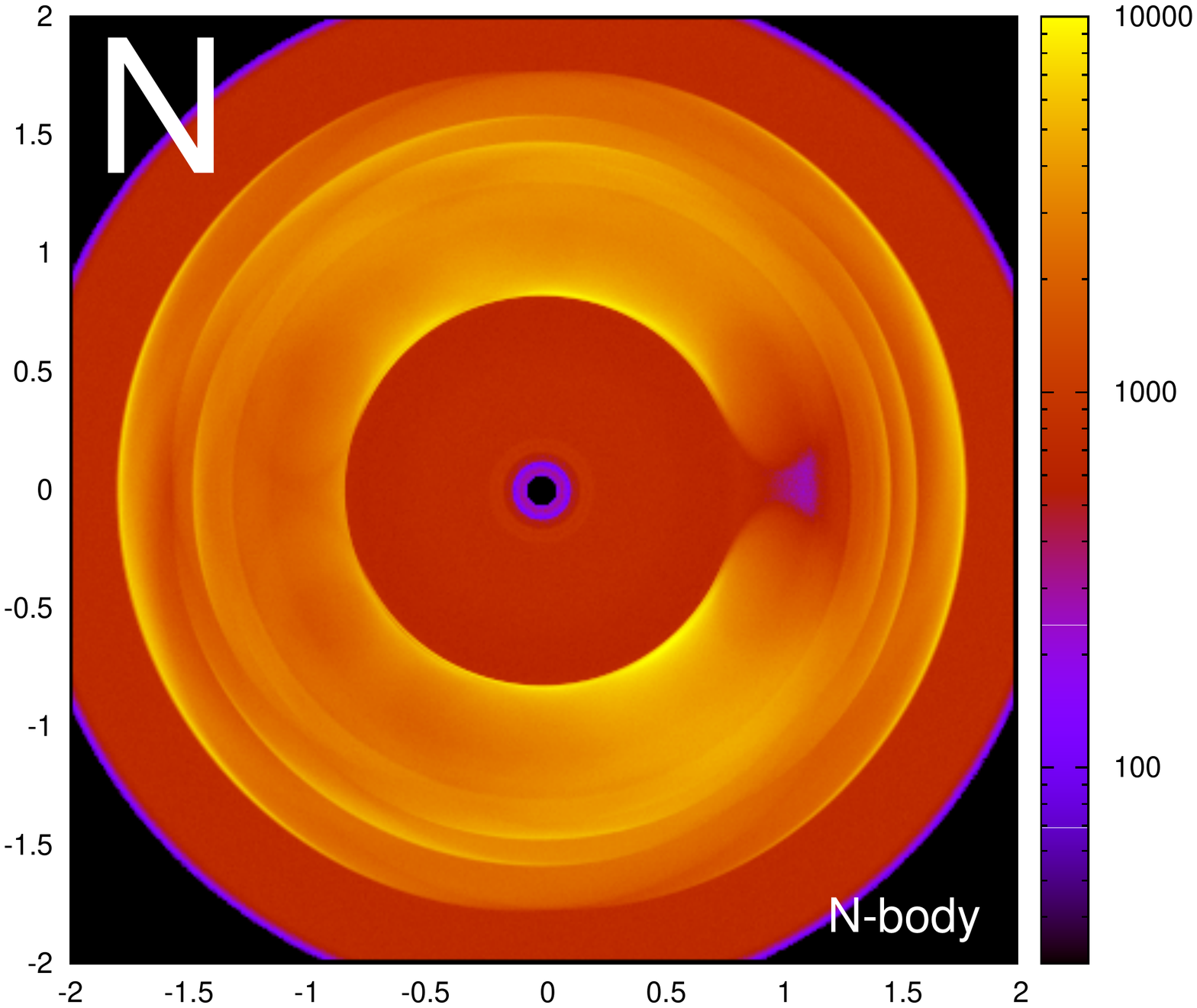}}
  \subfigure{\includegraphics[width=0.24\textwidth,trim = 100 100 100 100]{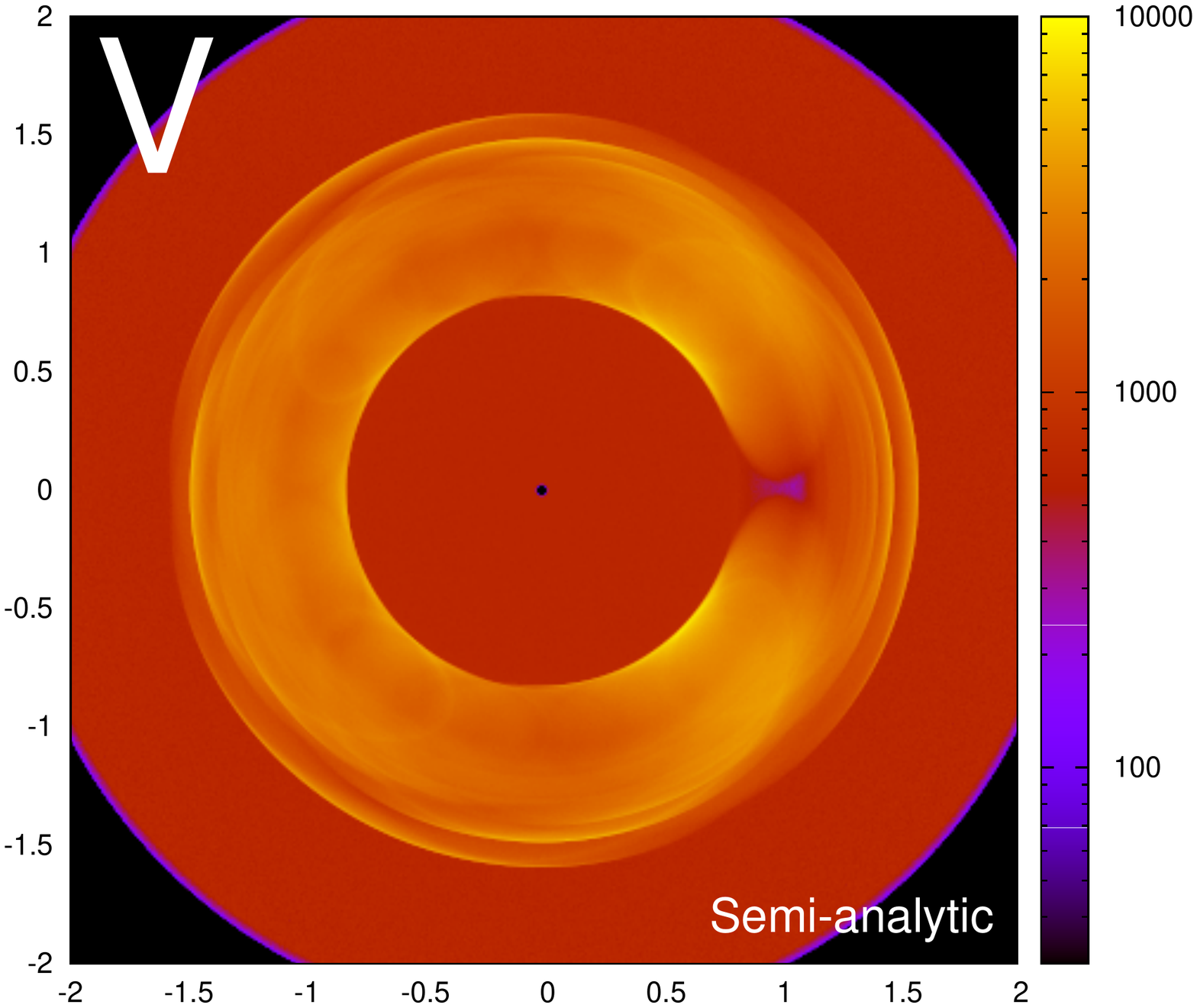}}
  \subfigure{\includegraphics[width=0.24\textwidth,trim = 100 100 100 100]{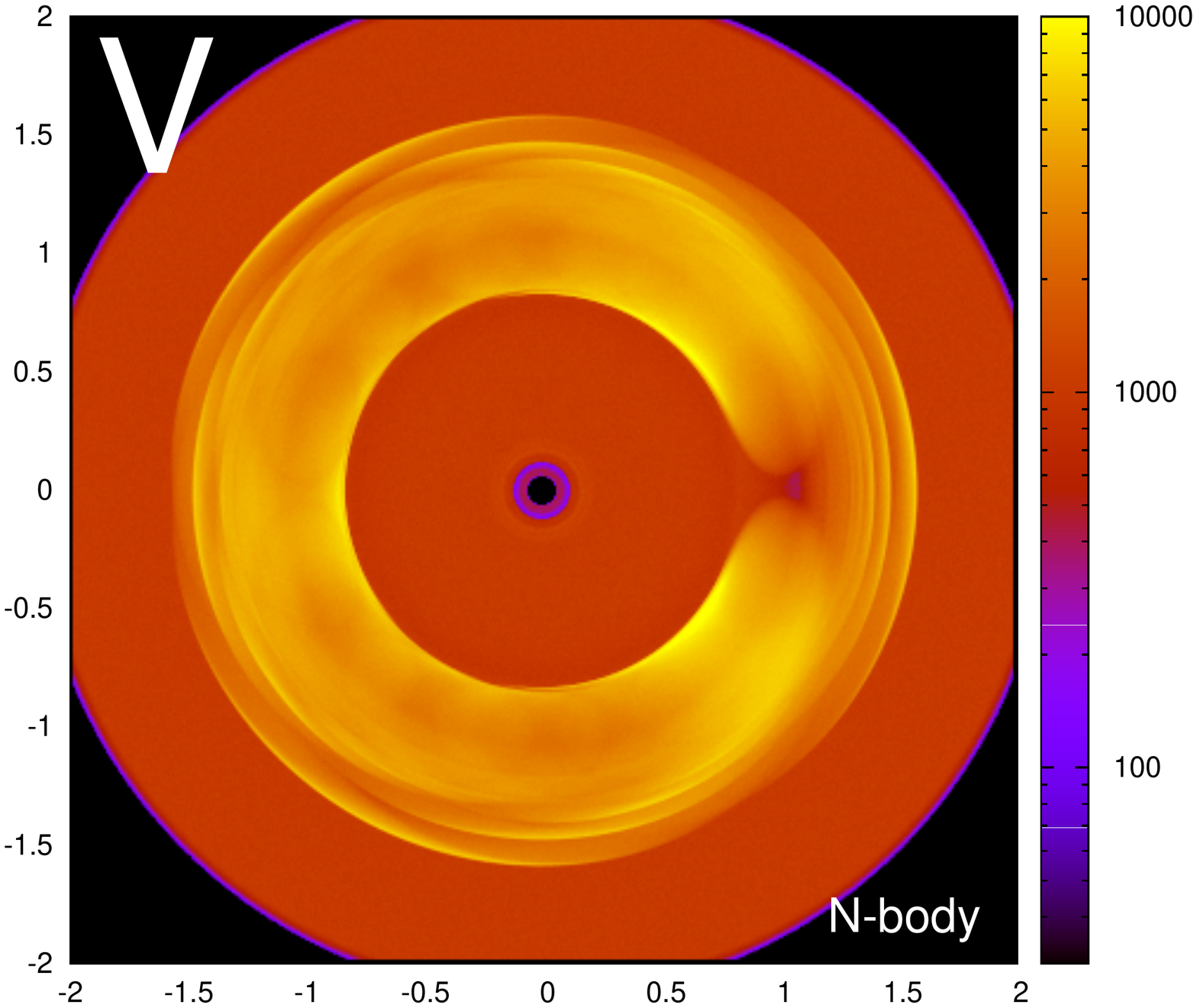}} \\
  \subfigure{\includegraphics[width=0.24\textwidth,trim = 100 100 100 100]{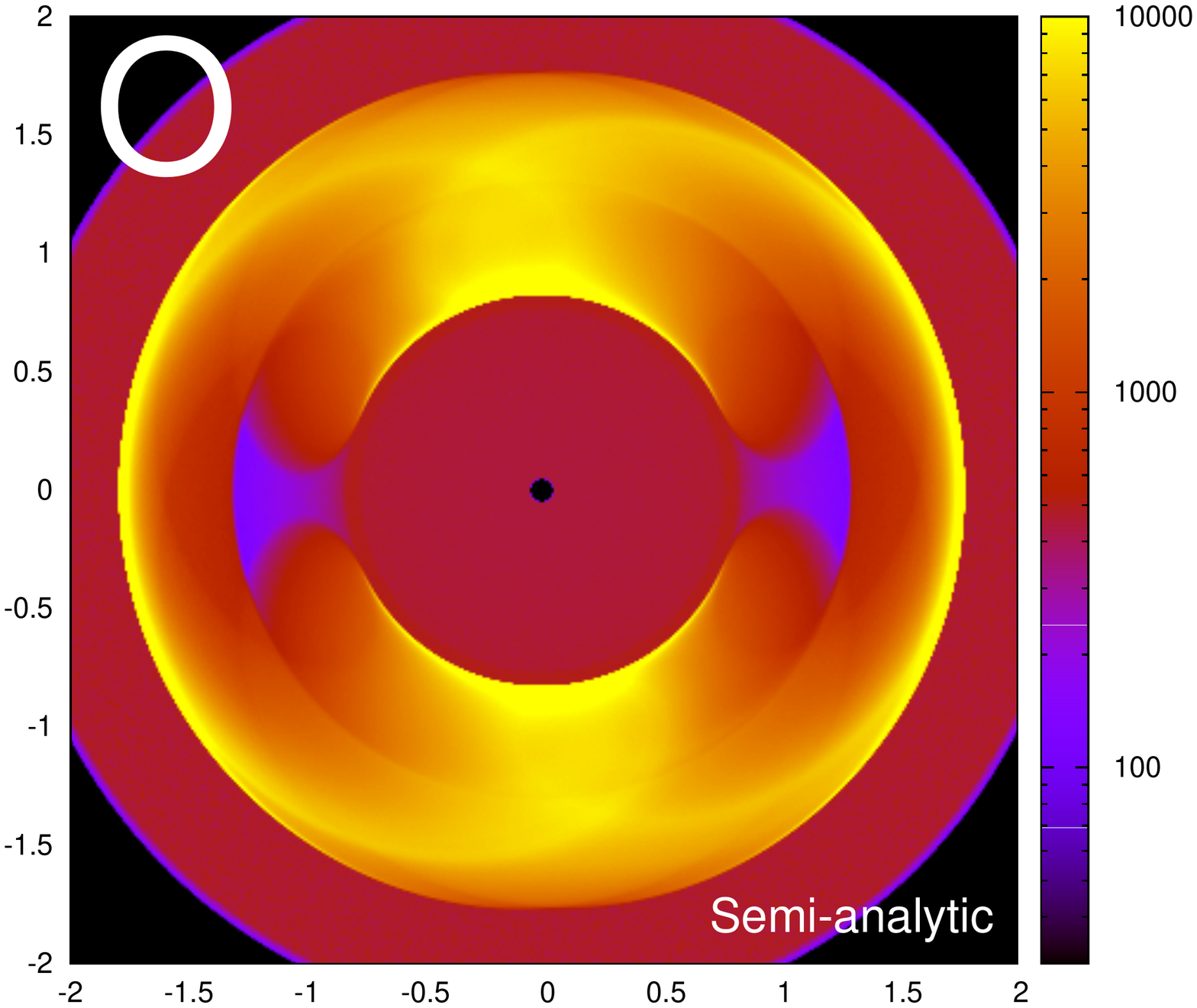}}
  \subfigure{\includegraphics[width=0.24\textwidth,trim = 100 100 100 100]{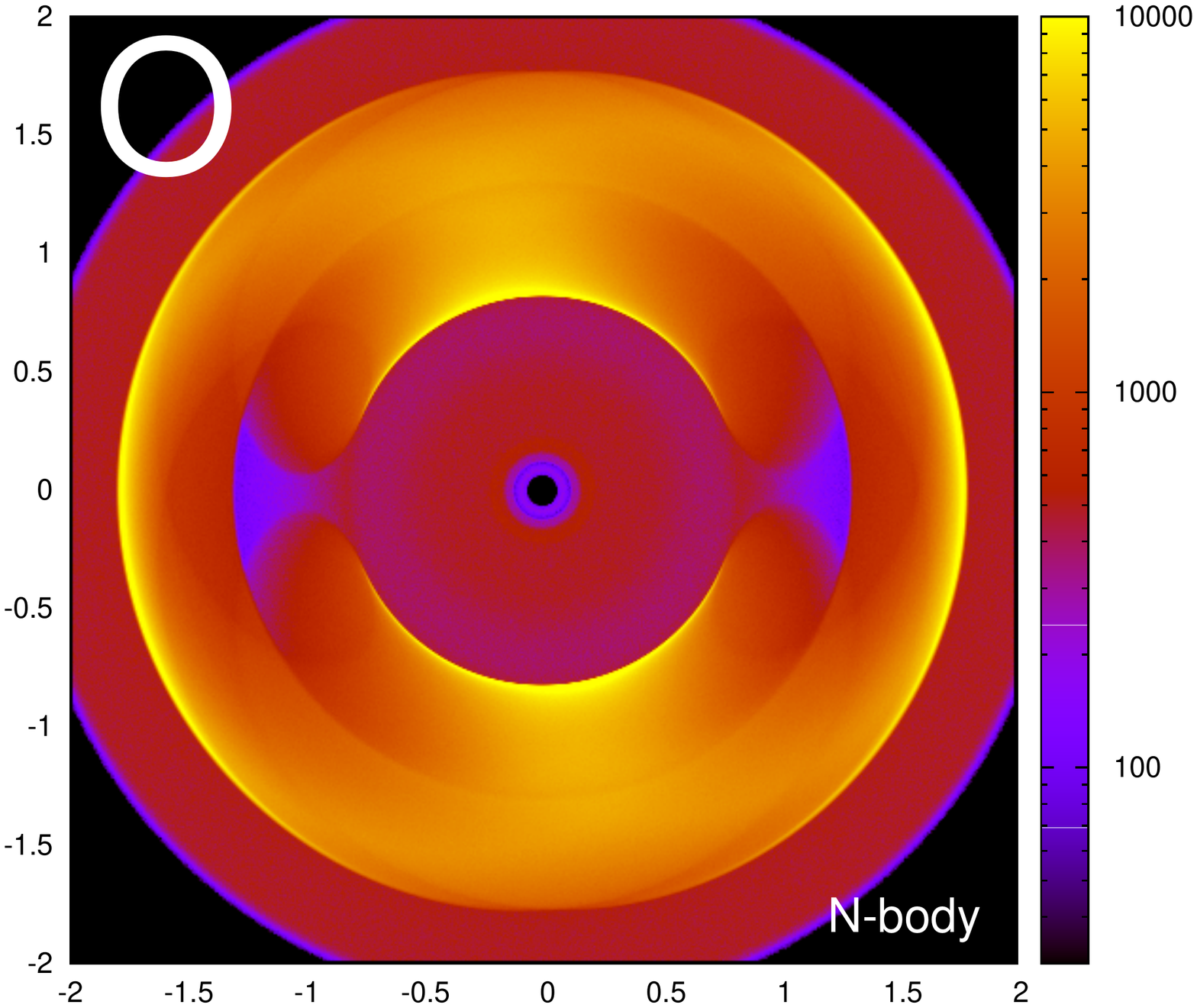}}
  \subfigure{\includegraphics[width=0.24\textwidth,trim = 100 100 100 100]{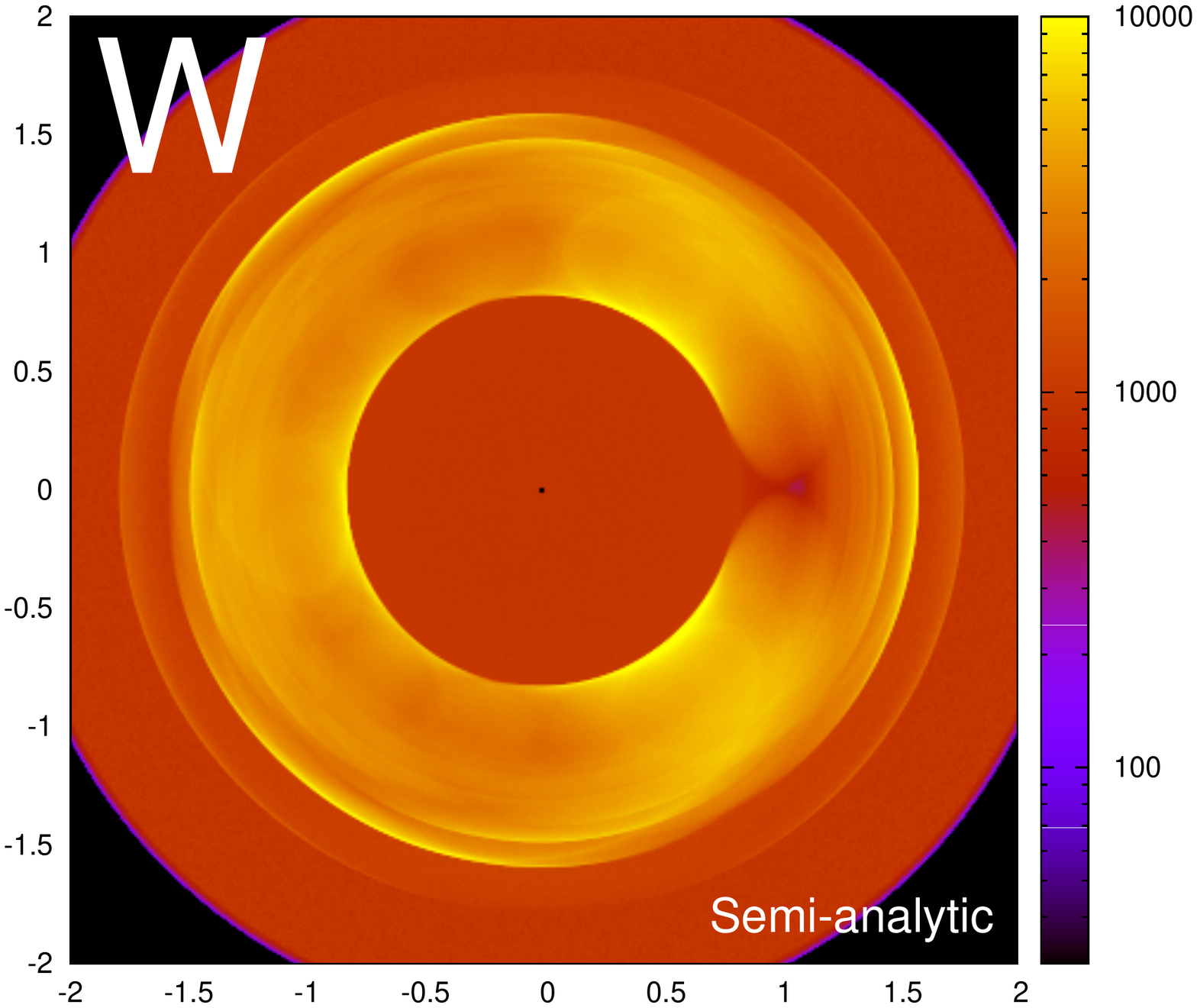}}
  \subfigure{\includegraphics[width=0.24\textwidth,trim = 100 100 100 100]{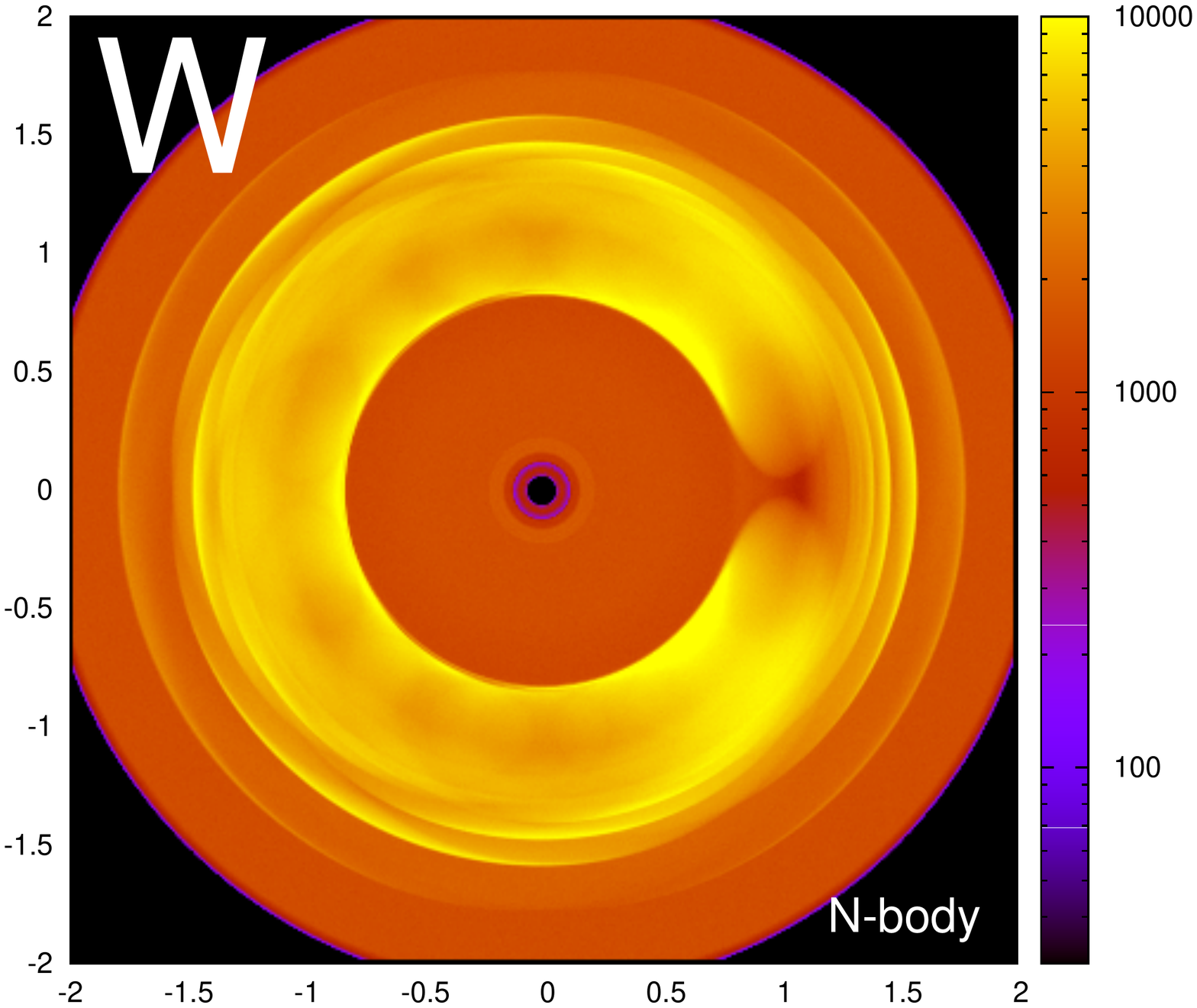}} \\
  \subfigure{\includegraphics[width=0.24\textwidth,trim = 100 100 100 100]{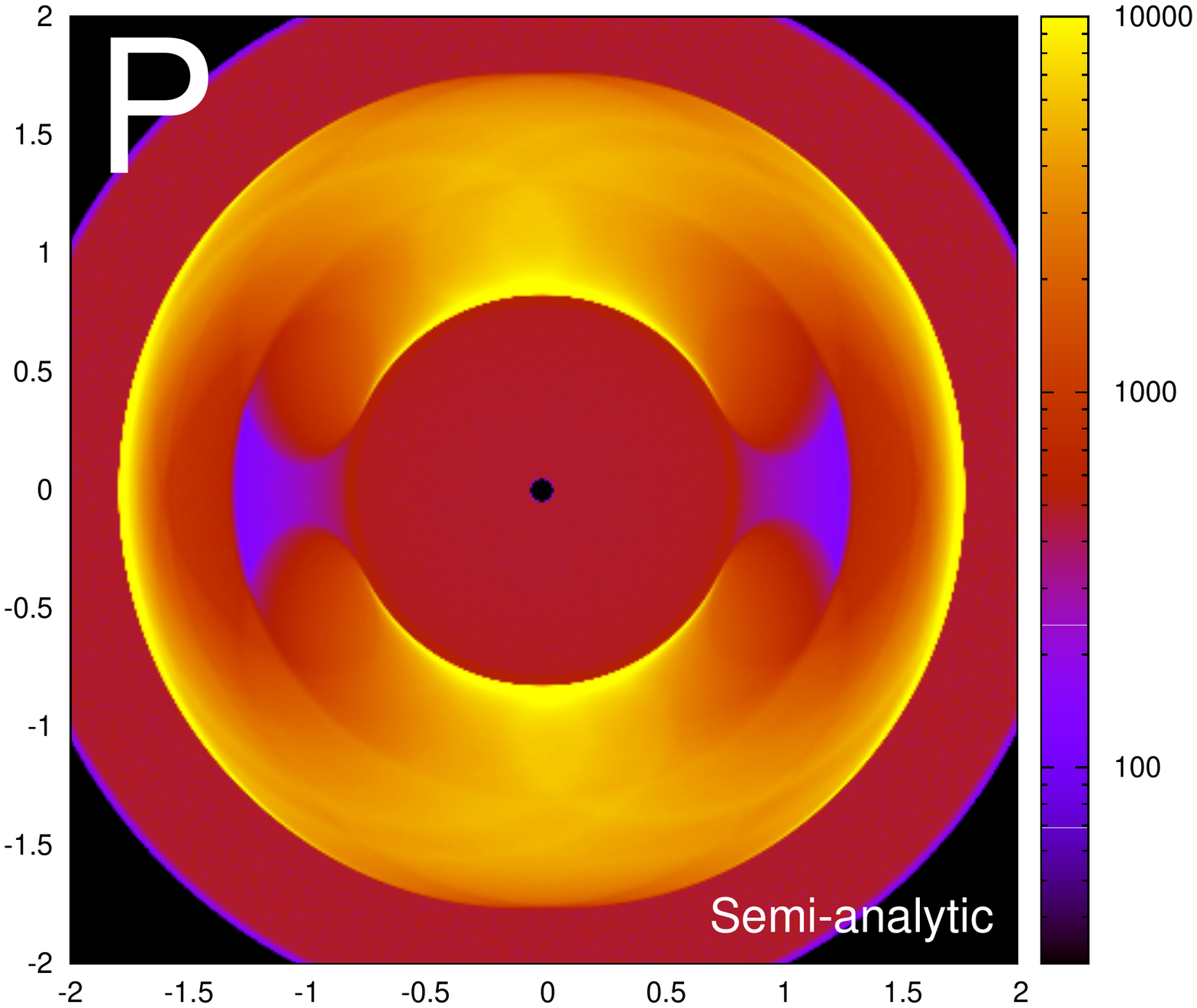}}
  \subfigure{\includegraphics[width=0.24\textwidth,trim = 100 100 100 100]{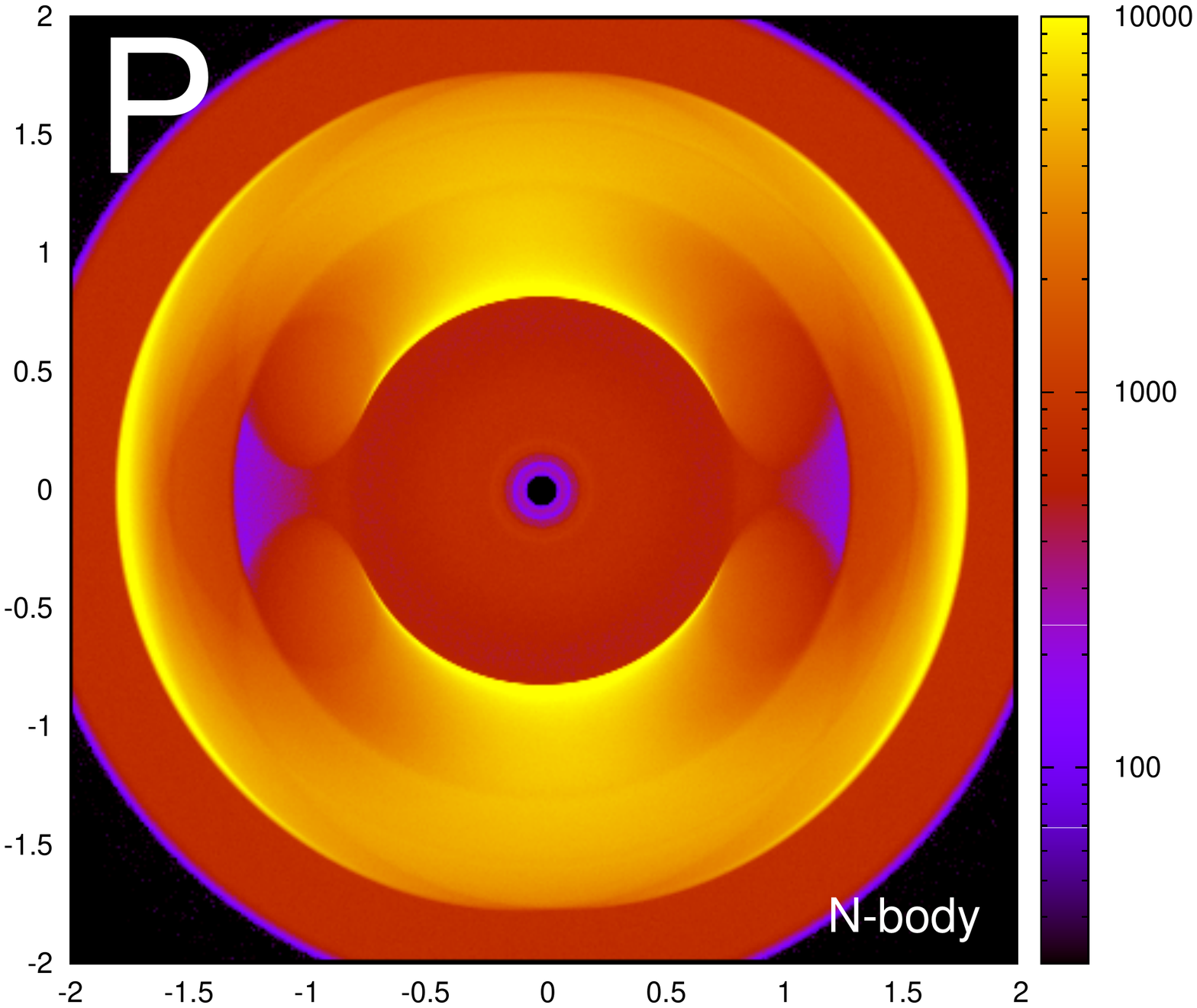}}
  \subfigure{\includegraphics[width=0.24\textwidth,trim = 100 100 100 100]{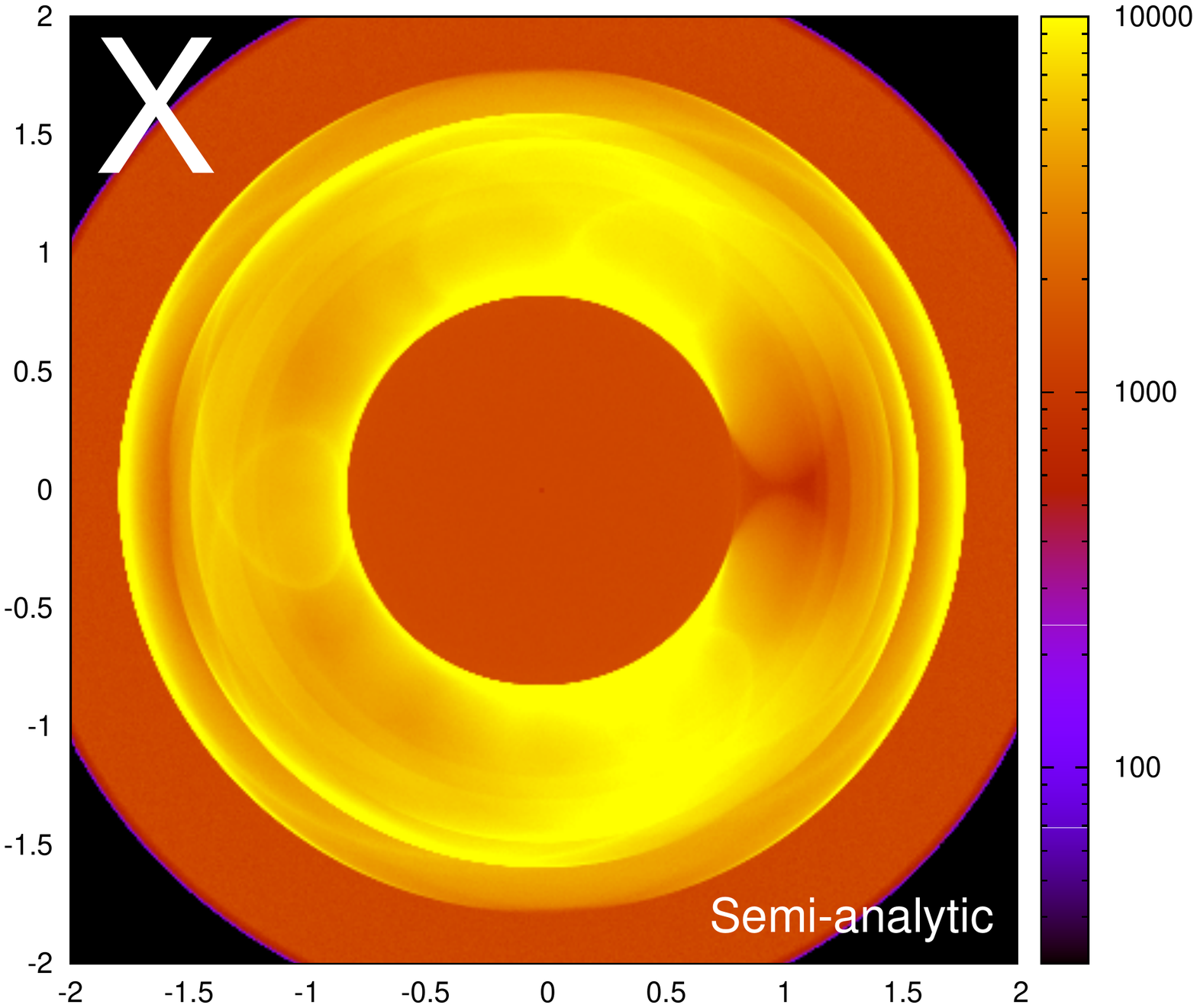}}
  \subfigure{\includegraphics[width=0.24\textwidth,trim = 100 100 100 100]{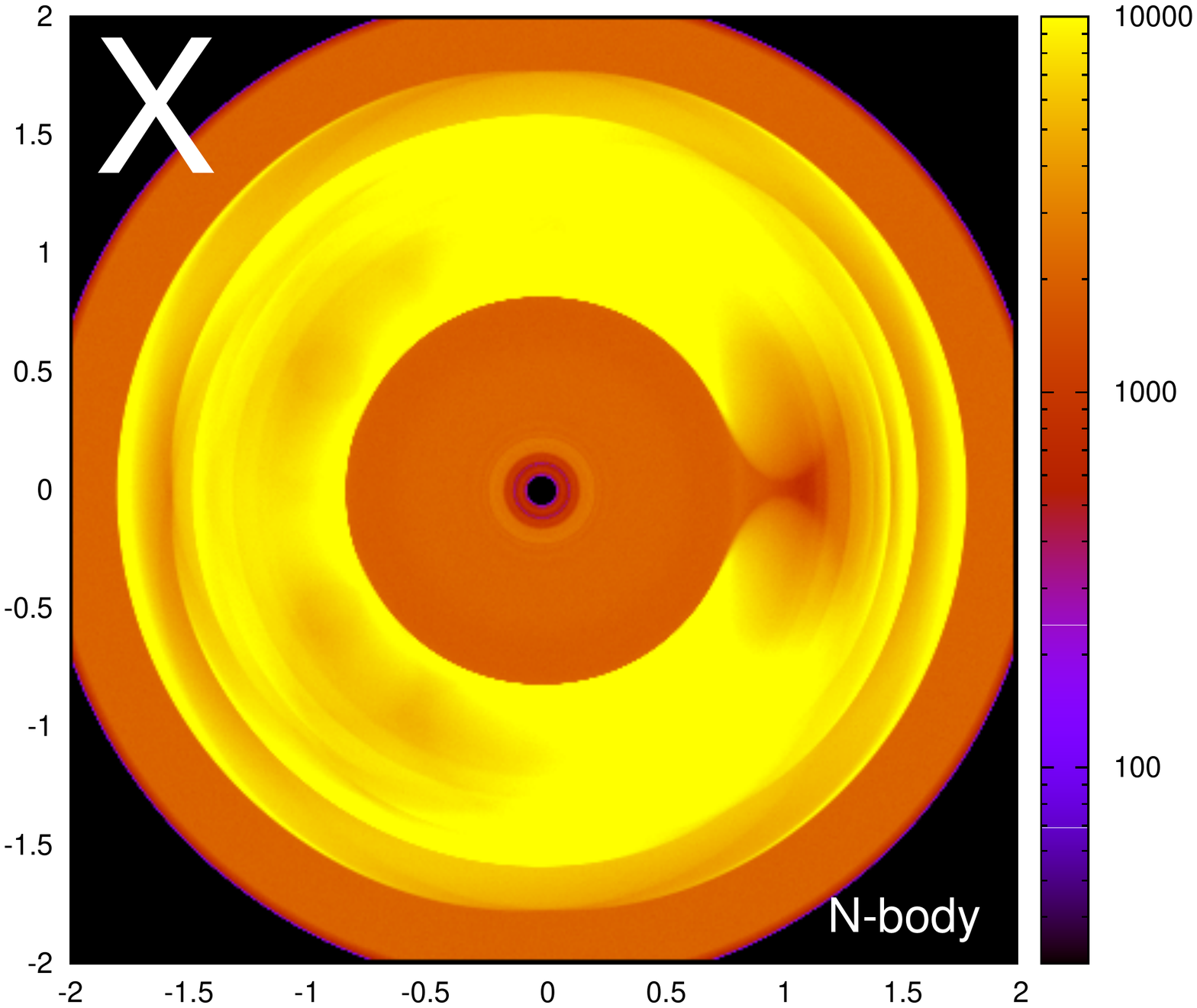}} \\
  \subfigure{\includegraphics[width=0.24\textwidth,trim = 100 100 100 100]{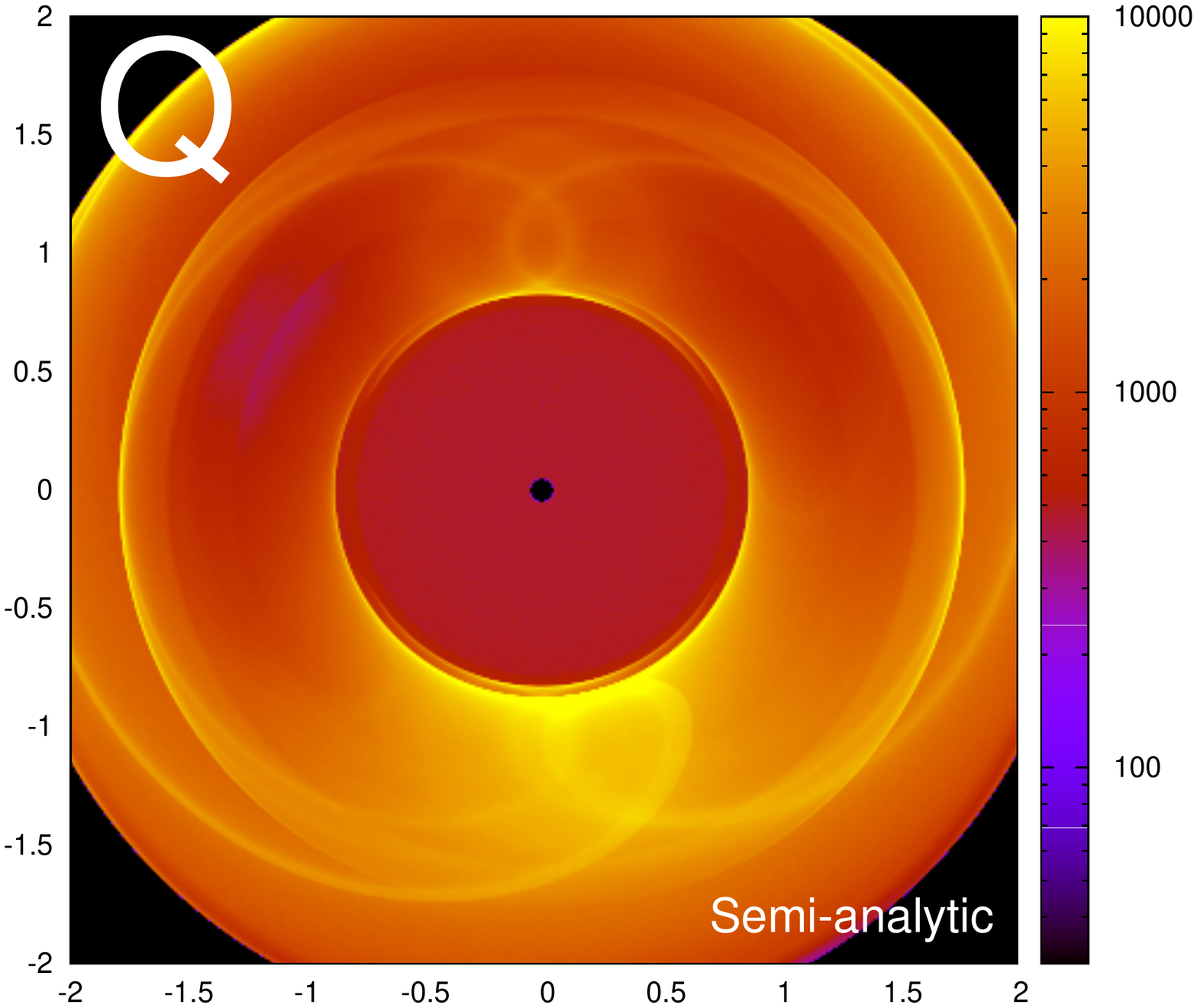}}
  \subfigure{\includegraphics[width=0.24\textwidth,trim = 100 100 100 100]{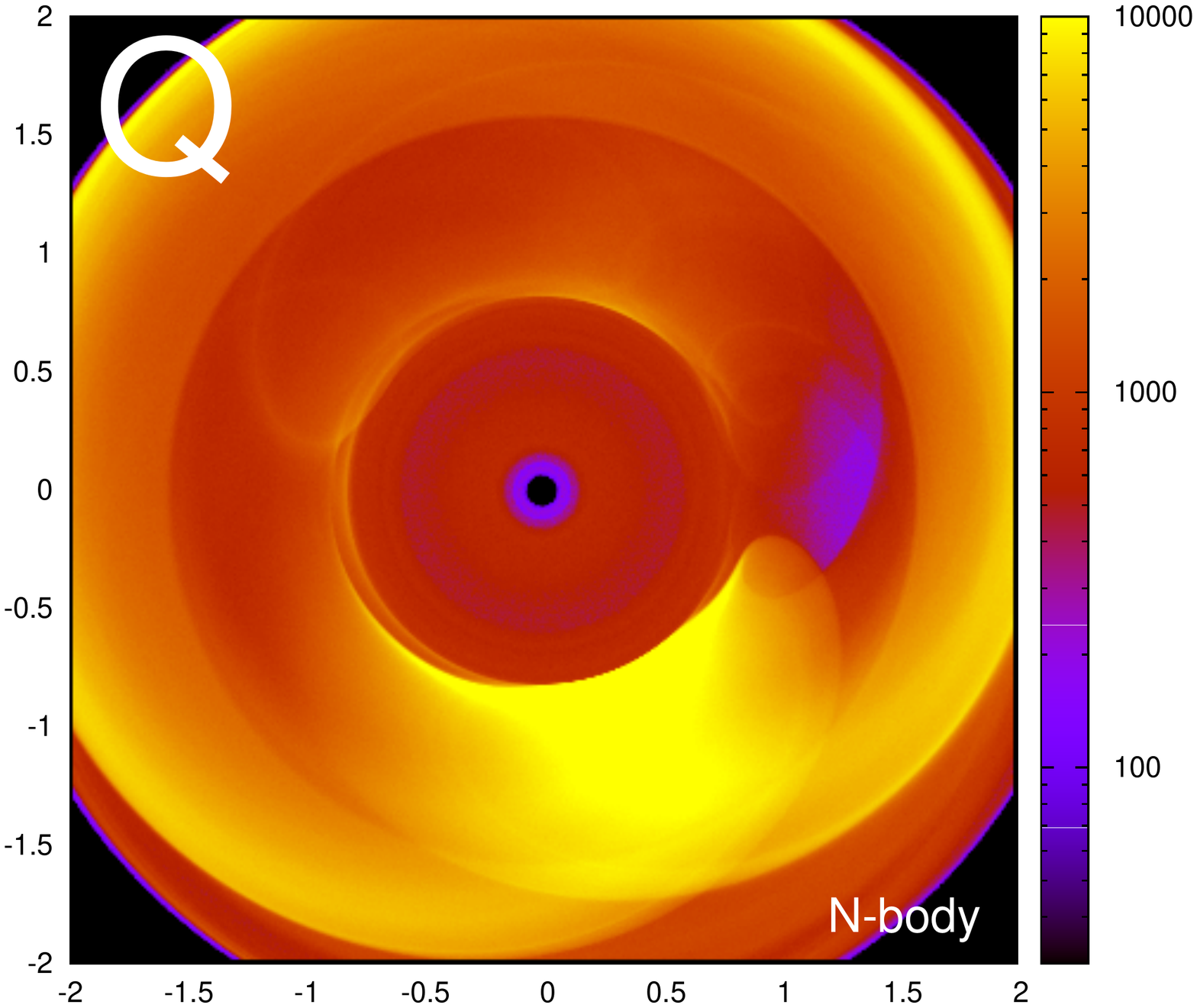}}
  \subfigure{\includegraphics[width=0.24\textwidth,trim = 100 100 100 100]{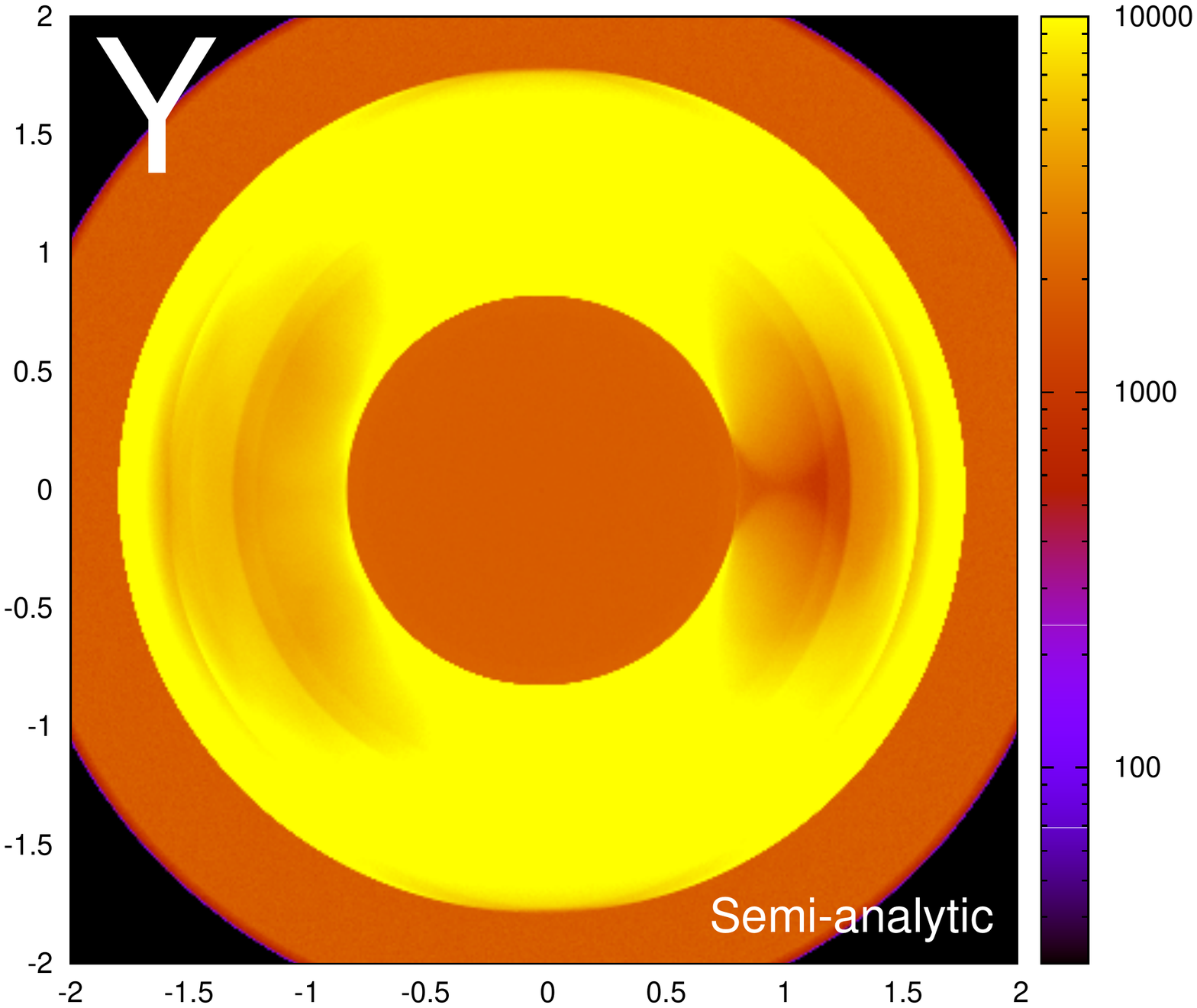}}
  \subfigure{\includegraphics[width=0.24\textwidth,trim = 100 100 100 100]{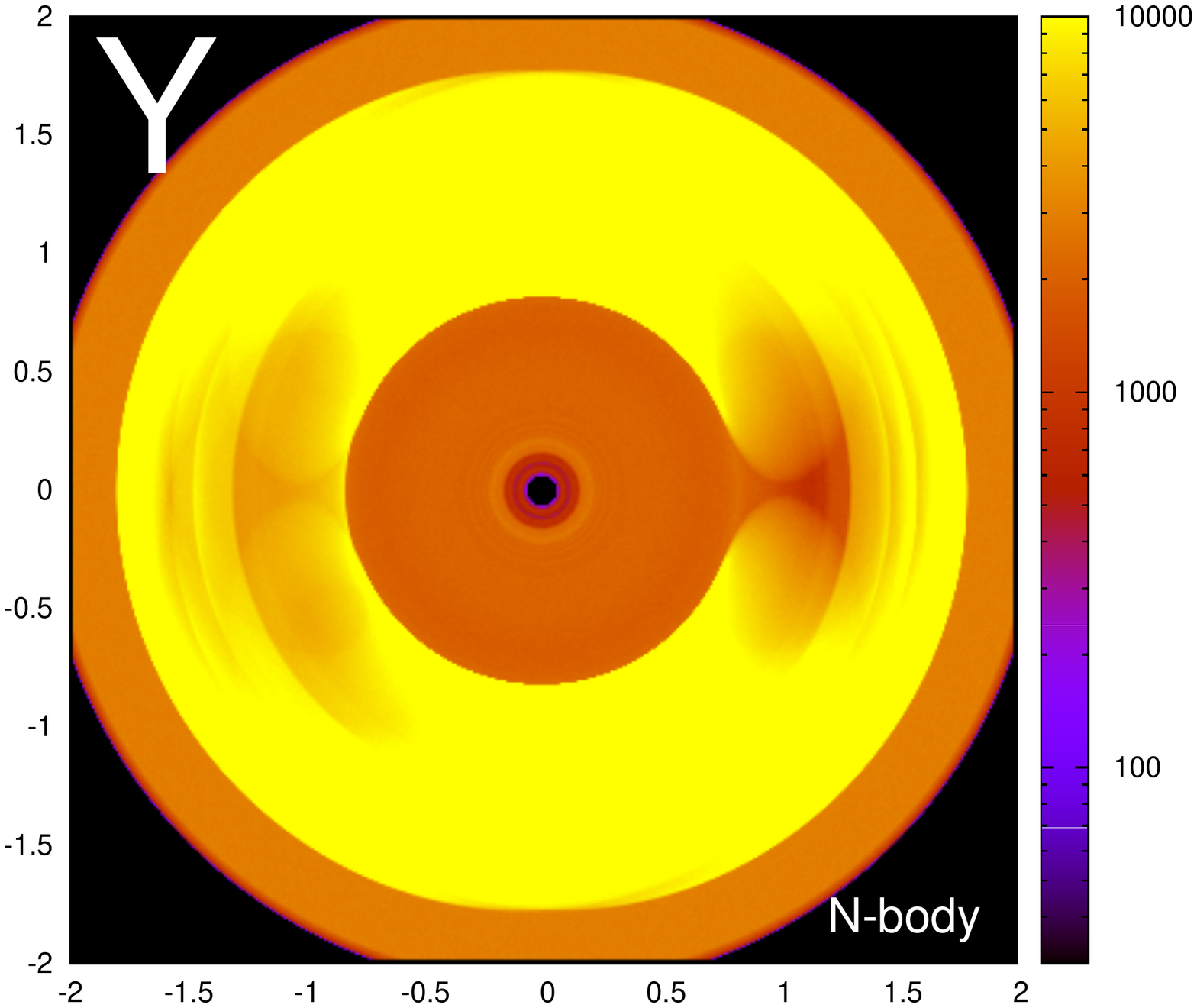}} \\
  \subfigure{\includegraphics[width=0.24\textwidth,trim = 100 100 100 100]{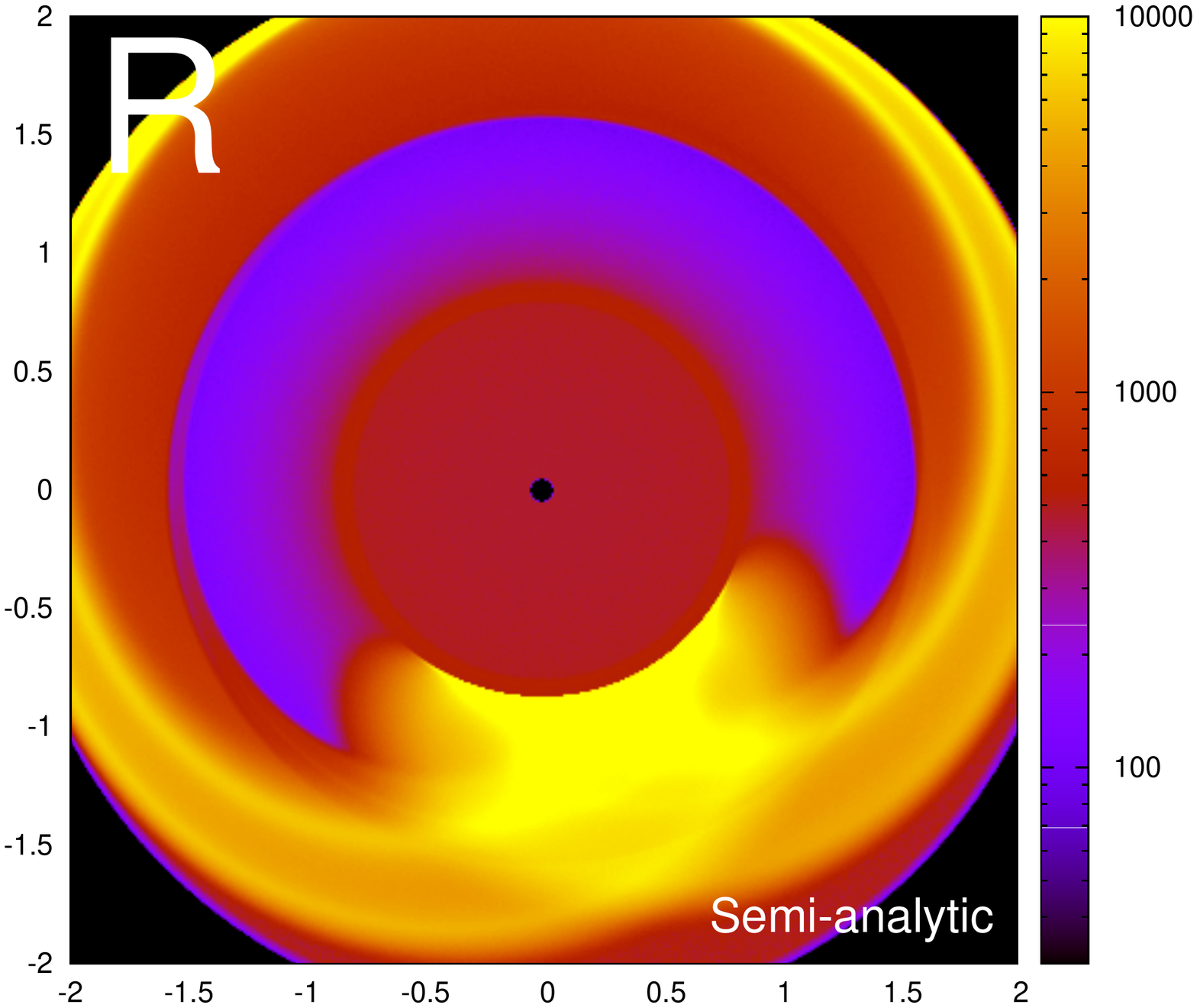}}
  \subfigure{\includegraphics[width=0.24\textwidth,trim = 100 100 100 100]{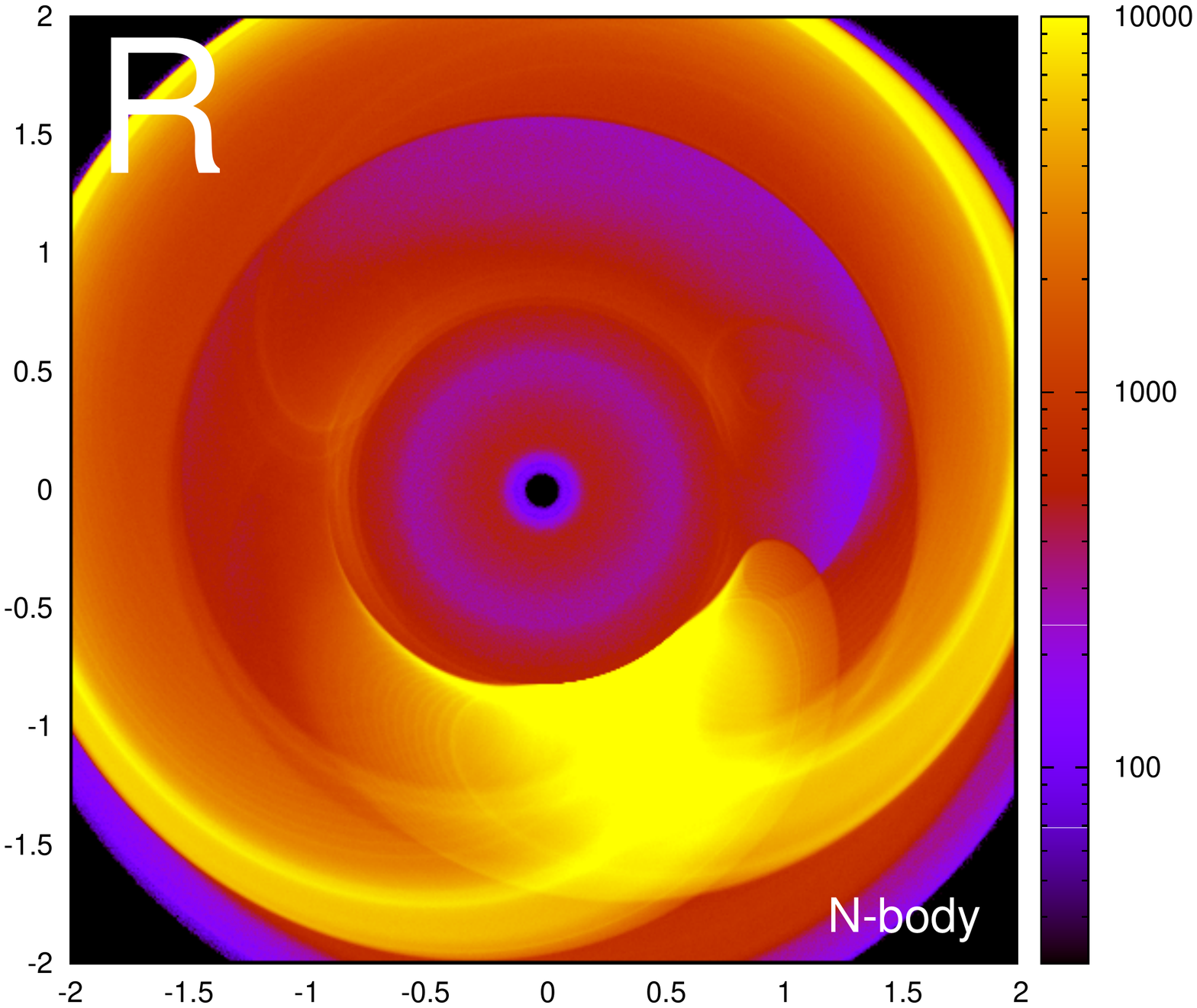}}
  \subfigure{\includegraphics[width=0.24\textwidth,trim = 100 100 100 100]{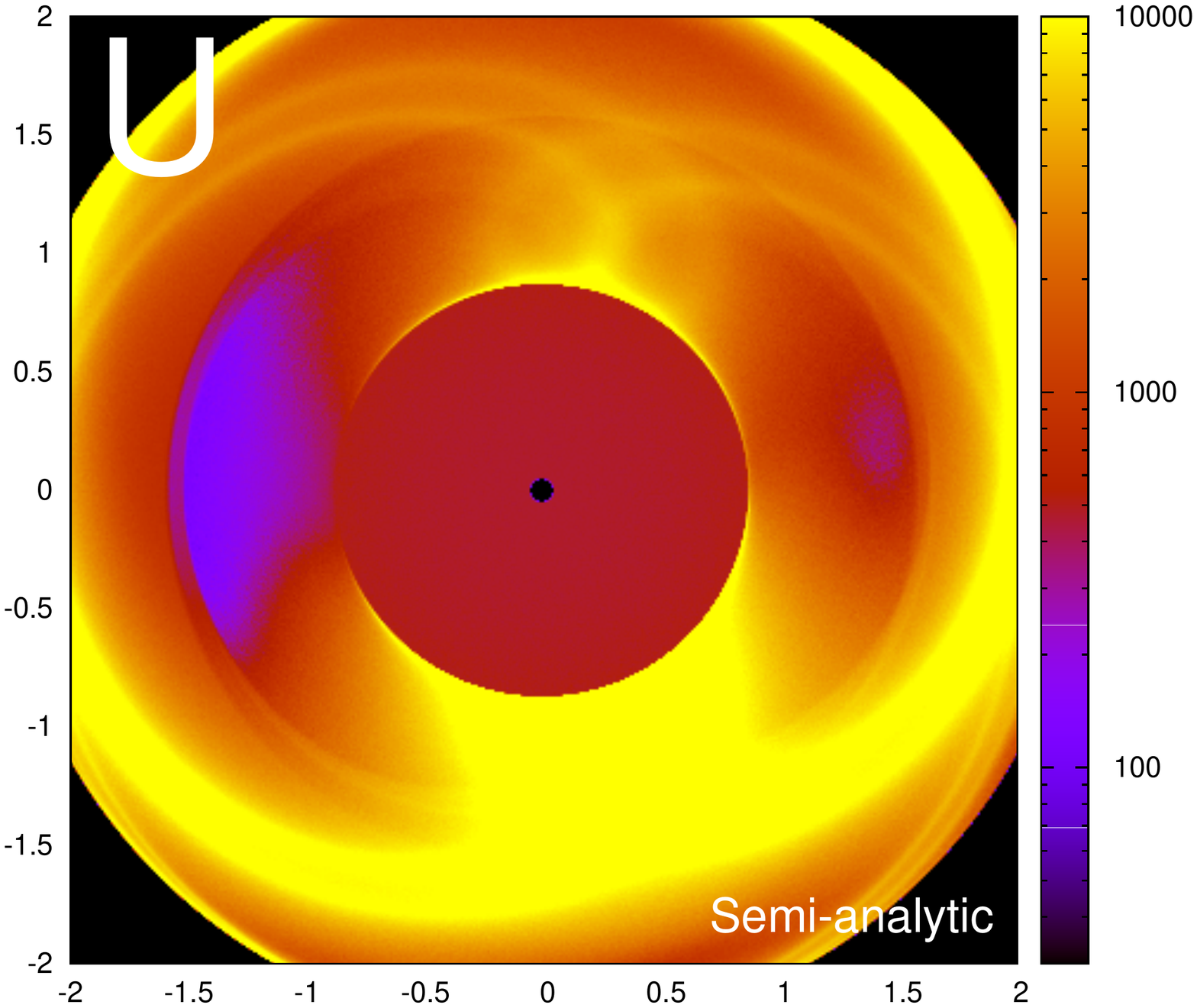}}
  \subfigure{\includegraphics[width=0.24\textwidth,trim = 100 100 100 100]{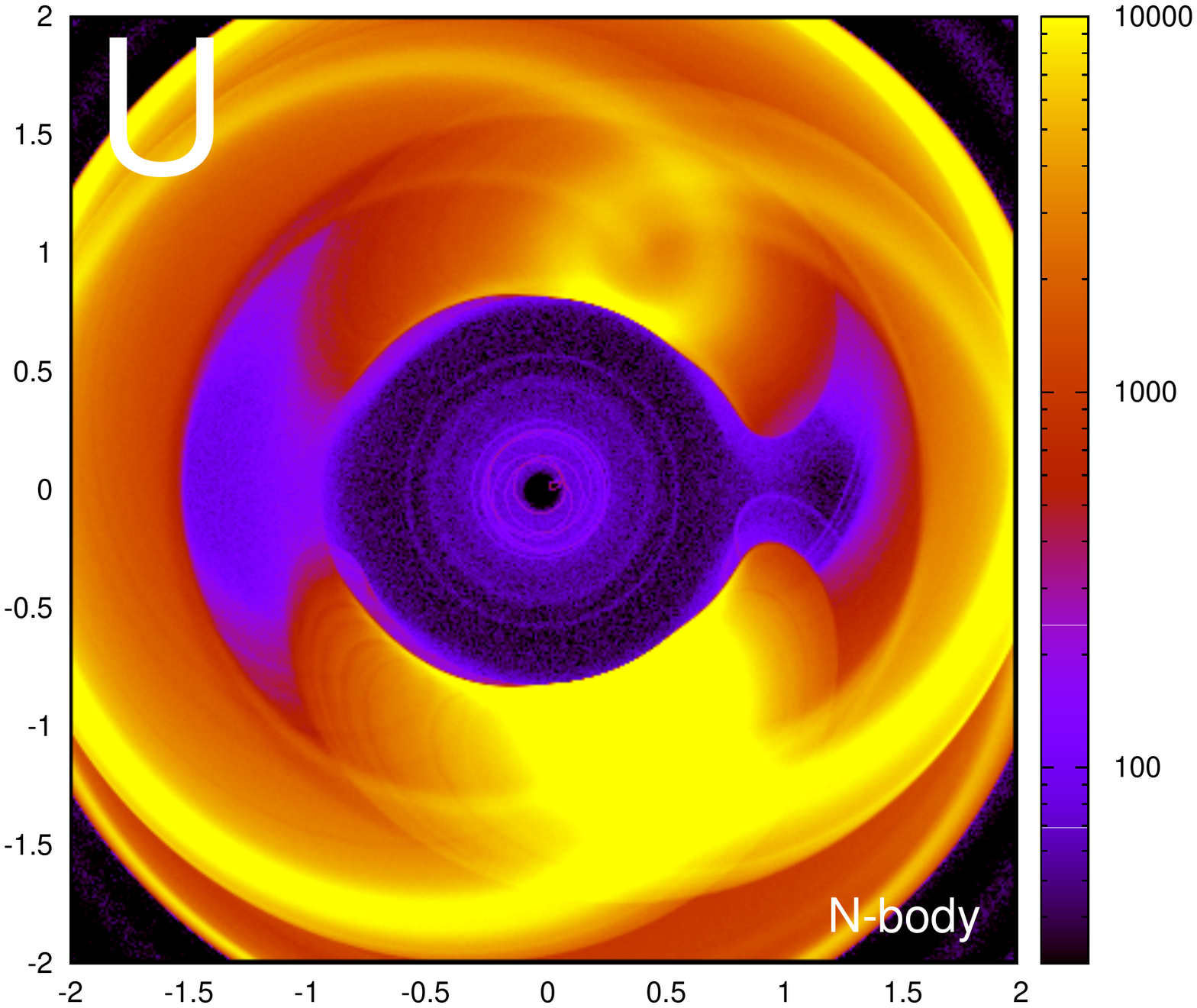}} \\
  \subfigure{\includegraphics[width=0.24\textwidth,trim = 100 100 100 100]{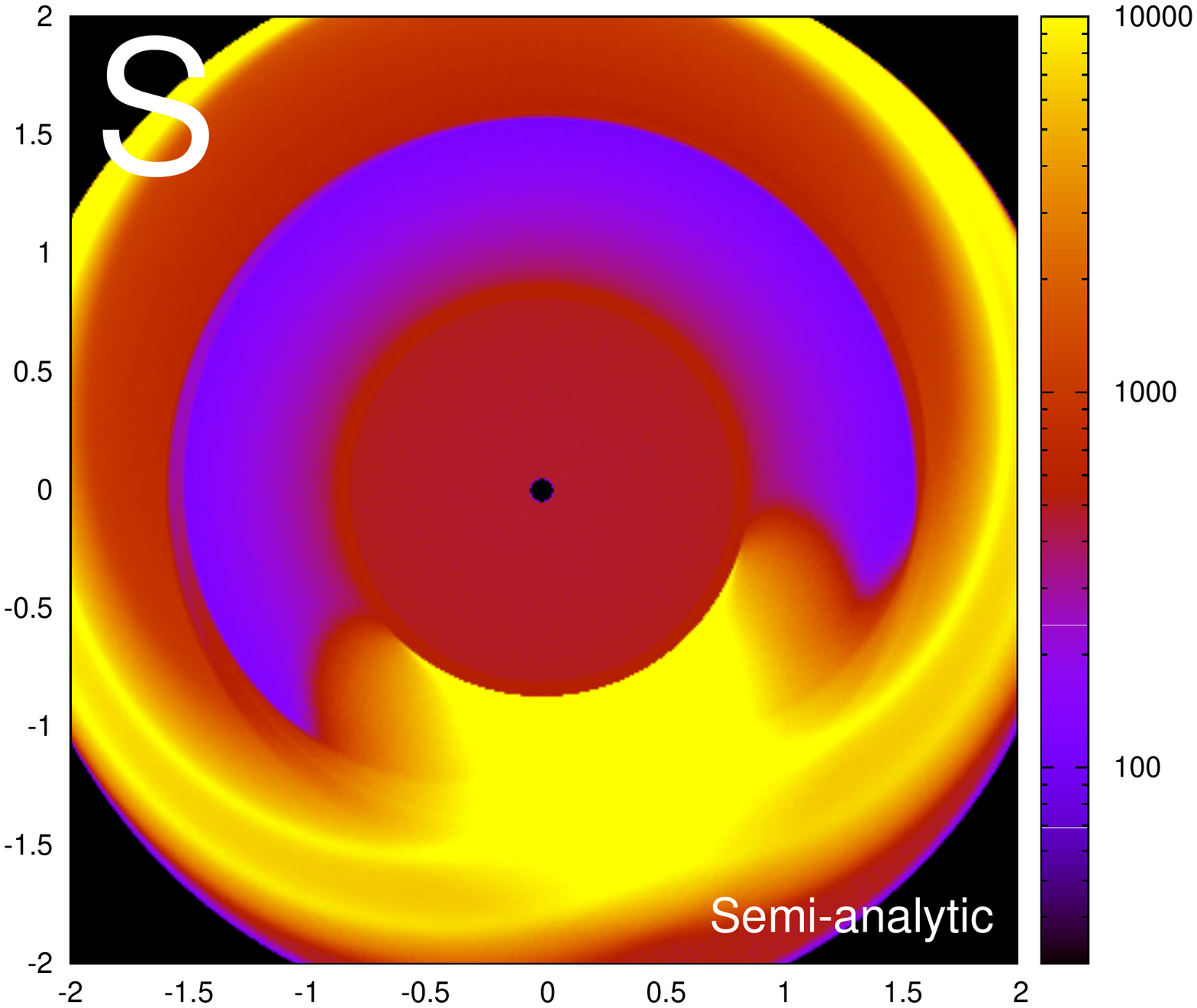}}
  \subfigure{\includegraphics[width=0.24\textwidth,trim = 100 100 100 100]{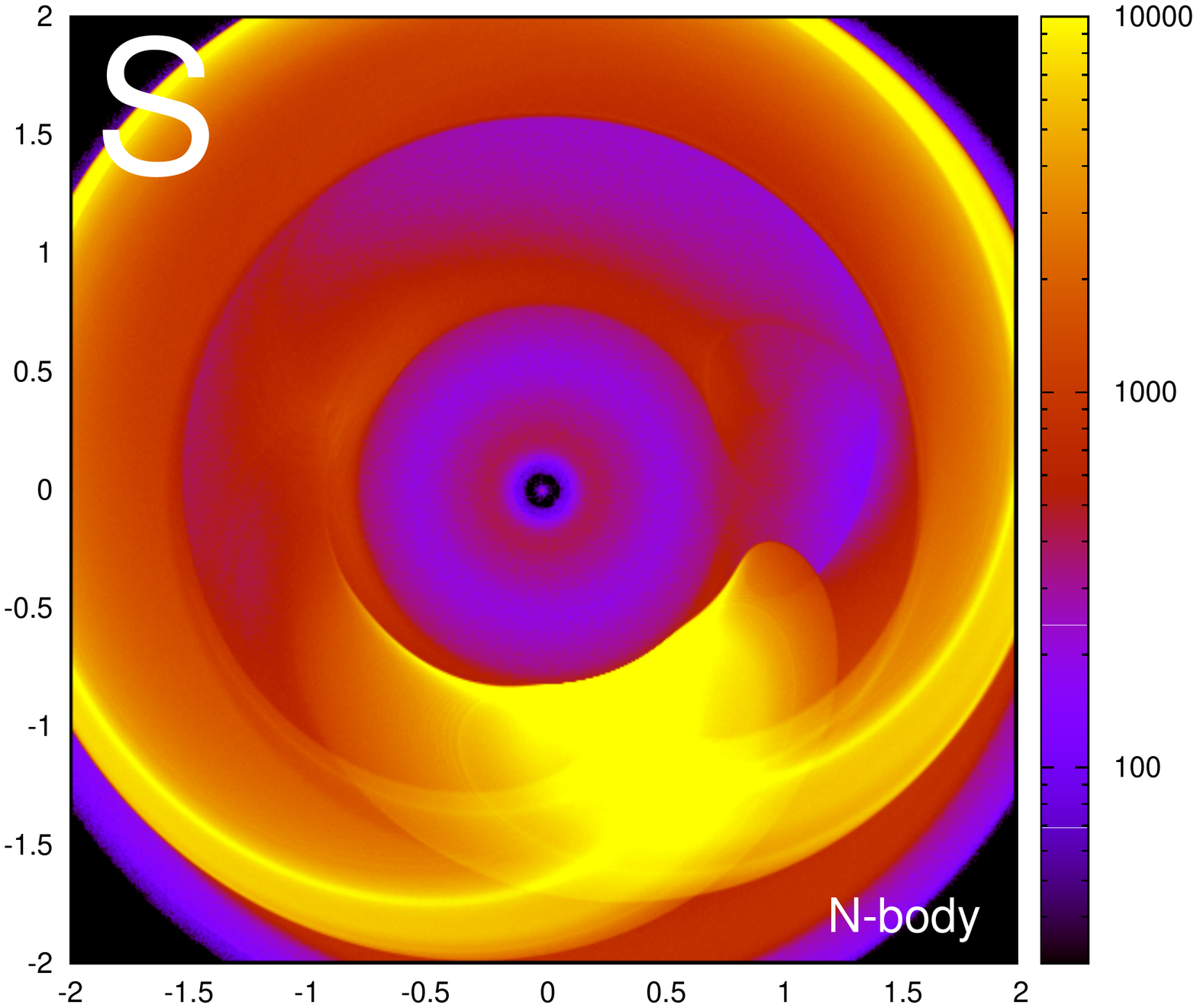}}
  \subfigure{\includegraphics[width=0.24\textwidth,trim = 100 100 100 100]{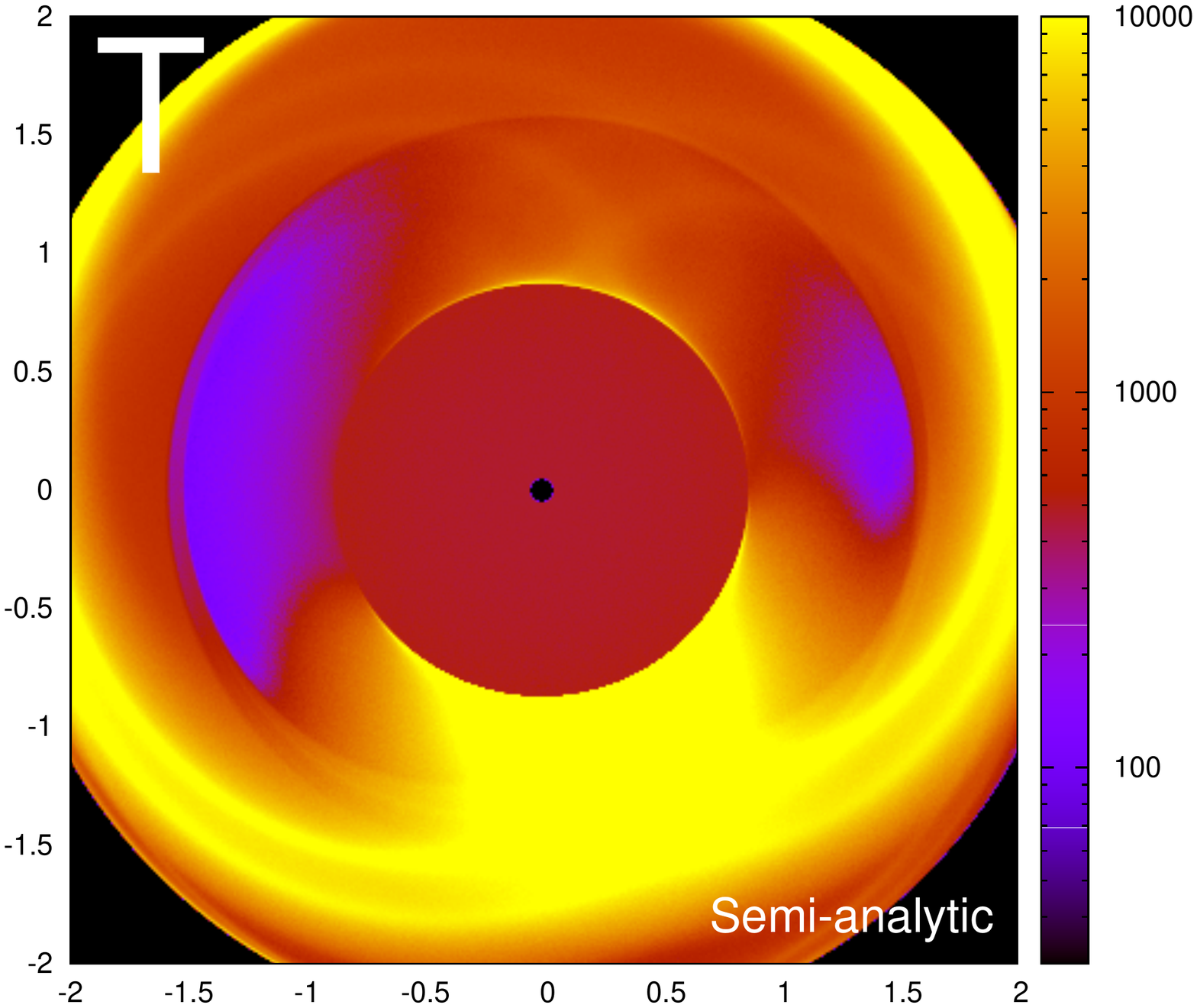}}
  \subfigure{\includegraphics[width=0.24\textwidth,trim = 100 100 100 100]{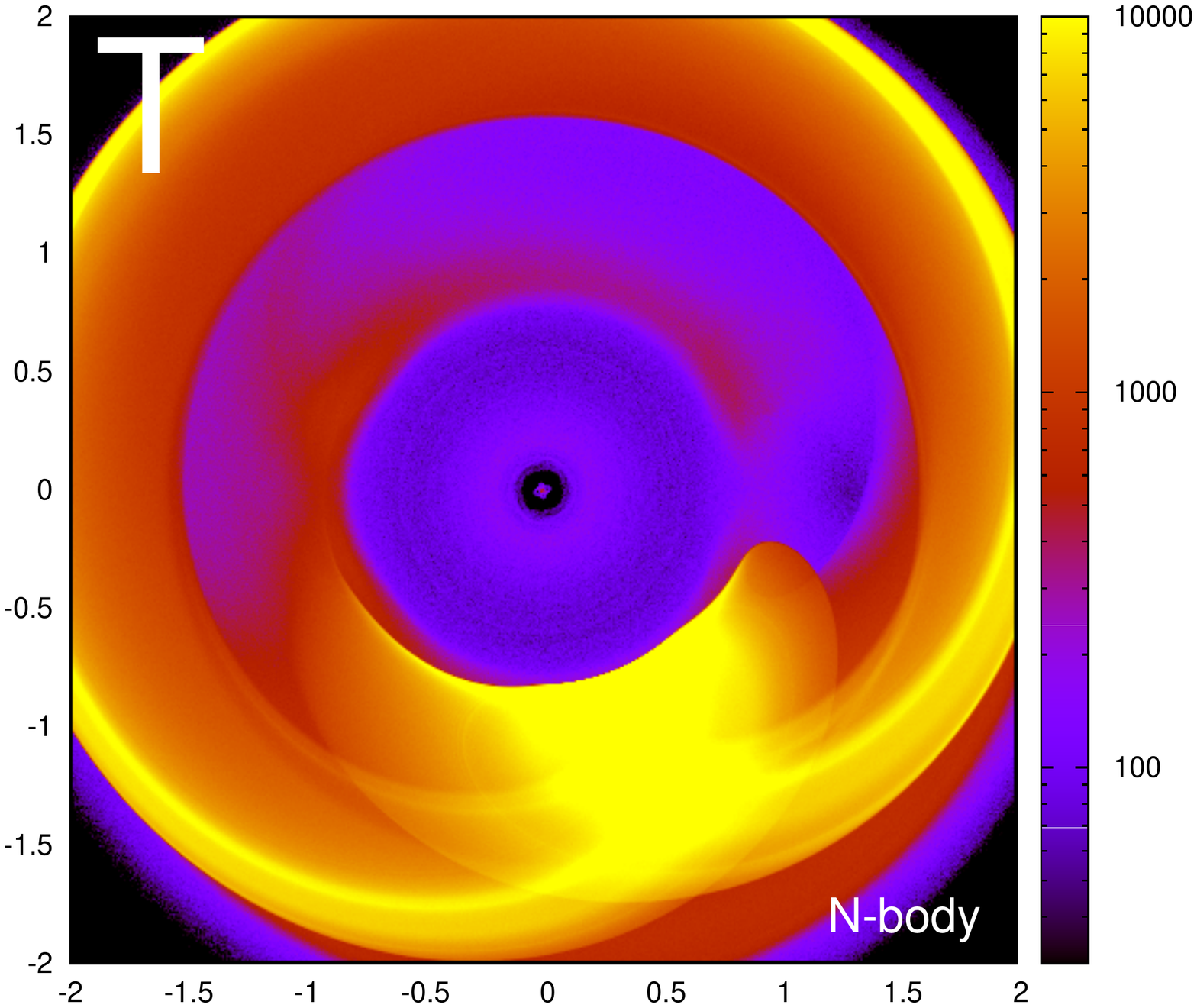}} \\
  \caption{Simulations B,N-S (variations of $m_p$)~left, and Simulations B, V-Y (variations of $a_p$)~right (and T-U, variations of $m_p$), bottom right.  In all plots, 10000 particles are each plotted once every $10^3\times\beta_{\rm{PR}}^{-0.5}a^{1.5}$~days, in a grid of 400 by 400 cells covering from $-2a_p$~to $2a_p$.  To produce the N-body plot, we rotate the dust grain around the star so the planet is located at (1,0).  All the subsequent disk images are produced in the same fashion.}
  \label{fig:allplots2}
\end{figure*}

We plot comparisons of the N-body disks of table \ref{tab:sims} to images produced using this semi-analytic prescription in figures \ref{fig:allplots} and \ref{fig:allplots2}.  N-body disk images were produced by sampling the location of the N-body particles every $10^3\beta_{\rm{PR}}^{-\frac{1}{2}}a_p^{1.5}$~days, where $a_p$~is measured in au.  To produce an image, at each time we rotated the dust grain around the star so the planet is located at (1,0).  The semi-analytic disks are very good matches for the N-body disk, with increasing discrepancies for higher mass planets.  In these cases, captures into the 2:1 resonance are not reproduced as well, as the grains move between the eccentricity and inclination resonances, which we do not model.  Additionally, while there is reasonable agreement with our model in the parameter space we probe with $N$-body simulations, as the calibration of capture into the $\left(u\right)$~and $\left(l\right)$~$2:1$~resonances in equation \ref{eq:twooneprobability}~is calibrated over that range, it is unclear how far it can be extrapolated outside this range.

\subsection{Collisional production}
\label{subsec:collisionalproduction}
\begin{figure}
  \centering
  {\includegraphics[width=0.48\textwidth,trim = 50 50 200 480, clip]{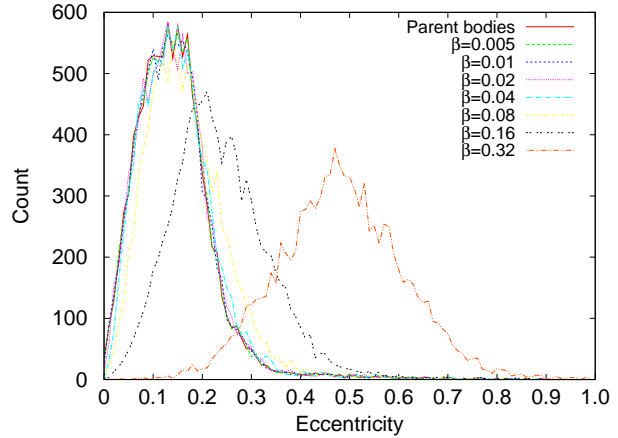}}
  \caption{Initial eccentricity of dust particles compared to a parent population.  The parent population's eccentricity distribution is that of the first ten thousand numbered asteroids.  The distance from the sun at which the collision takes place ($r$)~is chosen uniformly between periapse and apoapse; a real population would be expected to be biased towards periapse \citep{2010MNRAS.402..657W}.  The two distributions are roughly the same at low $\beta_{\rm{PR}}$, and diverge when $\beta_{\rm{PR}} \sim e_b$.}
\label{fig:dustinitialbeta}
\end{figure}

A real disk will contain particles with a variety of initial eccentricities, and a variety of different $\beta_{\rm{PR}}$.  These two distributions must be modelled to produce a realistic disk.  Of note, the initial dust properties will differ from the parent bodies, owing to radiation pressure, so the dust grains will begin with a distribution of semimajor axes and eccentricities that differ from those of the parent bodies and are functions of $\beta_{\rm{PR}}$~ (i.e., $a_d\left(\beta_{\rm{PR}} \right)$~and $e_d\left(\beta_{\rm{PR}} \right)$).  If a collision occurs at $r$, neglecting the velocity at which dust is ejected from the collision, the dust orbital parameters relate to the parent body parameters ($a_b$~and $e_b$) as:
\begin{eqnarray} \nonumber
 a_d &=& \left(\frac{1}{\left(1-\beta_{\rm{PR}}\right)a_b} -\frac{2\beta_{\rm{PR}}}{\left(1-\beta_{\rm{PR}}\right)r}\right)^{-1} \\ 
 &\approx& \frac{1-\beta_{\rm{PR}}}{1-2\beta_{\rm{PR}}}a_b \\
 e_d &=& \left(1+2\beta_{\rm{PR}}\frac{1-e_b^2}{\left(1-\beta_{\rm{PR}}\right)^2}\frac{a_b}{r}-\frac{1-e_b^2}{\left(1-\beta_{\rm{PR}} \right)^2}\right)^{\frac{1}{2}} \\ \nonumber
     & \approx & \frac{\beta_{\rm{PR}}}{\left(1-\beta_{\rm{PR}} \right)}
\end{eqnarray}
where the approximate expressions assume the parent bodies' semimajor axis $a_b \approx r$~and the parent bodies' eccentricity $e_b \approx 0$.  This last result motivates the choice of $e_0 = \beta_{\rm{PR}}$~as a default case.  Given the parent body population and the distribution of $r$~for generative collisions, the initial orbits of particles as a function of $\beta_{\rm{PR}}$~can be predicted \citep{2010MNRAS.402..657W} (e.g., Figure \ref{fig:dustinitialbeta} shows the expected dust
starting properties from collisions of asteroids in main belt in the Solar System.).

\section{Earth's Resonant Ring}
\label{sec:earthicandisk}

In addition to the general ring structure produced by resonantly trapped dust \citep{1998ApJ...508...44K}, a smaller pattern exists in which a `blob' of dust trails the Earth in its orbit \citep{1994Natur.369..719D,1995Natur.374..521R}.  Measurements of the North Ecliptic Pole sky brightness by the Spitzer Space Telescope, which is on an orbit that causes it to slowly drift away backwards along the Earth's orbit, have measured the azimuthal structure of this blob in 8$\mu$m emission \citep{2010Icar..209..848R}.  

\begin{figure}
  \centering
  {\includegraphics[width=0.49\textwidth,trim = 50 50 0 50, clip]{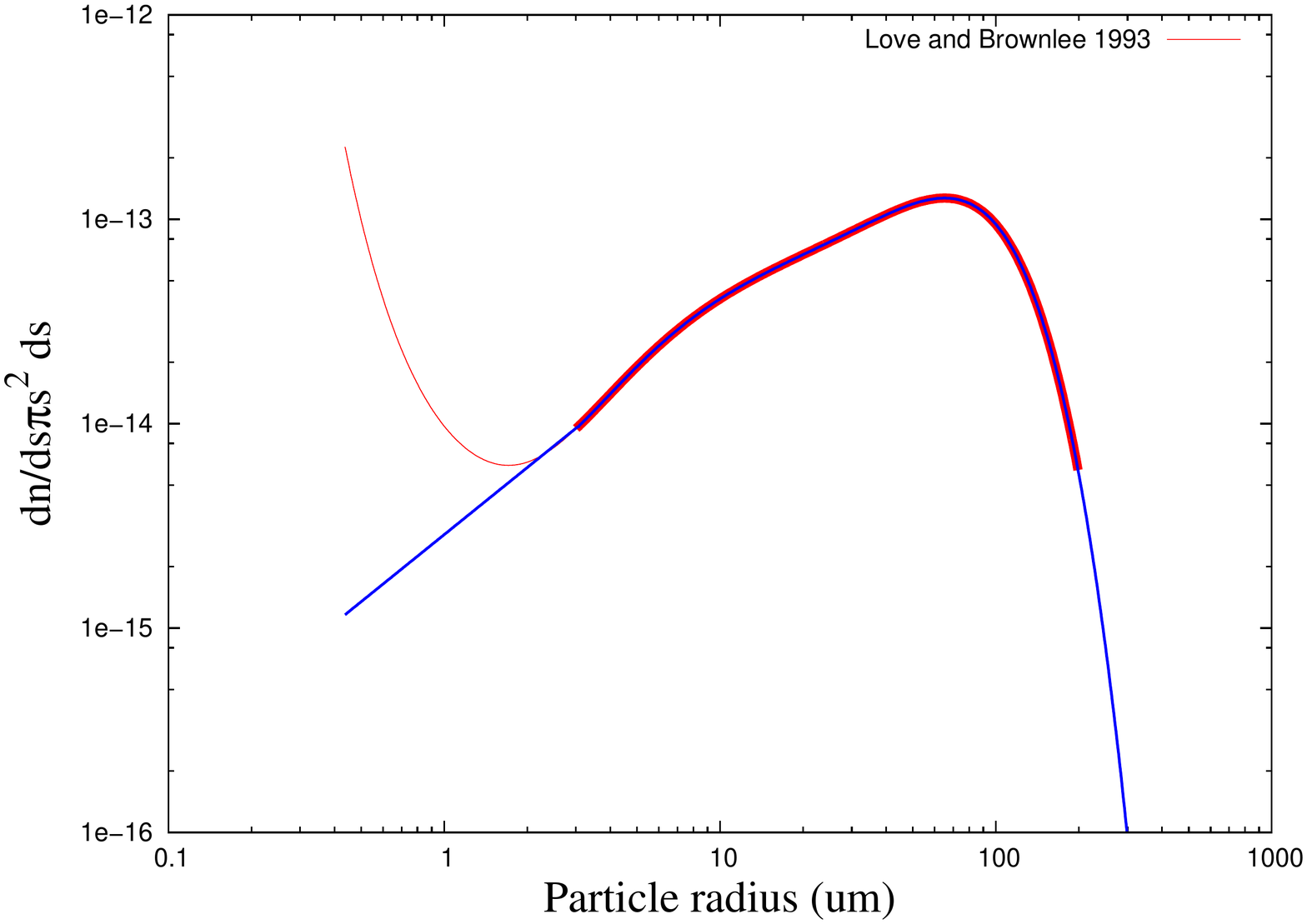}}
  \caption{Best fit size distribution of dust grains near the Earth, as measured by \citet{1993Sci...262..550L} (thin red line).  The thick red line segment is the size distribution over which their measurements were made.  As the upturn below $\sim 3 \mu \rm{m}$~is not attested from the data, we use an alternate size distribution at small sizes, the blue line (of intermediate thickness).}
  \label{fig:lovebrownlee}
\end{figure}

We apply our model to the structure of the trailing blob.  For the orbits of the parent bodies, we consider the first 10000 asteroids with orbital elements taken from the Minor Planet Centre's Orbit (MPCORB) Database.  We sample the size distribution at $\beta_{\rm{PR}} =$ 0.45, 0.4, 0.35, 0.3, 0.25, 0.18, 0.13, 0.088, 0.07, 0.063, 0.044, 0.031, 0.022, 0.016, 0.013, 0.011,~and $0.0055$.  The relative components are calculated separately, and added together to produce the spatial density of small grains.
This requires a size-number distribution of grains, and here we employ the result of \citet{1993Sci...262..550L}, who counted the craters from dust impacts on the space-facing side of the Long Duration Exposure Facility, and converted that to a size distribution of dust using experimental data on the impacts of dust grains onto aluminium targets and assumptions about the impact velocity.  We equate particle sizes with $\beta_{\rm{PR}}$~per equation \ref{eq:stobeta}~assuming $\rho = 2.5~\rm{g}~\rm{cm}^{-3}$.  Here a concern exists; \citet{1993Sci...262..550L} only measured crater sizes between $20 \mu\rm{m}$~and $1400 \mu \rm{m}$, corresponding to  $\beta \lesssim 0.07$~or $s \gtrsim 3\mu \rm{m}$.  At smaller sizes, their best fit size distribution employs a polynomial which turns up strongly in a way that may not be physical; we instead affix a power-law distribution between $s = 3 \mu\rm{m}$~and the blow-out size (figure \ref{fig:lovebrownlee}).

\begin{figure}
  \centering
  {\includegraphics[width=0.49\textwidth,trim = 50 50 0 50, clip]{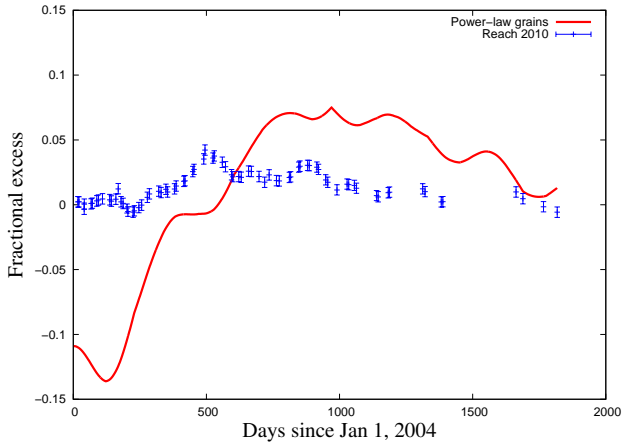}}
  \caption{Zodiacal cloud luminosity expected to be observed by Spitzer from our model of asteroidally produced dust (thick red line), compared to measurements by \citet{2010Icar..209..848R} (blue points).  A constant background with an annual sinusoidal modulation has been subtracted off; the remaining flux is normalised such that the component subtracted off has a mean amplitude of 1.  The model clearly does not match the data.
  }
  \label{fig:spitzerbad}
\end{figure}

With this we can then predict the light curve Spitzer would see on its actual path using our model disk.  We obtained Spitzer's trajectory from JPL's HORIZONS system \citep{1996DPS....28.2504G}.  Luminosities are calculated by treating the dust grains as spherical black bodies\footnote{With most of the emitting area in large grains (figure \ref{fig:lovebrownlee}), this should be fairly reasonable. For the smallest grains, this may get the temperature of the smallest grains wrong by less than a factor of 2 \citep{1993prpl.conf.1253B}.  Using the grain models of \citet{1999A&A...348..557A}, as implemented by \citet{2002MNRAS.334..589W}, we estimate the total flux as a function of size might be uncertain by a factor of a few at the smallest sizes, which would need to be included in a more detailed model.  But as they contribute relatively little flux, we do not attempt to model that here.}, and integrating the expected emission along the line of sight of Spitzer.  To compare with measurements of the trailing blob, we apply the same approach as \citet{2010Icar..209..848R}; a constant amplitude signal, with a sinusoidal modulation is fit to the data, and subtracted off.  We plot the residuals in figure \ref{fig:spitzerbad}.  The predicted amplitude of the signal is then much stronger than the observations.

\begin{figure}
  \centering
  {\includegraphics[width=0.49\textwidth,trim = 50 50 0 50, clip]{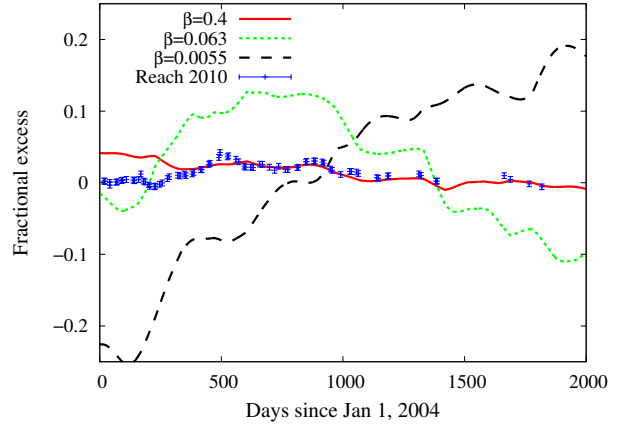}}
  \caption{Zodiacal cloud luminosity expected to be observed by Spitzer from our model of asteroidally produced dust of three different mono-size distributions ($\beta_{\rm{PR}} = 0.4, 0.063,$~and $0.0055$), compared to measurements by \citet{2010Icar..209..848R} (blue points).  A constant background has been subtracted off; the remaining flux is normalised such that the component subtracted off has an amplitude of 1.  From this, we can infer that the addition of more small grains, or grains that are not trapped into resonance for other reasons, could bring the model in figure \ref{fig:spitzerbad} into alignment with the data.
  }
  \label{fig:spitzerexample}
\end{figure}

\begin{figure}
  \centering
  {\includegraphics[width=0.49\textwidth,trim = 50 50 0 50, clip]{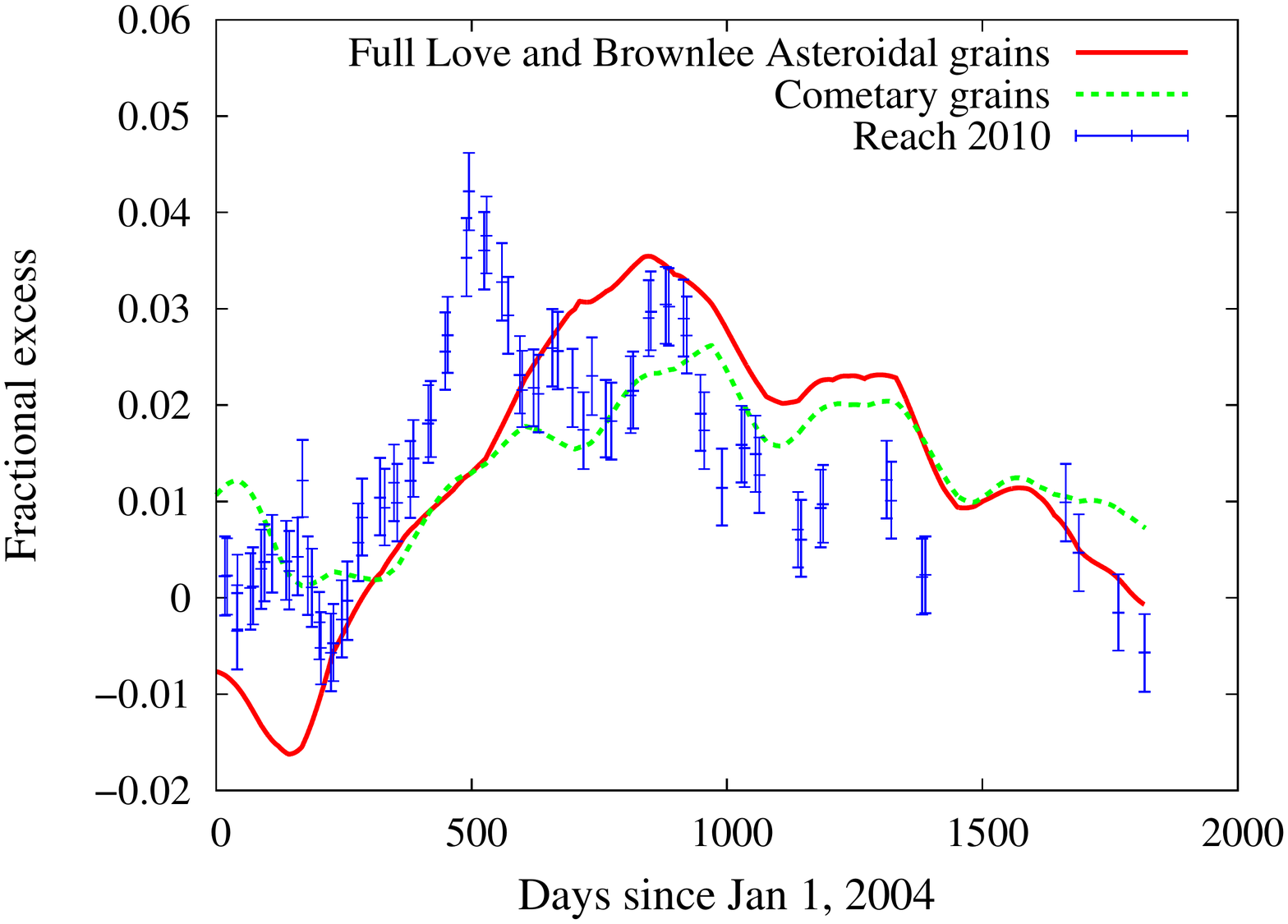}}
  \caption{Zodiacal cloud luminosity expected to be observed by Spitzer using the uncorrected dust size distribution fit of \citet{1993Sci...262..550L} assuming only asteroidally produced dust (solid red line), and using the corrected dust size distribution fit of \citet{1993Sci...262..550L}, but assuming 80\% of dust is produced by comets (dotted green line), compared to measurements by \citet{2010Icar..209..848R} (blue points).  The cometary model assumes all dust starts with the orbit of Halley's comet, and all dust is released at pericentre.  A constant background has been subtracted off; the remaining flux is normalised such that the component subtracted off has an amplitude of 1.  The model produces rough agreement with the data.
  }
  \label{fig:spitzergood}
\end{figure}

Detailed inspection of the contributions from different $\beta_{\rm{PR}}$~finds that the disagreement comes mostly from the larger grains, which are trapped at low $j$ (figure \ref{fig:spitzerexample}, and recall figure \ref{fig:allplots}).  This produces a large dip at the Earth, that rises to a flat level at later times.  As such, the model can be brought into rough agreement with the data by the addition of a smooth disk component.  We plot two such examples in figure \ref{fig:spitzergood}.  In one example, we simply use asteroidal grains, but do not correct the \citet{1993Sci...262..550L} polynomial fit to the size distribution at small sizes where the size distribution extrapolation is suspect.  In this case, the high migration rate of the high $\beta_{\rm{PR}}$~grains prevents their capture (recall figure \ref{fig:grid} - such grains have high $dB/dt$).  In another, as it has been suggested the zodiacal light may come mostly from cometary grains \citep{2010ApJ...713..816N}, we assume $80\%$~of the zodiacal light comes from cometary grains, which we model using Halley's comet as the parent body orbit, with all grains released at pericentre.  These grains are not captured due to their high eccentricity (again recall figure \ref{fig:grid} - such grains have high $J_0$).  Because of the high initial eccentricity, grains with $\beta_{\rm{PR}} > 0.01$~are produced with $e > 1.0$, and so only grains with $\beta_{\rm{PR}} \leq 0.01$~are used.  However, a more complicated population of parent bodies and ejection velocities would be needed for the cometary component to reproduce the \citet{1993Sci...262..550L} size distribution, or a model in which asteroids and comets contribute differently to the size distribution at different sizes.  Thus for now, this should be considered only a demonstration of principle, and we defer a more detailed model to a later paper.

\begin{figure}
  \centering
  {\includegraphics[width=0.49\textwidth,trim = 100 50 50 50, clip]{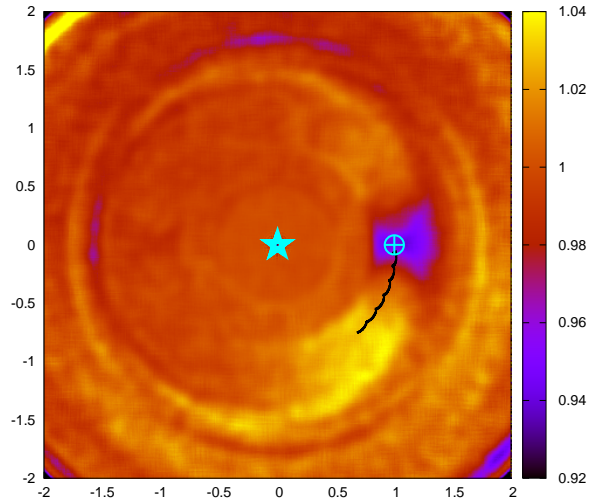}}
  \caption{
    Disk flux relative to the azimuthal average at the same radius in our 80\% cometary, 20\% asteroidal grains model.  The positions of the Sun and the Earth are plotted with $\bigstar$~and $\oplus$~respectively.  The path of Spitzer is plotted, showing its movement into Earth's trailing blob.  This plot also reveals that the trailing/leading asymmetry is the main asymmetry in the neighbourhood of the Earth, although a large asymmetry is just visible at $\sim 2.2$~au, the apocentre of low $\beta_{\rm{PR}}$~particles in the 2:1 resonance. 
  }
  \label{fig:spitzerpath}
\end{figure}

To aid in the visualisation of the structure of the ring and Spitzer's path through it, we plot the azimuthal asymmetries of an 80\% cometary grains and 20\% asteroidal grains disk in figure \ref{fig:spitzerpath}, along with the path of the Spitzer Space Telescope.

\section{Discussion}
\label{sec:conclusions}
Exozodiacal dust is a possible source of obscuration for future missions attempting the direct imaging of terrestrial exoplanets \citep{1999tpf..book.....B,2010A&A...509A...9D}.  However, capture of dust into resonance can allow for the detection of a planet's presence, even where direct imaging is impossible \citep{2003ApJ...588.1110K}.  Quick generation of disk images is likely to be key to determining planetary properties from the appearance of a resonantly captured ring.  In this work, we present a semi-analytic approach for calculating the evolution of dust particles moving past a planet due to PR drag.  This allows dust evolution to be calculated on the PR time-scale, rather than the orbital timescale, which allows orders of magnitude faster calculations than traditional N-body methods.  We compare this approach to results generated from N-body simulations, and find them to be an excellent match for terrestrial and super-Earth type planets ($m \lesssim 10 M_{\oplus}$), which radial velocity results \citep{2011arXiv1109.2497M} and transit results \citep{2010Sci...330..653H} find to be the most prevalent types of planets.  Larger planets capture many dust grains into the chaotic 2:1 resonance, for which our approach produces a reasonable match.  

We apply this model to dust produced in the Asteroid belt that is resonantly trapped as it passes the Earth.  Using the measured size distribution of dust grains near the Earth, the model does not match the observations, requiring an additional smooth component to the Zodiacal cloud to bring the model into agreement with the data.  This smooth component can be produced by comets, for instance, which release dust grains with much higher initial eccentricities.  Although other models can be considered, this could be taken as support for other lines of evidence that suggest the Zodiacal cloud is mostly produced by comets, not asteroids \citep{2010ApJ...713..816N}.

Application of this model is likely to find great utility in exozodiacal systems, where the decrease in computation time can make it more feasible to address the large parameter space uncertainly associated with the unknown orbits of parent bodies, and the unknown orbits and masses of planet.  This improvement in computation time for studying the dynamical aspects of the evolution will also make it feasible to consider more realistic prescriptions for the collisional evolution of the debris as it evolves from the source location.  While it was possible to ignore this for our study of the Solar system because the size-mass distribution of dust at the Earth has been measured directly, this would have to be calculated for the simulations for extrasolar systems.  We intend to address this question in an upcoming work, and explore the application of this model to measuring disk structure with repeated LBTI observations of zodiacal light, or possible direct imaging by future missions.

\section{Acknowledgements}
We thank James Harrod for assistance in setting up the project, Grant Kennedy for useful discussions, and the anonymous referee for useful suggestions.  AS and MW are supported by the European Union through ERC grant number 279973.  AJM acknowledges support from Spanish grant AYA 2010/20630, grant number KAW 2012.0150 from the Knut and Alice Wallenberg foundation, and the Swedish Research Council (grant  2011-3991).

\bibliographystyle{mn2e}
\bibliography{dustcapture}

\appendix

\section{Modifications to Mustill \& Wyatt 2011 for Radiation Forces}

To adapt MW11 for radiation pressure, we non-dimensionalise as in MW11 to put time in units of the planet's mean motion and length in units of the planet's semi-major axis.

The Poincare variables for the particle will take the form
\begin{eqnarray}
\Lambda&=&\sqrt{Ba}\\
\Gamma&=&\Lambda\left(1-\sqrt{1-e^2}\right)\sim\Lambda e^2/2,
\end{eqnarray}
for some value of $B$. The Keplerian part of the Hamiltonian will take the form
\begin{equation}
\mathscr{H}_\mathrm{Kep}=-\frac{A}{2\Lambda^2}
\end{equation}
for some $A$. These will not be the same as in the $\beta_{\rm{PR}}=0$ case because the radiation pressure changes slightly the location of the resonances, and hence the coefficients in MW11, but $A$ as a function of $\beta_\mathrm{PR}$ must tend to 1 as $\beta_\mathrm{PR}\to0$. We can exploit the fact that the equations of motion must be canonical, so that the mean motion
\begin{equation}
n=\frac{\partial\mathscr{H_\mathrm{Kep}}}{\partial\Lambda}=\frac{A}{\lambda^3},
\end{equation}
but we know that also
\begin{equation}
n=\sqrt{(1-\beta_\mathrm{PR})/a^3}
\end{equation}
and so
\begin{equation}\label{eq:AB}
A^2B^{-3}=1-\beta_\mathrm{PR}.
\end{equation}

Now we expand the Keplerian Hamiltonian. Let $\Lambda=\Lambda_0$ at the resonance location
\begin{equation}
 a_0=(1-\beta_\mathrm{PR})^{1/3}\left(\frac{j+1}{j}\right)^{2/3},
\end{equation}
so
\begin{equation}
 \Lambda_0=B^{1/2}(1-\beta_\mathrm{PR})^{1/6}\left(\frac{j+1}{j}\right)^{1/3}.
\end{equation}
Expand about the resonance location with $\Lambda=\Lambda_0+I$
\begin{equation}
 \mathscr{H}_\mathrm{Kep}=-\frac{A}{2\Lambda_0}+\frac{A}{\Lambda_0^3}I-\frac{3A}{2\Lambda_0^4}I^2
\end{equation}
and drop the first term as it is constant. Then
\begin{eqnarray} \nonumber
 \mathscr{H}_\mathrm{Kep}&=&\frac{A}{2}B^{-3/2}(1-\beta_\mathrm{PR})^{-1/2}\frac{j}{j+1}I \\
  &&- \frac{3A}{2B^2}(1-\beta_\mathrm{PR})^{-2/3}\left(\frac{j}{j+1}\right)^{4/3}I^2. 
\end{eqnarray}

The resonant term will have the form
\begin{equation}
 \mathscr{H}_\mathrm{res}=r\Gamma^{1/2}\cos[(j+1)\lambda-j\lambda_\mathrm{pl}-\varpi]
\end{equation}
for some $r$, with $\lambda$ and $\varpi$ the mean longitudes and longitudes of periastron. We need to set the value of $r$.  Lagrange's Equations for a first order resonance give \citep{2000ssd..book.....M}
\begin{equation}
 \frac{de}{dt}=-\frac{\mu e}{na_0^3}(f_{31}-\frac{a_0^2}{2})\sin[(j+1)\lambda-j\lambda_\mathrm{pl}-\varpi],
\end{equation}
For symplecticity we require
\begin{equation}
 \frac{d\Gamma}{dt}=-\frac{\partial\mathscr{H}_\mathrm{res}}{\partial{\theta}}=r\Gamma^{1/2}\sin\theta.
\end{equation}
But from the definition of $\Gamma$ we also have
\begin{equation}
 \frac{d\Gamma}{dt}=\Lambda e\frac{de}{dt},
\end{equation}
and substituting in $de/dt$ and equating coefficients gives
\begin{equation}
 r=-2^{1/2}B^{7/4}\mu f_{31}a_0^{-5/4}A^{-1}+2^{-1/2}B^{7/4}a_0^{3/4}\mu A^{-1},
\end{equation}
which is Eq A7 of MW11 with some extra factors of $A$ and $B$.

Here we impose $A=1$, since it is always possible to scale a Hamiltonian so long as time is also rescaled to maintain symplecticity, and here we have the correct time scaling via the mean motion. Fixing $A=1$ gives $B=(1-\beta_\mathrm{PR})^{-1/3}$ and hence the full Hamiltonian
\begin{eqnarray} \nonumber
 \mathscr{H}&=&\frac{(1-\beta_\mathrm{PR})^{\frac{3}{2}}}{a_0^{3/2}}I-\frac{3(1-\beta_\mathrm{PR})^{\frac{2}{3}}}{2a_0^2}I^2\\ \nonumber
 &&-\left[\frac{2^{\frac{1}{2}}\mu f_{31}(1-\beta_\mathrm{PR})^{-\frac{7}{12}}}{a_0^{\frac{5}{4}}}-2^{-\frac{1}{2}}(1-\beta_\mathrm{PR})^{-\frac{7}{12}}a_0^{\frac{3}{4}}\mu \right] \\ 
 &\times& \Gamma^{1/2}\cos[(j+1)\lambda-j\lambda_\mathrm{pl}-\varpi]
\end{eqnarray}
which is the revised Eq A14 from MW11. Next we make the usual transformations
\begin{eqnarray}
 \Gamma=J_1,&&I=jJ_1+J_2,\\
 \theta_1=(j+1)\lambda-j\lambda_\mathrm{pl}-\varpi,&&\theta_2=\lambda .
\end{eqnarray}
The Hamiltonian becomes
\begin{equation}
 \mathscr{H}=-aJ_1^2+bJ_2-rJ_1^{1/2}\cos\theta_1 ,
\end{equation}
where
\begin{equation}
 a=\frac{3}{2}(1-\beta_\mathrm{PR})^{1/3}\alpha_0^2 ,
\end{equation}
and
\begin{equation}
 b=\frac{1}{2}(1-\beta_\mathrm{PR})^{1/2}\alpha_0^{3/2}(j+1)-3(1-\beta_\mathrm{PR})^{2/3}\alpha_0^2J_2.
\end{equation}
This latter determines the migration rate.

We now make the usual transformations
\begin{eqnarray}
 \theta^\prime&=&\pi-\theta ,\\
 J^\prime&=&XJ_1 ,\\
 t^\prime&=&Yt ,\\
 b^\prime&=&Zb ,\\
 \mathscr{H}^\prime&=&-W\mathscr{H} ,
\end{eqnarray}
where comparison of coefficents and enforcement of the canonical condition on the equations of motion requires
\begin{eqnarray}
 W&=&a^{1/3}r^{-4/3} ,\\
 X&=&a^{2/3}r^{-2/3} ,\\
 Y&=&a^{1/3}r^{2/3} ,\\
 Z&=&-a^{-1/3}r^{-2/3} ,
\end{eqnarray}
or, explicitly,
\begin{eqnarray} \nonumber
 W&=&3^{1/3}2^{-1/3}(j+1)^{2/3}(1-\beta_\mathrm{PR})\alpha_0^{-1}\mu^{-4/3}\times \\
 &&\left[2^{1/2}f_{31}-2^{-1/2}\alpha_0^{-2}\mathbb{I}(2:1)\right]^{-4/3} ,\\ \nonumber
 X&=&3^{2/3}2^{-2/3}(j+1)^{4/3}(1-\beta_\mathrm{PR})^{5/6}\alpha_0^{1/2}\mu^{-2/3}\times \\
 &&\left[2^{1/2}f_{31}-2^{-1/2}\alpha_0^{-2}\mathbb{I}(2:1)\right]^{-2/3} ,\\ \nonumber
 Y&=&3^{1/3}2^{-1/3}(j+1)^{2/3}(1-\beta_\mathrm{PR})^{-1/6}\alpha_0^{3/2}\mu^{2/3}\times \\
 &&\left[2^{1/2}f_{31}-2^{-1/2}\alpha_0^{-2}\mathbb{I}(2:1)\right]^{2/3} ,\\ \nonumber
 Z&=&-3^{-1/3}2^{1/3}(j+1)^{-2/3}(1-\beta_\mathrm{PR})^{1/6}\alpha_0^{-3/2}\mu^{-2/3}\times \\
 &&\left[2^{1/2}f_{31}-2^{-1/2}\alpha_0^{-2}\mathbb{I}(2:1)\right]^{-2/3}, 
\end{eqnarray}
where $\mathbb{I}(2:1) = 1$~if $j = 1$, and $0$~otherwise.

With $J$~and $b$~now defined, we can calculate the initial value of $J$, $J_0$, and the migration rate $db/dt$, which allows us to apply the capture probabilities, non-capture eccentricity kicks, and the distributions of initial libration widths, found by MW11.

\begin{figure*}
 \centering
  \subfigure{\includegraphics[width=0.49\textwidth,trim = 0 0 120 350, clip]{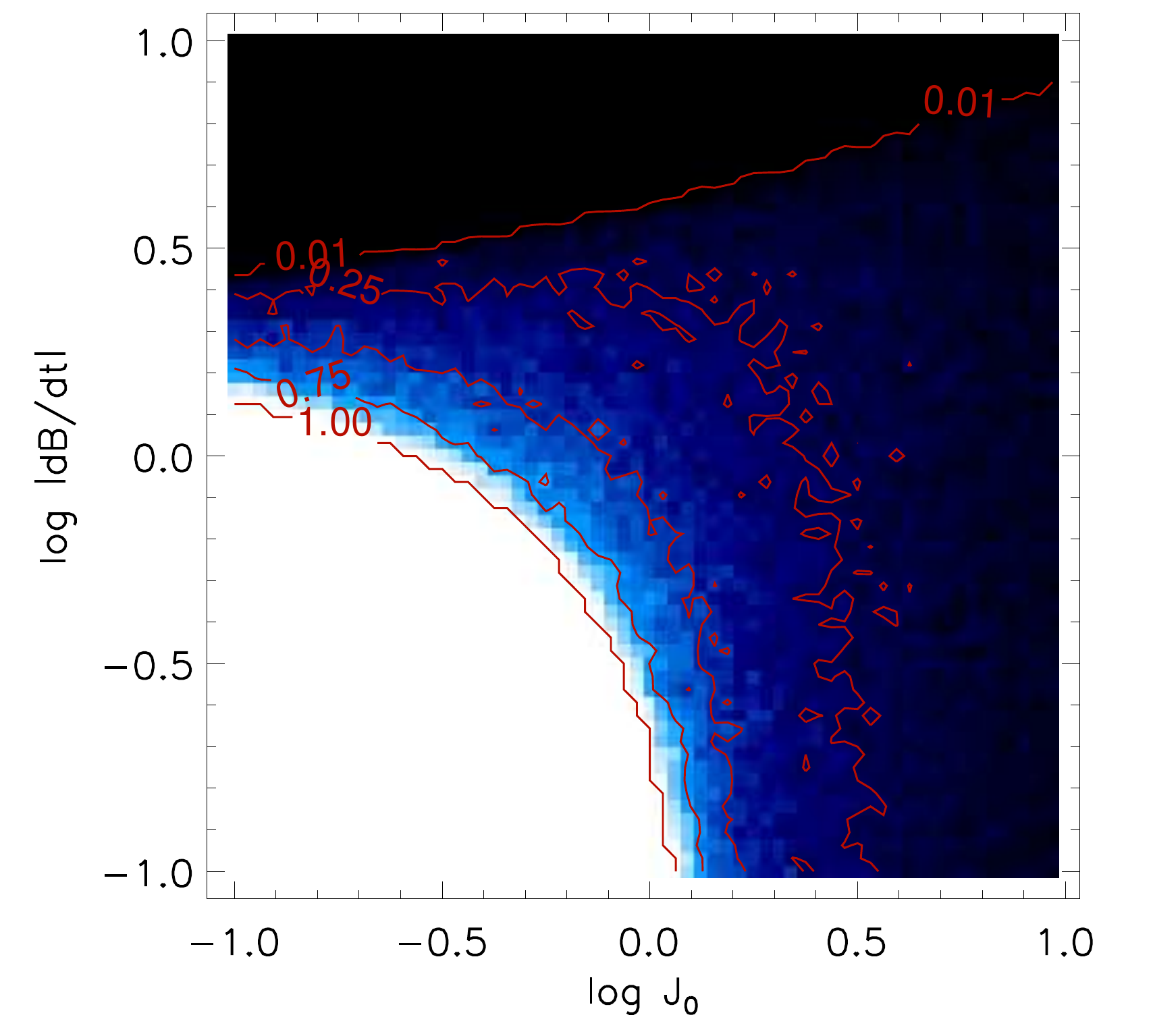}}
  \subfigure{\includegraphics[width=0.49\textwidth,trim = 0 0 120 350, clip]{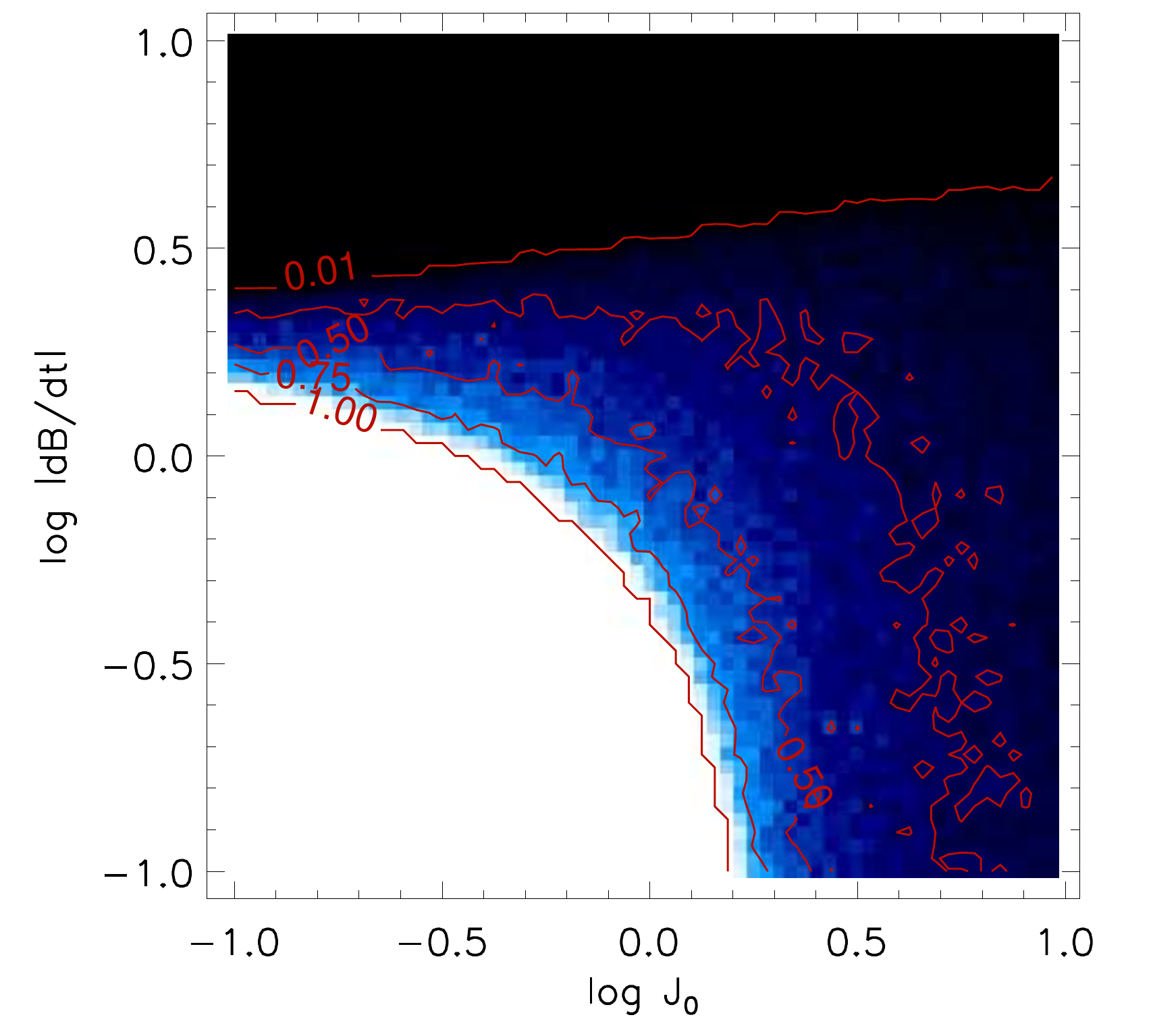}}
  \caption{Capture probability grids incorporating eccentricity damping with $k=5\times10^{-5}$ (left) and $k=0.02$ (right).}
  \label{fig:p-damp}
\end{figure*}

A further possible improvement over the MW11 model is the incorporation of eccentricity damping into the Hamiltonian model. Under PR drag, $\dot e/e=1.25\dot a/a$. Since the change in $a$ when crossing a resonance is small, eccentricity does not decay greatly during the capture process; nevertheless, we include this effect for completeness. Eccentricity damping is incorporated into the model by the addition of a term
\begin{eqnarray}
 \dot J&=&\frac{5}{2}\frac{\dot a}{a}\frac{dt}{dt^\prime} ,\\ \nonumber
 &=&\frac{5}{2l_jg_j}\left(\frac{m_\mathrm{pl}}{m_\oplus}\right)^{2/3}\times \\
 &&\left(\frac{M_\star}{M_\odot}\right)^{-2/3}\frac{a_\mathrm{pl}}{au}(1-\beta_\mathrm{PR})^{-2/3}\dot b J ,\\
 &=&k\dot bJ .
\end{eqnarray}
This introduces a third dimension to the parameter space. Rather than explore this thoroughly, we integrated two grids in $J_0-db/dt$ space, one at $k=5\times10^{-5}$ and one at $k=0.02$. These roughly correspond to the 2:1 resonance for $m_\mathrm{pl}=0.1m_\oplus$ and $1000m_\oplus$ respectively. Despite the large difference in damping strengths, the capture probability is only weakly dependent on this (Fig~\ref{fig:p-damp}). The eccentricity damping shifts the contours of the plot to the left, as some momentum $J\propto e^2$ is lost prior to resonance passage.  As we do not notice a significant effect, we do not add this complexity to our model.

\end{document}